\documentclass{emulateapj}
\usepackage{graphicx}
\usepackage{subfigure}

\newcommand{\secpoint}{\mbox{$''\mskip-7.6mu.\,$}}

\newcommand{\lya}{Ly$\alpha$}
\newcommand{\elyc}{$\epsilon_{{LyC}}$}
\newcommand{\elyca}{$\epsilon_{LyC}^{LBG}$}
\newcommand{\elycb}{$\epsilon_{LyC}^{LAE}$}

\newcommand{\fdr}{$\eta$}
\newcommand{\elbg}{$\eta_{LBG}$}
\newcommand{\elae}{$\eta_{LAE}$}
\newcommand{\fesc}{$f_{esc}^{LyC}$}
\newcommand{\fuv}{$f_{esc}^{UV}$}

\shorttitle{LYMAN-CONTINUUM EMISSIVITY AT $z=3.1$}
\shortauthors{NESTOR ET AL.}

\begin{document}

\title{A Refined Estimate of the Ionizing Emissivity from Galaxies at
  $z\simeq3$: Spectroscopic Follow-up in the SSA22a Field\altaffilmark{1}}

\author{\sc Daniel B. Nestor, Alice E. Shapley\altaffilmark{2},
  Katherine A. Kornei}
\affil{Department of Astronomy, University of California,
Los Angeles, 430 Portola Plaza, Los Angeles, CA 90095}

\author{\sc Charles C. Steidel}
\affil{California Institute of Technology, MS 105--24, 
Pasadena, CA 91125}

\author{\sc Brian Siana}
\affil{Department of Physics and Astronomy, University of California,
Riverside, CA 92521}

\altaffiltext{1}{Based, in part, on data obtained at the 
W.M. Keck Observatory, which is operated as a scientific 
partnership among the California Institute of Technology, 
the University of California, and NASA, and was made 
possible by the generous financial support of the W.M. Keck 
Foundation.}
\altaffiltext{2}{David and Lucile Packard Fellow}

\begin{abstract}
We investigate the contribution of star-forming galaxies to the
ionizing background at $z\sim 3$, building on previous work based on
narrowband (NB3640) imaging in the SSA22a field. We use new
Keck/LRIS spectra of Lyman break galaxies (LBGs) and
narrowband-selected \lya\ emitters (LAEs) to measure redshifts for
16 LBGs and 87 LAEs at $z > 3.055$, such that our NB3640 imaging
probes the Lyman-continuum (LyC) region. When we include the existing set
of spectroscopically-confirmed LBGs, our total sample with $z>3.055$
consists of 41 LBGs and 91 LAEs, of which nine LBGs and 20 LAEs are
detected in our NB3640 image. With our combined imaging and
spectroscopic data sets, we critically investigate the origin of
NB3640 emission for detected LBGs and LAEs. We remove from our
samples 3 LBGs and 3 LAEs with spectroscopic evidence of
contamination of their NB3640 flux by foreground galaxies, and
statistically model the effects of additional, unidentified
foreground contaminants. The resulting contamination and
LyC-detection rates, respectively, are $62\pm13$\% and $8\pm3$\% for
our LBG sample, and $47\pm10$\% and $12\pm2$\% for our LAE sample.
The corresponding ratios of non-ionizing UV to LyC flux-density,
corrected for intergalactic medium (IGM) attenuation, are
$18.0^{+34.8}_{-7.4}$ for LBGs, and $3.7^{+2.5}_{-1.1}$ for LAEs. We
use these ratios to estimate the total contribution of star-forming
galaxies to the ionizing background and the hydrogen photoionization
rate in the IGM, finding values larger than, but consistent with,
those measured in the \lya\ forest. Finally, the measured UV to LyC
flux-density ratios imply model-dependent LyC escape fractions of
$f_{esc}^{LyC} \sim 5-7$\% for our LBG sample and $f_{esc}^{LyC}
\sim 10-30$\% for our fainter LAE sample.
\end{abstract}

\keywords{galaxies: high-redshift -- intergalactic medium -- 
cosmology: observations -- diffuse radiation}

\section{INTRODUCTION}    \label{sect:intro}

Identifying the sources of the radiation that re-ionized the
intergalactic medium (IGM) at redshift $z\ga 6$ is a key challenge for
observational cosmology.  Massive stars
in star-forming galaxies are considered the most
likely source of the Lyman-continuum (LyC) photons necessary for
reionization.  QSOs, while also efficient producers of ionizing radiation,
appear to be too rare at high redshift to produce sufficient ionizing
flux \citep{hopkins2007,cowie2009}.   Unfortunately, direct observation of ionizing radiation
from high-redshift star-forming galaxies is not possible, as the
Universe remains opaque to 
LyC photons above redshifts $z\sim 4$ due to the abundance of residual
neutral gas in the IGM.   Thus, empirical constraints on the sources
responsible for reionization come primarily from (i) the
determination of the non-ionizing UV luminosity function of the
highest-redshift galaxies currently observable, and
(ii) direct measurements of ionizing radiation from
lower-redshift analogs of the $z \ga 6$ galaxies
responsible for reionization. 

Progress has recently been made in both of the above approaches.
Bright, color-selected $z\sim 6$ - 7 candidate galaxies have recently 
been spectroscopically confirmed
\citep[e.g.,][]{pentericci2011,ono2012,schenker2012}.  Searches for fainter, line-emitting
galaxies using narrowband 
imaging techniques have uncovered significant samples of Ly$\alpha$
emitters (LAEs) at redshifts as high as $z\sim7$
\citep[e.g.,][]{krug2012,ota2010,ouchi2010}.  
At even higher redshift,
deep near-IR observations of the Hubble Ultra Deep Field using the
WFC3 camera on the Hubble Space Telescope ($HST$),
combined with optical $HST$/ACS data, have revealed populations of
redshift $z\sim 7-10$ galaxies \citep[e.g.,][]
{bouwens2010,bunker2010,lorenzoni2011,mclure2011,vanzella2011,wilkins2011,oesch2012,trenti2012}.
Among the results of
such studies, the emerging $z\sim8$ luminosity functions
suggest that there are too few bright galaxies
to reionize the IGM, implying that relatively low-mass galaxies are
important contributors to the ionizing flux budget.  

The effort to unambiguously identify low-redshift analogs
of the galaxies responsible for reionization has been difficult.
Observations to directly detect ionizing
continuum escaping from galaxies at $0 < z \la 2$ have been unsuccessful
\citep[e.g., ][]{malkan2003,cowie2009,grimes2009,siana2010,bridge2010}.  Thus, searches have
turned to the highest redshifts at which the opacity to 
LyC photons through the IGM, $\tau_{LyC}$, is $\la 1$, i.e., $z\sim3$.  The
expected rest-frame UV flux levels are low at such
redshifts, requiring deep observations to identify sources and to
detect escaping LyC emission.  The low flux levels also complicate the
interpretation of presumed detections of LyC flux, as the sky surface
density of faint foreground sources is large ($\sim 75$ arcmin$^{-1}$), leading to
the possibility of contamination from lower-redshift sources.  With these
caveats in mind, there
have been several reported detections of escaping LyC
flux from galaxies at $z\sim3$ through both spectroscopy of Lyman
break galaxies (LBGs) \citep{steidel2001,shapley2006}
 and narrowband imaging of LBGs and
LAEs \citep{iwata2009,nestor2011,vanzella2011}.  

In \citet{nestor2011}, we searched for 
escaping LyC flux from $z\ge 3.055$ LBGs and photometrically-selected LAE
candidates in the SSA22a field, which contains a large over-density of
galaxies at $z=3.09$ \citep{steidel1998}.  In that paper we
reported the detection of six $z \ge 3.09$ LBGs and 27 candidate
$z\simeq3.09$ LAEs through a narrowband filter that is opaque to
non-ionizing flux from sources at $z\ga3.055$, and thus probes light
blueward of the Lyman limit for sources at $z\simeq3.09$.  We interpreted these
detections  as direct evidence of escaping LyC flux, which we in turn used to 
estimate the comoving ionizing emissivity, \elyc, at $z\sim3$.  Our
primary conclusions were that the contribution to \elyc\ from star-forming
galaxies at $z\sim3$ exceeds, but is roughly consistent with, given our
uncertainty, that expected from
determinations of the photoionization rate in the \lya\ forest
\citep[e.g.,][]{bolton2007,faucher2008}, and that 
the bulk of this contribution comes from relatively faint sources
($M_{AB} \ga -20$) such as those that compose our sample of LAEs.  

Our ability to
constrain \elyc, however, was limited by our fairly small sample of LBG detections,
which is very likely to contain contamination by foreground
interlopers \citep[see, e.g.,][]{vanzella2012}.  
As our seeing-limited observations restricted our ability to account for
contamination by foreground galaxies in individual systems, we
used a statistical approach for the
sample as a whole.  Furthermore, many of the 
results in \citet{nestor2011} were based on the sample of LAEs, the
majority of which were not spectroscopically confirmed.  
Each of these limitations can be alleviated with
follow up spectroscopy.  Therefore, in order to increase the size of
our LBG sample, we obtained spectra of additional color 
selected LBG candidates lacking prior spectroscopic redshift
confirmation.  We also obtained 
spectra of as many of our narrowband-selected LAE candidates as
possible, in order to confirm their redshifts.  These data have the additional
benefit that, in most cases, we can directly search for evidence of
foreground contamination in individual objects.  We also obtained
particularly deep spectra of many of the sources having apparent LyC detections, 
which allow us to perform a detailed analysis (e.g., of the spatial distribution) of the detected
\lya, non-ionizing UV, and LyC fluxes.

The organization of this paper is as follows.  In
Section~\ref{sect:data} we describe the observations and 
reduction of the data used in this study.  We update our LBG and LAE
samples based on the results of 
our new spectroscopy in Section~\ref{s:sample}.  In Section~\ref{s:det}
we discuss individual LBGs and LAEs with presumed LyC detections.  In
Section~\ref{s:monte} we describe our techniques for statistically
accounting for foreground contamination and attenuation of LyC flux by
the IGM.  We present updated estimates of \elyc\ and discuss LyC escape fractions
in Section~\ref{s:results}, and summarize our results in Section~\ref{s:summary}.
Throughout the paper all magnitudes are in the AB system, and we
assume a cosmology with $\Omega_m=0.3$, $\Omega_{\Lambda}=0.7$, 
and $H_0= 70$~km~s$^{-1}$~Mpc$^{-1}$.

\section{OBSERVATIONS AND DATA REDUCTION}    \label{sect:data}

\subsection{Photometric Observations and Sample} \label{sect:data-phot}

Our analyses make use
of deep multiband imaging available in the SSA22a field.  These data,
which are described in 
detail in \citet{nestor2011}, include ground-based broad $B$-, $V$-
and $R$-band images and narrowband images with effective wavelengths
at $\lambda \sim$3640\AA\ 
and $\lambda \sim$4980\AA\ (hereafter NB3640 and NB4980, respectively).
The NB3640 and part of the NB4980 data were obtained using the
Keck/LRIS imaging spectrograph, while the broadband imaging and the
remainder of the NB4980 data were obtained with the Subaru/Suprime-Cam.
Additionally, archival $HST$/ACS F814W imaging is available for
70\% of the field.  

The NB3640
filter has a central wavelength of 3635\AA\ and FWHM of 100\AA.  For
sources at $z\simeq3.09$, NB3640 samples the rest-frame spectral range
$\lambda \simeq 875 - 900$\AA.  It is opaque to wavelengths longward of
the redshifted Lyman 
limit for sources above $z\simeq3.055$ and therefore provides a clean
probe of escaping LyC emission for such galaxies.
The effective wavelength of the broad $R$-band filter ($\lambda \simeq
6510$\AA) corresponds 
to rest-frame $\lambda \simeq 1600$\AA\ at $z\simeq3.09$.  
The NB3640$-R$ color is accordingly a measure of the non-ionizing to
ionizing UV flux-density ratio, $F_{UV}/F_{LyC}$, for galaxies with 
$z\ge3.055$.  The ACS/WFC F814W filter has an effective wavelength of
$\lambda \simeq 8090$\AA\ and also probes the non-ionizing UV
continuum at $z\sim3.09$.
The footprint of our NB3640 image contains 109 color-selected LBG
candidates, of which 28 (including 2 QSOs) have
previous spectroscopically confirmed redshifts $z\ge3.055$, as well as 41 LBG
candidates without previous spectroscopic confirmation.  The footprint
also contains 110 LAE candidates identified by
\citet{nestor2011}. 

LAE candidates were selected using the $B$, $V$, and NB4980 data.
The NB4980 filter covers redshifted HI \lya\  $\lambda1216$\AA\
for galaxies with $3.05 \lesssim z \lesssim3.12$.  We created a linear
combination of the broad $B$ and $V$ images after scaling to a common
photometric zero-point, such that our so-called ``$BV$" image has an
effective wavelength of $\lambda\simeq 4980$\AA.   The
$BV-\mathrm{NB4980}$ color is therefore a measure of \lya\ equivalent width for galaxies
with $3.05 \lesssim z \lesssim3.12$.  The LAEs in the sample described by
\citet{nestor2011} were required to have NB4980 $\le26$ and
$BV-\mathrm{NB4980} \ge0.7$, which corresponds to a \lya\ rest-frame equivalent
width (REW) of $\gtrsim20$\AA.  We also subtracted the $BV$
image from the NB4980 image to create a so-called ``$LyA$'' image,
which was used to define the centroids of \lya\ emission from the
LAE candidates.   

\subsection{Spectroscopic Observations} \label{sect:data-spec}

We performed spectroscopy in the SSA22a field using both
shallow and deep observations. The purpose of the shallow observations
was to obtain spectroscopic redshifts for LBG and LAE photometric
candidates, while the deep observations were intended to provide
detailed information for objects with NB3640 detections.
Multi-object spectroscopy was performed using the blue arm
of the LRIS dichroic spectrograph on Keck~I \citep{oke1995,steidel2004}.
Nine shallow slit masks were observed over the course
of five observing runs in 2009 June, 2009 September, 
2010 July, 2010 August, and 2011 May. A single
deep mask was observed in 2010 August. Typical
exposure times for the shallow masks were 5400 seconds
(ranging from 3600 to 6270 seconds), while the deep mask
was observed for 23700 seconds. For most runs, the 
conditions were photometric, with seeing ranging
from 0\secpoint5 to 0\secpoint7. In 2011 May,
when additional data were collected for one of the shallow 
masks, conditions featured variable cloudiness and
seeing ranging from 0\secpoint8 to 1\secpoint0.

The primary targets for shallow masks were LBG and LAE
photometric candidates without previously-determined
spectroscopic redshifts. LBGs with spectroscopic redshifts
were added as filler targets on the mask. A total of 50 LBGs and 114 LAEs
were targeted on the shallow masks. Slits were centered
on the detections in the NB4980 (i.e., Ly$\alpha$) for
LAEs, and in the $R$-band (i.e., rest-frame UV continuum)
for LBGs. Seven out of nine shallow masks were observed
using a 300 line~mm$^{-1}$ grism blazed at 5000~\AA,
along with the ``d680" dichroic beam splitter, sending
light with wavelengths bluer than $\sim 6800$\AA\
to the blue arm of LRIS. The spectral resolution
for these masks was $R=530$. The two remaining shallow masks
were observed at higher spectral resolution using
a 600 line~mm$^{-1}$ grism blazed at 4000~\AA,
one with the ``d560" dichroic (splitting the incoming
light beam at $\sim 5600$~\AA), and the other 
with the ``d680" dichroic. The spectral
resolution for these two masks was $R=1200$.

The primary targets for the deep mask were LBGs and LAEs
with NB3640 detections. A sample of 4 LBGs and 13 LAEs
with NB3640 detections were targeted on the deep mask
(along with one LBG and 4 LAEs lacking NB3640 detections,
but added as filler). Slits for objects with NB3640 detections
were centered on the detections in the NB3640
image (i.e., LyC for objects at $z\geq 3.055$), 
while the NB4980 and $R$-band detections were used,
respectively, for centering the slits for the filler LAEs 
and LBG without NB3640 detections.
The deep mask was observed using a 400 line~mm$^{-1}$ grism blazed at 3400~\AA,
along with the ``d680" dichroic beam splitter, sending
light with wavelengths bluer than $\sim 6800$\AA\
to the blue arm of LRIS. The spectral resolution for this
mask was $R=700$.

The data were primarily reduced using IRAF tasks,
with scripts designed for cutting up the multi-object
slit-mask images into individual slitlets, flat-fielding
using spectra of the twilight sky, rejecting cosmic
rays, subtracting the sky background, averaging
individual exposures into final stacked two-dimensional
spectra, extracting to one dimension, wavelength and
flux calibrating, and shifting into the vacuum
frame. These procedures are described in detail
in \citet{steidel2003}. There were a couple of
notable differences in the data reduction
procedures used for this sample, relative
to the typical LBG reduction strategy. First,
we used a custom IDL script (N. Reddy 2010, private
communication) to rectify the curved slitlets before
performing any of the standard IRAF reduction tasks.
Since the majority of our targets are LAEs
with negligible continuum and a single emission
line, we required slit rectification in order to extract
a spectrum over a broad wavelength range 
at the location of the object indicated
by the isolated bright Ly$\alpha$ feature.
Likewise, for deep-mask spectra, the faintness of the continuum level in the LyC region precluded
a robust trace without rectification. We also followed
the procedures outlined in \citet{shapley2006} for background
subtraction of deep-mask spectra. Accordingly, to avoid potential
over-subtraction of the background, the object continuum location
was excluded from the estimate of the background fit at each dispersion
point.  We used the maximum possible number of pixels to fit the sky
emission for each object.  In practice, the widths of the sky regions
on either side of the continuum location depended
on the length of each slitlet and the position of the object along the
slit. 

For LAEs, redshifts were calculated from the observed centroid of the Ly$\alpha$
emission feature (rest-frame $\lambda_{Ly\alpha}=1215.67$~\AA). For LBGs,
emission redshifts were estimated from the observed centroid of Ly$\alpha$,
and absorption redshifts from the centroids of interstellar metal absorption features
when present.

Finally, as described in Siana et al.\ (in preparation), near-IR spectra
of several objects in the sample were obtained in 2011 August with
NIRSPEC \citep{mclean1998} on Keck~II.  These data
were collected for a separate project; we refer to these data here as,
in three cases, the near-IR spectra reveal emission 
features relevant to the interpretation of the NB3640 detections (see
Sections~\ref{s:lbg_contam} and \ref{s:aug96M16}).

\section{THE UPDATED LBG AND LAE SAMPLES}
\label{s:sample} 

The \citet{nestor2011} sample contained 26 LBGs with spectroscopic
redshifts $z\ge3.055$.  Our new data include spectra of 11 LBGs and 1 QSO
with previously determined redshifts, and 41
candidate LBG sources.  For each of 
the 12 re-observed galaxies, the redshifts determined from our new
data agree with the previous determinations within $\Delta z=0.012$.
The spectrum of one of the newly observed LBGs candidates reveals it
to be a star.  Of the 40 remaining galaxies, we were able to
determine redshifts for 26 objects by identifying \lya\ emission
and/or interstellar metal absorption features.  All 26 galaxies have
$z>2.45$, 16 of which have $z>3.055$.  Thus, our updated sample of LBGs with
redshifts $z\ge3.055$ contains 42 galaxies (26 previously known and 16 new).
For this sample, the NB3640 filter probes the redshifted
LyC spectral region.  Coordinates, redshifts, and $R$-band photometry
are listed in Table~\ref{t:one} for  
the 42 $z>3.055$ sources, and in Table~\ref{t:2} for the
10  $z<3.055$ newly confirmed LBGs.

\citet{nestor2011} identified 110 LAE candidates having
$BV-\mathrm{NB4980}\ge 0.7$ and $\mathrm{NB4980}\le26$.  Our new data include spectra
of 96 of the 110 LAE candidates.  Four of the LAEs for which we
did not acquire new spectra are also LBGs with previously determined
spectroscopic redshifts: C4, C28, M28 and MD23.  
The other 10 LAE candidates for which we did not acquire
a spectrum are relatively faint in NB4980 ($25.25 \la \mathrm{NB4980} \la 26$)
but are randomly distributed in $BV-\mathrm{NB4980}$ color.  Thus, this
incompleteness should not bias our results.  We detect an emission line in the
expected spectral region, $\approx 4935$\AA\ $- 5015$\AA, in the spectra of 
88 of the candidates.  One of the 88 is D3, which is also in our LBG sample with a
previous spectroscopic redshift, and another two, C9 and M13, were LBG candidates
that are now spectroscopically confirmed members of our LBG sample.
Thus, there are seven objects that appear in both our LBG and LAE samples.

In principle, some of the emission lines detected in our spectra could
be [OII]$\lambda3727$, or in some cases 
H$\beta$ or [OIII]$\lambda5007$, from very faint lower-redshift
systems.  However, 76 of the 88 (86\%) of the sources with a
line detection have spectra with sufficient spectral coverage that,
if the detected line 
was [OII], H$\beta$, or [OIII] at lower redshift, at least one of the
other rest-frame optical features should have been detected as well.  In no cases
do we detect such corresponding lines.  Furthermore, in a similar
LAE survey at $z=3.1$ in the Extended Chandra Deep Field--South,
\citet{gronwall2007} argue that, given the relatively small volume
covered by the narrowband filter at $z=0.34$ relative to $z=3.1$ and
the rarity of [OII] emitting galaxies with REW above their threshold
(REW$>60$\AA\ at $z=0.34$), their level of
contamination by low-$z$ [OII] emitters is negligible.  As our filter is
$\sim60$\% broader than that used by \citet{gronwall2007}, we
recognize a slight possibility of misidentification of a small number
of emission lines.  However, we continue with the assumption that all 88 
of the lines that we determined to be HI \lya\ are correctly
identified as such.  One of these spectroscopically confirmed objects, LAE034, has 
$z=3.044$.  At this redshift some non-ionizing UV flux will contribute
to the NB3640 detection, and therefore we excluded it from our sample.  

\begin{figure}
\epsscale{1.15}
\centering
\plotone{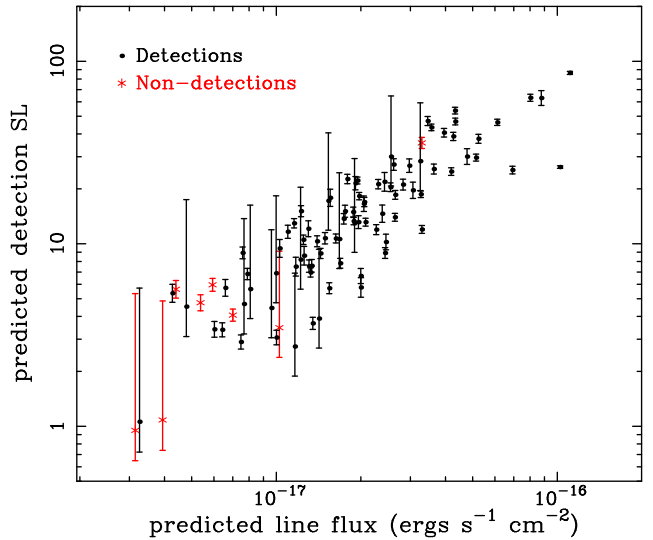}
\caption{\small
The predicted detection significance level ($SL$) for emission
lines in the spectra of our LAE candidates.  The $SL$ values are calculated using the
photometrically-estimated line fluxes and the noise properties of the spectra.
They do not account for slit losses or the possibility of resolved
lines, and are thus upper limits.  The predicted $SL$ values for spectra
with and without a detected emission line are represented by the black
dots and red stars, respectively.  Only one of our non-detections,
LAE020, has a relatively high predicted $SL$ value.  LAE020
appears very diffuse in the $LyA$ image, however, and thus its
spectrum is expected to suffer from large slit losses. 
\label{f:rewsl}}
\epsscale{1.}
\end{figure}

 Of the eight candidates for which we did not detect
any emission line in the $\approx 4975 \pm 40$\AA\ spectral region,
seven are relatively faint in NB4980, and the other is very
diffuse in the $LyA$ image with an extent significantly larger than
the slit width.   Thus, these objects may also be
\lya\ emitting galaxies at $z\ge3.055$ that have insufficient line
fluxes and/or sufficient slit losses such that the \lya\ emission line
is not detectable in our data.  In order to quantify the expected
detection significance levels of \lya\ emission lines in our
spectra, we predicted the \lya\ line fluxes
for each LAE candidate using the measured $BV$ and NB4980 magnitudes.
We then measured the noise
properties of the calibrated spectra in the wavelength
interval corresponding to the width of the NB4980 filter in order to  
assess the minimum detectable line fluxes
assuming unresolved lines.  In practice, this method will underestimate the
detection limit in some of our spectra, as we do not quantitatively
account for slit losses or the possibility of resolved line profiles.  

\begin{figure*}
\epsscale{1.0}
\plottwo{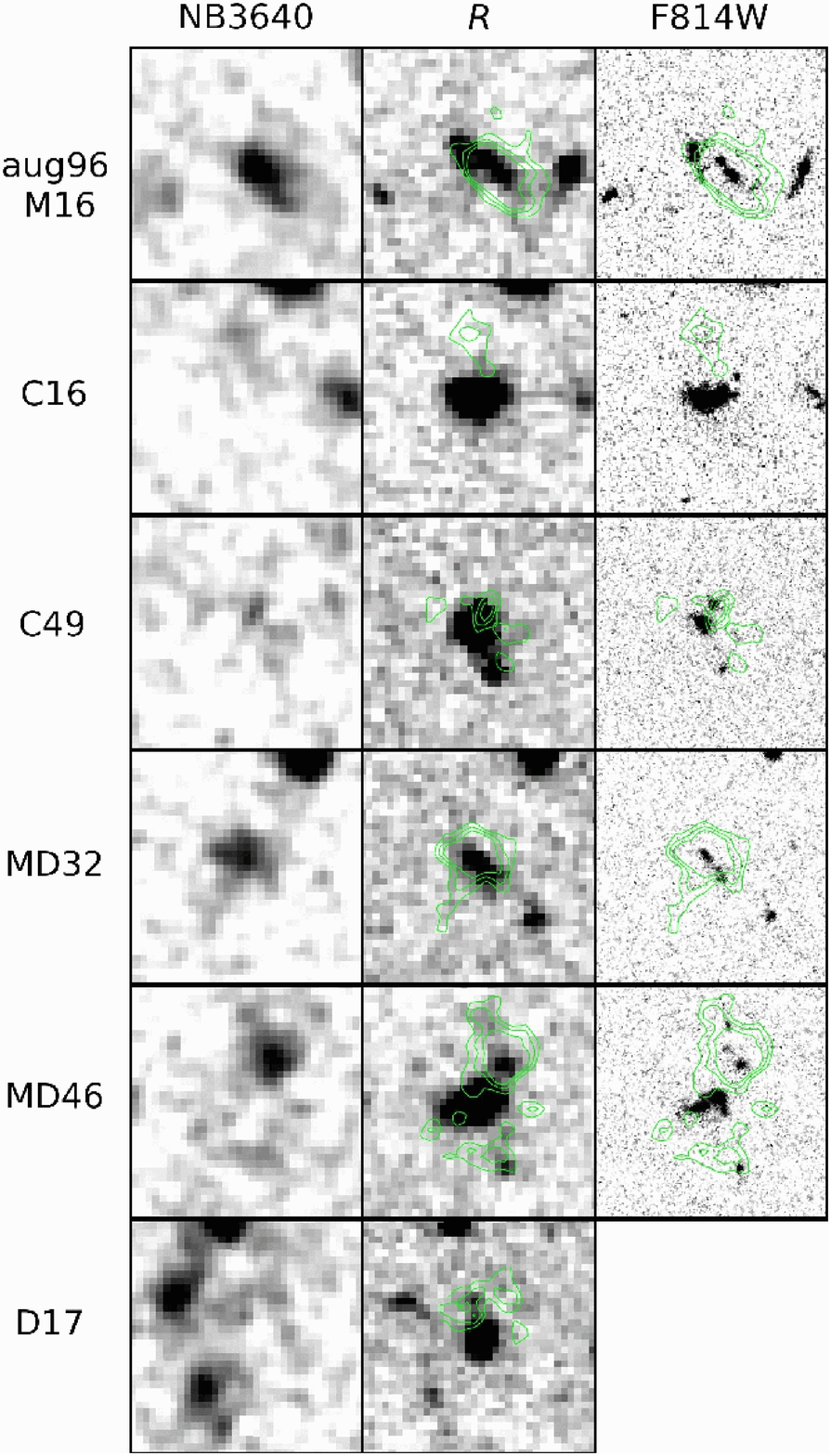}{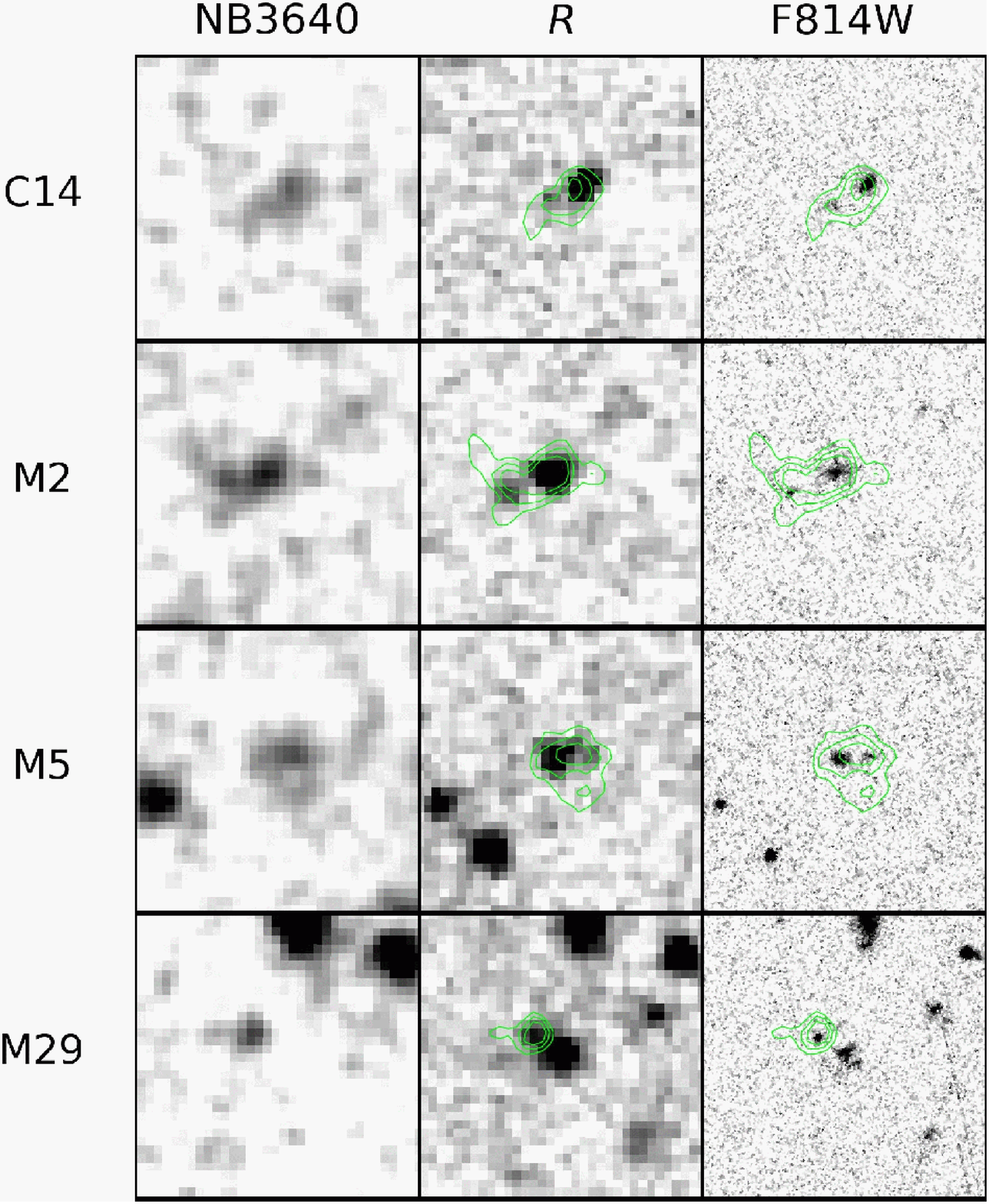}
\caption{\small 
LBGs with $z\ge3.055$ detected in NB3640.  Images are centered on the
  $R$-band centroid and span 7\arcsec $\times
  7$\arcsec.  The orientation is such that north is up and east is to the left.  At these
  redshifts, the NB3640 filter (image shown here after smoothing by
  a Gaussian kernel with FWHM$=$0\secpoint5)  
  samples the rest-frame LyC, while the $R$ and F814W
  filters sample the rest-frame non-ionizing UV continuum.  The green
  contours indicate NB3640 flux levels.  Left: LBGs
  discussed in \citet{nestor2011}.  Right: Newly confirmed $z\ge3.055$ LBGs.
\label{f:lbgs}}
\epsscale{1.}
\end{figure*}

Figure~\ref{f:rewsl} shows the
predicted detection significance level ($SL$), i.e., 
the ratio of the photometrically-predicted line flux to the
approximated line flux uncertainty, for each LAE candidate observed.
For objects with multiple spectra having different resolutions, we
show the smallest $SL$ value if \lya\ is detected, and the largest
$SL$ value if it is not.  We do not consider spectra in
which the slit did not cover the position of the 
$LyA$ flux, as is the case for some of the deep-mask spectra for which
the slit was positioned over the offset NB3640 flux.  
All of the detections are consistent with having predicted $SL >3$,
with a median value of $SL=13$.   
Seven of the eight objects without a detected emission line
have, relative to those with detections, small upper-limits to their predicted detection
significance, $SL\lesssim6$, with a median value of $SL=4$; the other
object is the diffuse system (LAE020) mentioned above. 
Thus, we can not rule the non-detections out as being $z\simeq3.09$ LAEs.  
Additionally, considering the high success
rate ($\ge91$\%) of our photometric selection at identifying
$z\ge3.055$ LAEs, it is likely that most of the 10 candidates for
which we have no spectroscopic data are also at $z\ge3.055$. 
Nonetheless, we
conservatively removed the 18 candidate LAEs without spectroscopic
confirmation from the statistical sample discussed in 
this work, leaving a spectroscopic LAE sample of 91 galaxies 
spanning $3.057 \le z \le 3.108$.   We summarize our photometric and
spectroscopic results for the LAE sample in Table~\ref{t:3}.

\citet{nestor2011} also identified an additional 20 ``faint sample'' LAE candidates with
$26< \mathrm{NB4980} \le26.5$ and $BV - \mathrm{NB4980} \ge 1.2$,
corresponding to LAE IDs $111 - 130$.  We obtained spectra
for 18 of these faint LAE candidates.  We were unable to determine
redshifts for two such systems (again, in data with low $SL$
values).  All of the other 16 systems were determined to have $3.069
\le z \le 3.111$.  The faint sample LAEs have larger photometric
uncertainties than the main sample LAEs and were selected with
slightly different photometric criterea \citep{nestor2011}.  Thus, they are not
included in the statistical analyses presented in this work.
We include them in Table~\ref{t:3} for completeness.

\section{SYSTEMS WITH NB3640 DETECTIONS}
\label{s:det}
With our updated spectroscopic sample of LBGs and LAEs in place, we
now consider the set of objects with detections in our NB3640 image.
Of the 26 $z\ge3.055$ LBGs discussed in \citet{nestor2011}, six are
detected in NB3640.  Four of our 16 newly confirmed high-redshift LBGs are
also detected in NB3640: C14, M2, M5, and M29.  Their NB3640-, $R$-, and $HST$
F814W images are shown in Figure~\ref{f:lbgs}, with the contours
corresponding to 28.81, 28.06 and 27.62~mag~arcsec$^{-2}$ in the
NB3640 image.
All four of the newly confirmed LBGs with NB3640
detections appear clumpy in 
the $HST$/ACS images, with multiple discrete regions of non-ionizing
continuum.  The NB3640 flux appears to cover all of the 
clumps in C14, M2 and M5.  In M29, the NB3640 flux appears to be 
associated with only the more compact clump, which is $\simeq
0$\secpoint75 to the northeast of a more extended region of
non-ionizing UV flux.  With the
addition of these four new NB3640 detections, we now have 10 NB3640
detections associated with 42 $z\ge3.055$ LBGs.  Their 
NB3640 magnitudes, the spatial offset between the $R$-band and
NB3640 flux centroids, and the inferred non-ionizing to ionizing UV
flux-density ratios are presented in Table~\ref{t:one}.

In our sample of 91 spectroscopically confirmed LAEs, 20 have NB3640
detections.  An additional 6 LAE candidates from
\citet{nestor2011} have NB3640 detections: one was not targeted in our
spectroscopy, and we were unable to confirm the redshifts of five other
candidates.  Thus, these six objects are not discussed here.  
For the 20 LAEs with associated NB3640 detections, we report in Table~\ref{t:3} the 
NB3640 magnitudes, spatial offsets between both $R$-band and LyA and NB3640 flux centroids,
and the inferred non-ionizing to ionizing UV flux-density ratios.

\begin{figure}
\epsscale{1.17}
\centering
\plotone{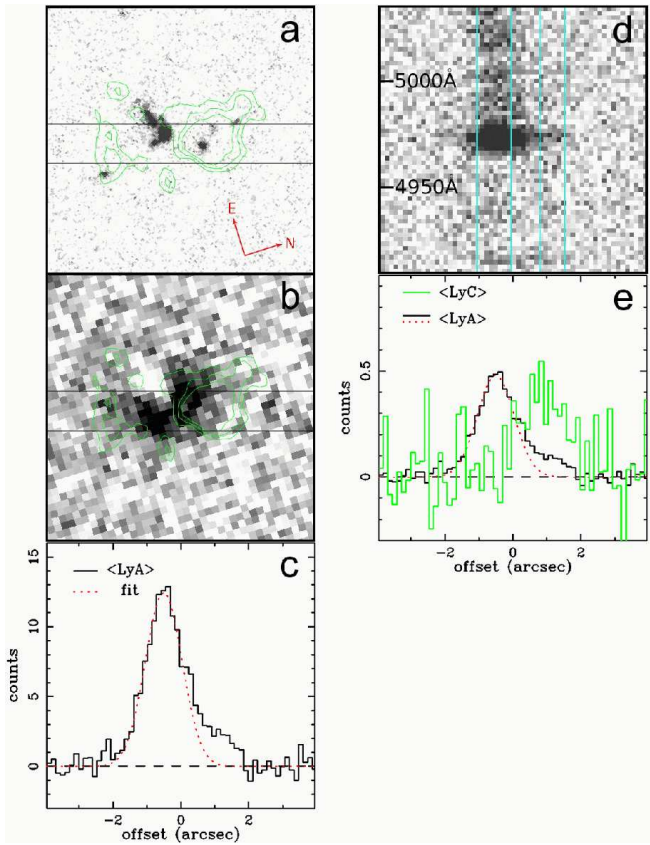}
\caption{\small 
The emission structure of MD46.  Left: The spatial extent of \lya\
emission compared to the rest-frame non-ionizing
UV continuum emission and putative LyC radiation (green contours).
Panel (a) shows the {\it HST}/ACS-F814W 
image (spatial resolution $\simeq0\secpoint1$) while panel (b)
is our ``$LyA$''  (seeing $\simeq0\secpoint8$) 
image.  The slit position for our deep-mask spectrum is indicated in
both images.   
Panel (c) displays the spatial (i.e., along the slit) extent
of the \lya\ line, averaged over (rest-frame) $\approx4$\AA\ of our
deep-mask spectrum and registered
to the images.  The dotted curve is a Gaussian fit to the profile,
excluding the excess at positive offset.  
Right: The relative spatial extents of LyC and \lya\ emission.
Panel (d) shows the two-dimensional spectrum in the redshifted \lya\ region.
The two extraction apertures applied in Figure~\ref{f:md46spec} are
indicated with cyan boxes.  In panel (e), we show the profile for the region of the
two-dimensional spectrum averaged in the spectral direction from
rest-frame $\lambda \simeq 875$\AA\ to $\lambda \simeq 910$\AA\ (green), as
well as that of the \lya\ 
region, re-scaled for ease of comparison (black).  
\label{f:md46}}
\epsscale{1.}
\end{figure}

\begin{figure}
\epsscale{1.17}
\centering
\plotone{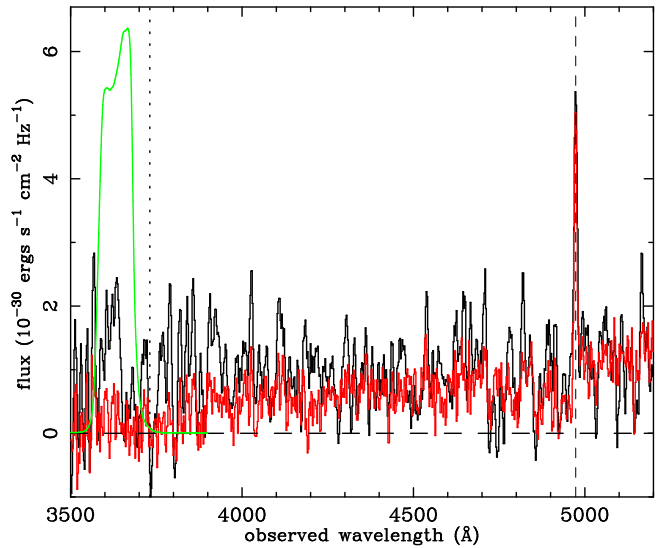}
\caption{\small 
The extracted deep-mask spectra of MD46.  We identify the
emission feature at $\lambda=4973$\AA\ as \lya\ at redshift $z=3.091$
(vertical dashed line).  The
corresponding Lyman break is marked with a vertical dotted line.  The black
spectrum, which has been boxcar smoothed (5 pixels), is the extraction
containing only the northwestern clump (see Figure~\ref{f:md46}), which
is spatially consistent with the detected NB3640 flux.  The red spectrum represents
the extraction that excludes the northwestern clump.  The red spectrum has been
scaled such that the average flux levels redward of \lya\ are
approximately equal.  Also shown in green, below the Lyman limit, is
the NB3640 filter transmission curve. 
\label{f:md46spec}}
\epsscale{1.}
\end{figure}

Below, we discuss in detail the combined spectroscopic and
imaging data sets for the individual objects in our updated samples
that were considered 
LyC detections in \citet{nestor2011}, as well as the newly
determined $z > 3.055$ LBGs.  The data are shown in
Figures~\ref{f:md46}$-$\ref{f:nb2082}.  $HST$/ACS-F814W images are
shown when available, $BV$ images are shown otherwise, and all
of the images span 8\arcsec\ per side.  Slit
positions are indicated by blue (deep mask) and/or black (shallow
mask) boxes.  Green contours represent NB3640 flux levels (see above).
The LAE images also contain
red contours, representing $LyA$ (see Section~\ref{sect:data-phot}) flux
levels.\footnote{Due to the large range of $LyA$ fluxes in our LAE sample,
the flux levels represented by the red contours vary from image to
image.}  Two-dimensional spectra, when shown, are registered in the spatial direction to match the
accompanying imaging and span $\simeq125$\AA\ in the spectral
direction.  Unless otherwise noted, one-dimensional spectra are shown
without smoothing or binning.  When multiple one-dimensional spectra
are shown, the additional spectra are offset for clarity and the
non-offset spectrum corresponds to the horizontal slit in the imaging
(by default, the deep-mask slit when available).

\subsection{LBGs}
\label{s:lbgdet}
In this section, we discuss the individual LBGs having NB3640
detections for which we have new data.  We begin with the four systems
which we retain as possible LyC-leaking galaxies.  We next discuss the
three LBGs for which we find evidence for the presence of a foreground
interloper in our new data.  We then discuss the special case of
aug96M16, and conclude the section with a summary of our LBG NB3640
detection sample.

\subsubsection{LBG Lyman-continuum Candidates}

\medskip
\centerline{\it MD46}
\smallskip

The LBG MD46 ($z=3.091$) has a complex multi-component morphology, which can be
clearly seen in Figure~\ref{f:md46}.  Panel (a) shows the
$HST$/ACS-F814W image, in which the flux appears 
to originate from at least three distinct clumps.  
Panel (b) shows the ground-based $LyA$ image.  The clump to the
east, which partly enters the slit, is spatially coincident with
negative flux in the $LyA$ image (corresponding to \lya\ absorption at
$z\simeq3.09$).  The bulk of the \lya\ emission appears to emanate
from in and around the central clump, while the bulk of the NB3640
flux avoids this central clump and is instead spatially coincident 
with the clump $\simeq$1\secpoint1 to the northwest, which also falls on
the slit.  Thus, it is important to determine if the northwestern clump is
also at $z\simeq3.09$.  

\begin{figure*}
\epsscale{0.9}
\centering
\plotone{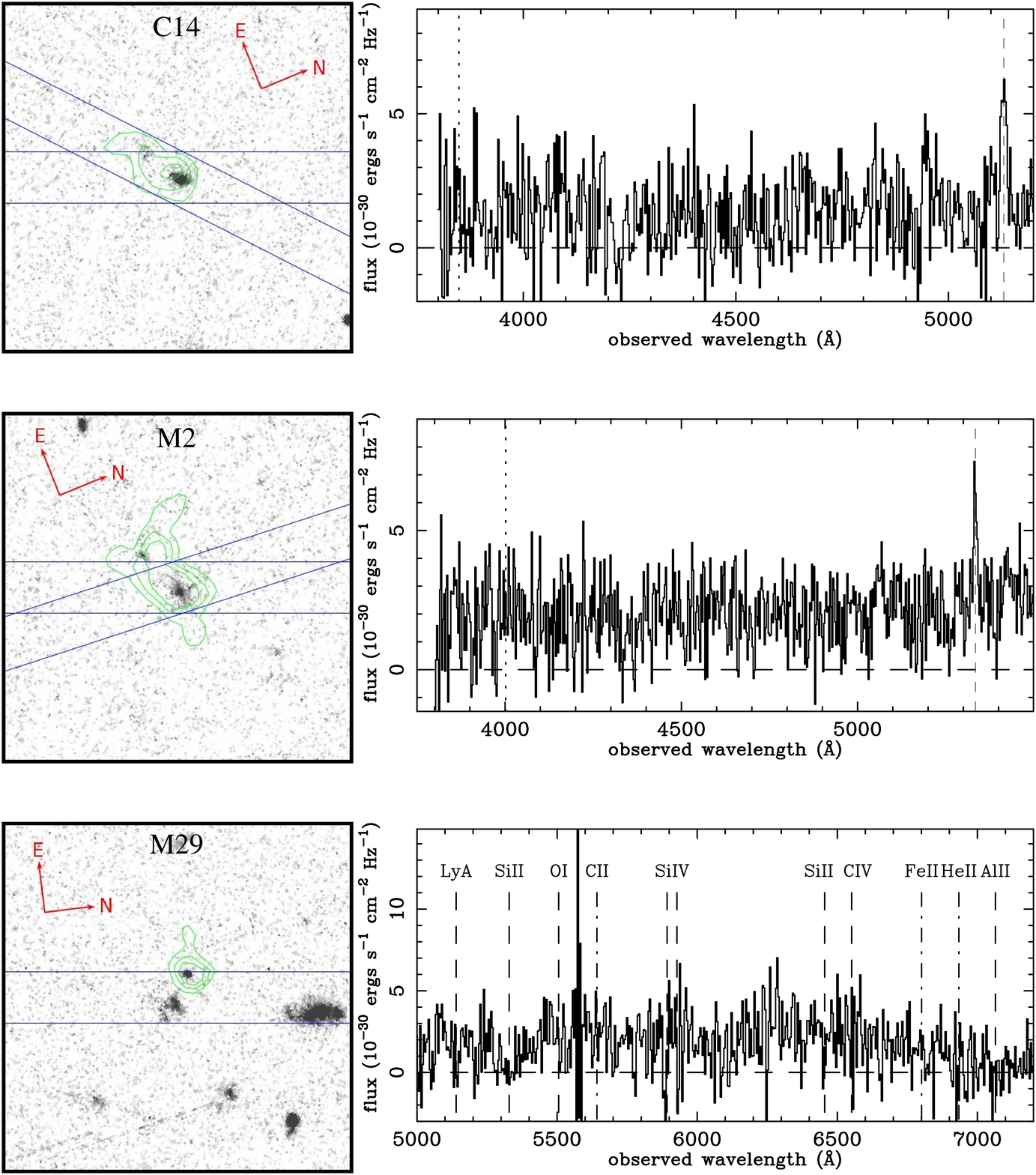}
\caption{\small 
$HST$/ACS F814W images (8\arcsec\ per side) and spectra of LBGs with NB3640
detections retained as possible LyC emitters.   The positions and sizes of the slits are indicated with
boxes, and green contours indicate NB3640 flux levels.  When present,
vertical dashed and dotted lines in the spectra indicate the position of the
redshifted \lya\ line and Lyman break, respectively.
Top: Image and shallow-mask
spectrum C14.  The other shallow-mask spectrum of  C14 (corresponding
to the diagonal slit) has a very low signal to noise ratio and is not
shown.  Middle: The F814W image and combined shallow-mask spectrum of M2.
Bottom: Image and
shallow-mask spectrum of M29.  The locations of several common interstellar
absorption features are marked with vertical dashed (detected) and dash-dotted
(non-detected) lines. 
\label{f:3lbgs1}}
\epsscale{1.}
\end{figure*}

\begin{figure*}
\epsscale{0.9}
\centering
\plotone{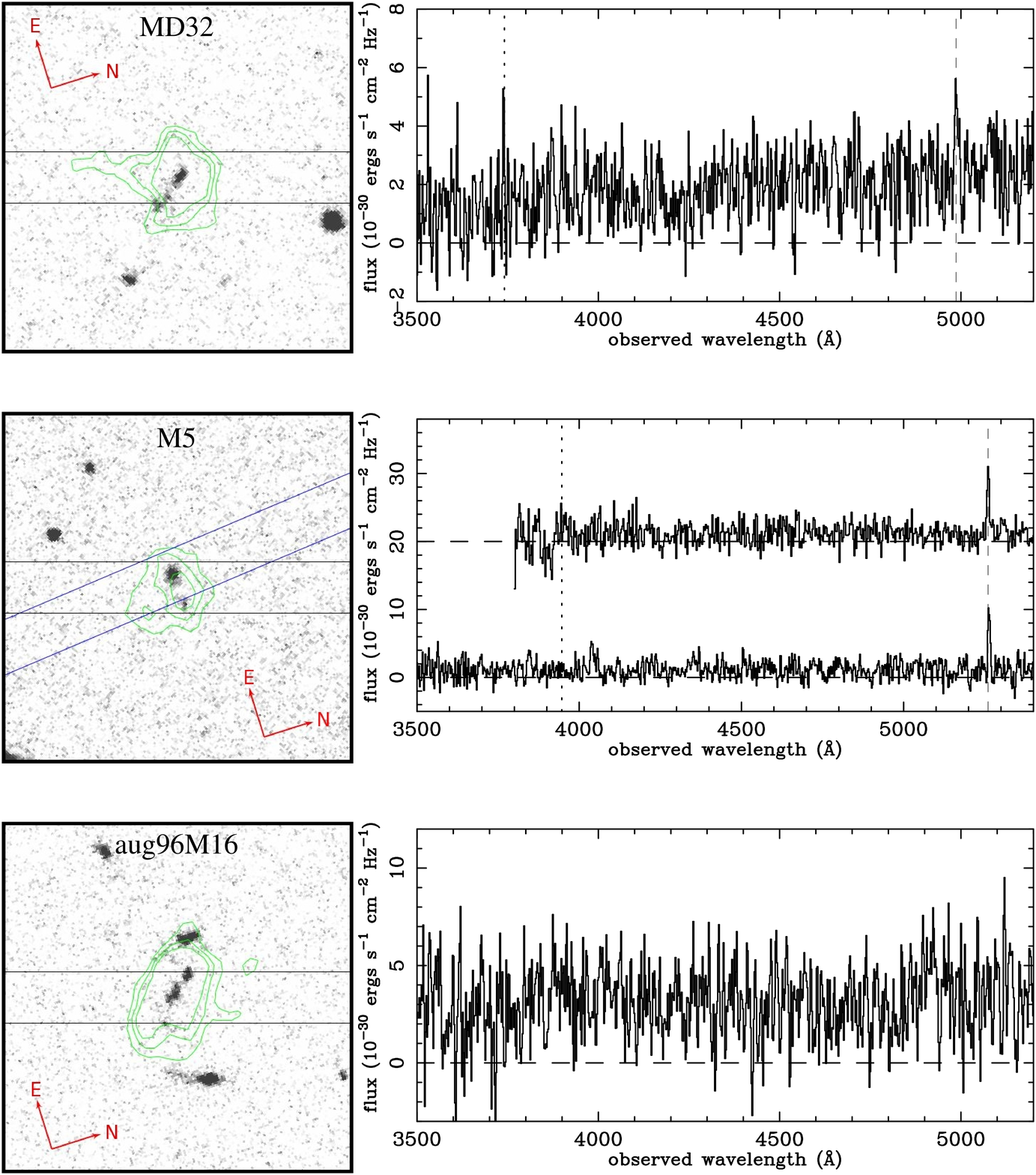}
\caption{\small 
As in Figure~\ref{f:3lbgs1}, but for systems with evidence of
contamination (top and middle) or with unconfirmed redshift (bottom).
Top: Image and deep-mask spectrum of MD32.
Middle: Image and shallow- (offset, upper) and deep-mask (lower) spectra of M5.  Bottom: Image of 
aug96M16 and the deep-mask spectrum.  We were unable to determine the
redshift of the region most closely associated with the NB3640 emission.  
\label{f:3lbgs2}}
\epsscale{1.}
\end{figure*}

\begin{figure*}
\epsscale{0.9}
\centering
\plotone{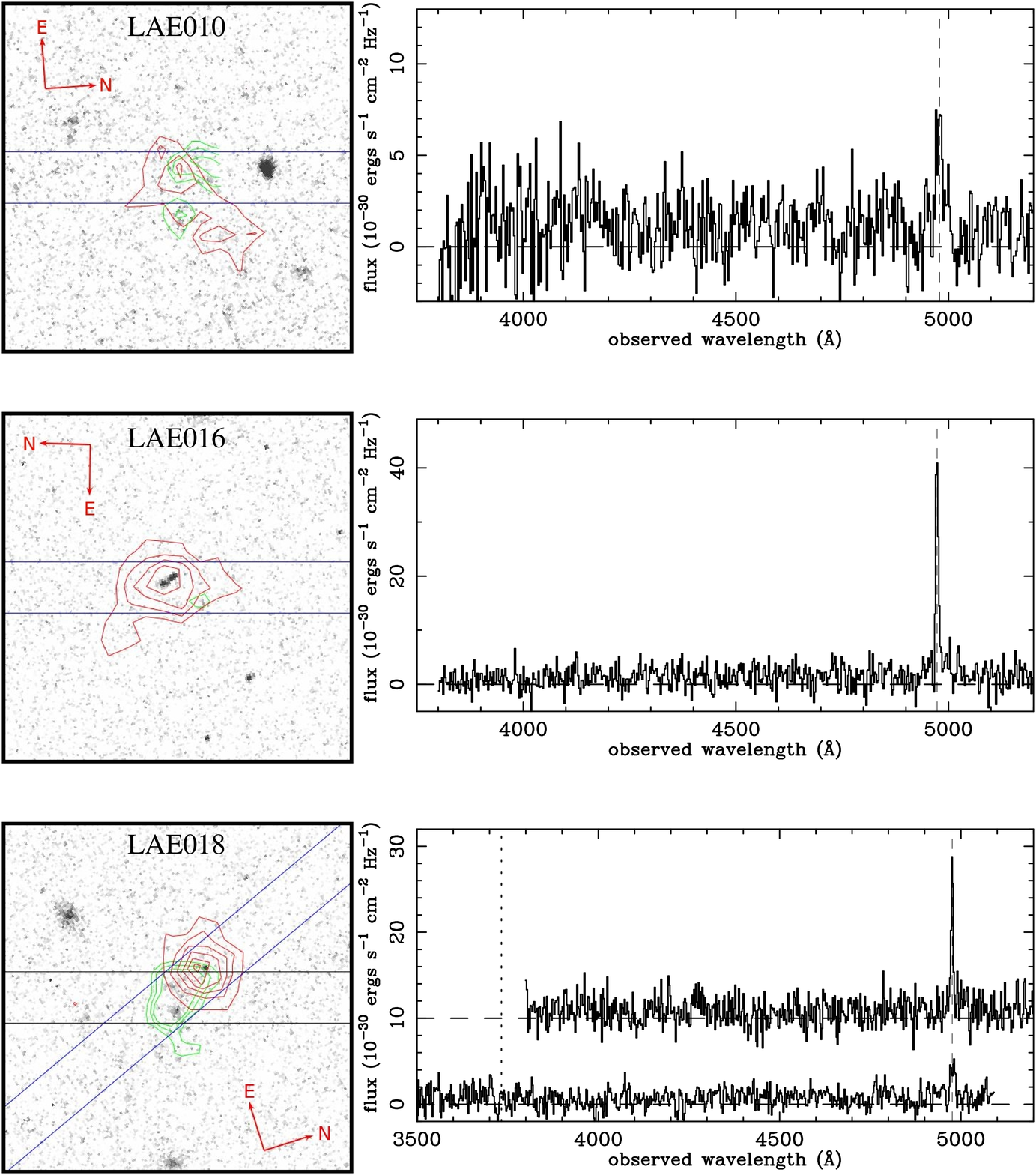}
\caption{\small 
The emission structures of (from top down) LAE010, LAE016, and LAE018.  
The left panels show the {\it HST}/ACS-F814W images.  The boxes indicate
the positions of all shallow- (blue) and deep- (black)
mask slits.  Green and red contours represent flux levels in the
NB3640 and $LyA$ images, respectively.
The right panels display the one-dimensional extracted
spectra.  
The vertical dashed/dotted lines indicate the position of the
redshifted \lya/Lyman limits, respectively.  
For LAE018, we show both the shallow- (offset, upper) and deep-mask
(lower) spectra.  
\label{f:3laes1}}
\epsscale{1.0}
\end{figure*}

\begin{figure*}
\epsscale{0.9}
\centering
\plotone{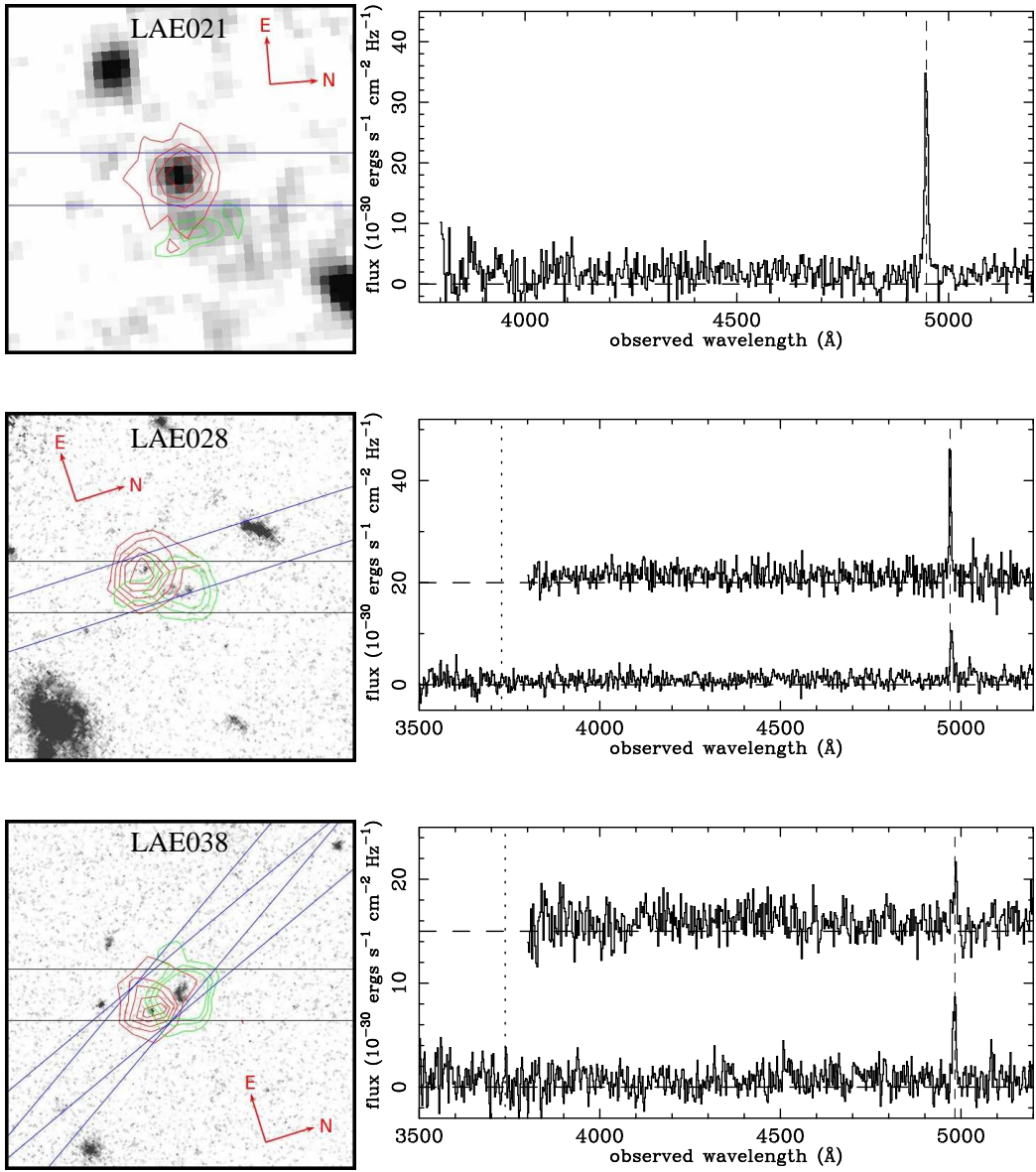}
\caption{\small 
Same as Figure \ref{f:3laes1}, but for LAE021, LAE028, and LAE038.
Our BV image is shown
for LAE021 as it does not have $HST$/ACS-F814W imaging.
For LAE028 and LAE038, we show both the shallow- (offset, upper) and deep-mask
(lower) spectra. 
\label{f:3laes2}}
\epsscale{1.0}
\end{figure*}

\begin{figure*}
\epsscale{0.9}
\centering
\plotone{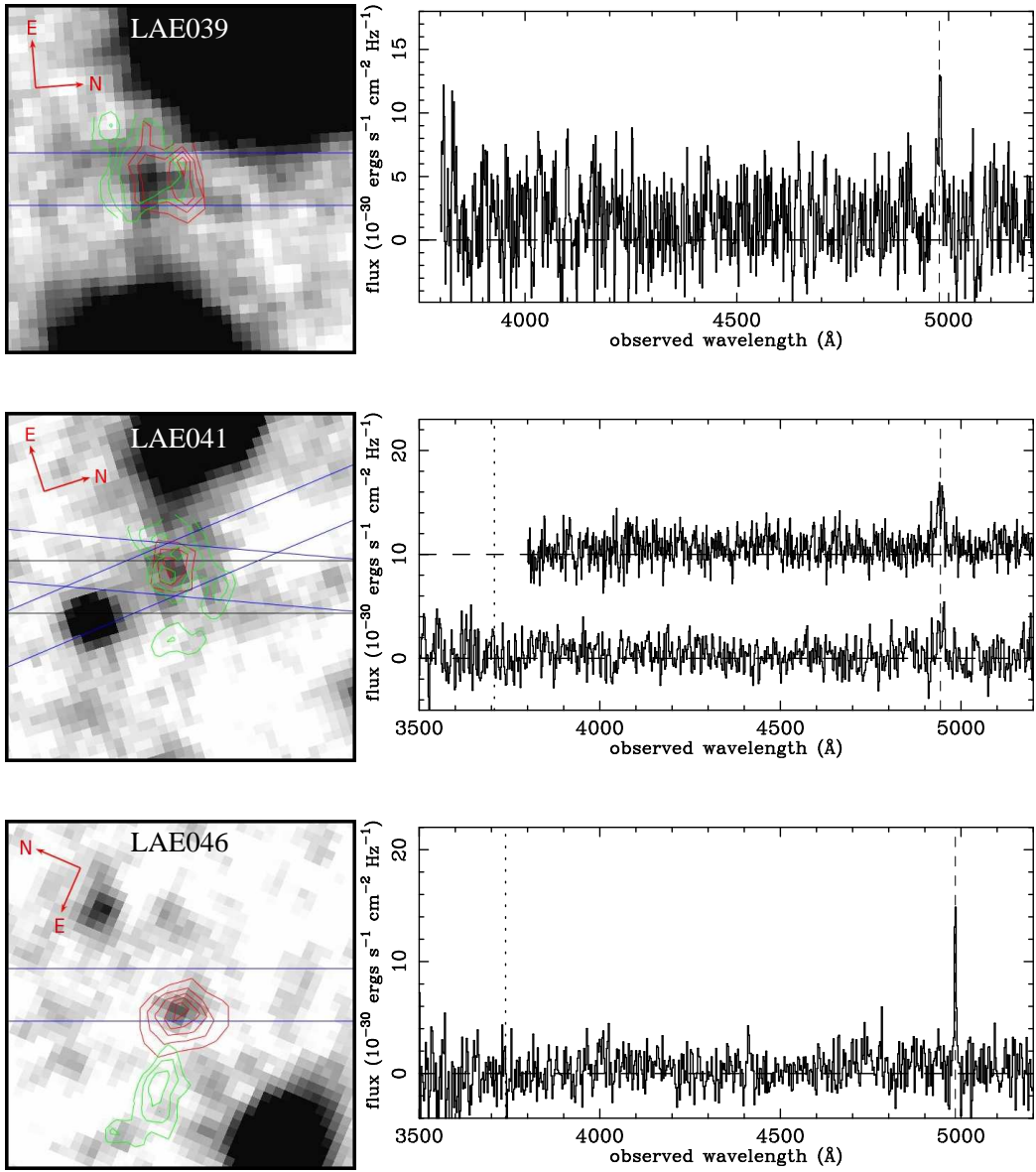}
\caption{\small 
Same as Figure \ref{f:3laes1}, but for LAE039, LAE041, and LAE046.
BV images are shown
as LAE039, LAE041 and LAE046 do not have $HST$/ACS-F814W imaging.
For LAE041, we show both the shallow- (offset, upper) and deep-mask
(lower) spectra.
 \label{f:3laes3}}
\epsscale{1.0}
\end{figure*}

\begin{figure*}
\epsscale{0.9}
\centering
\plotone{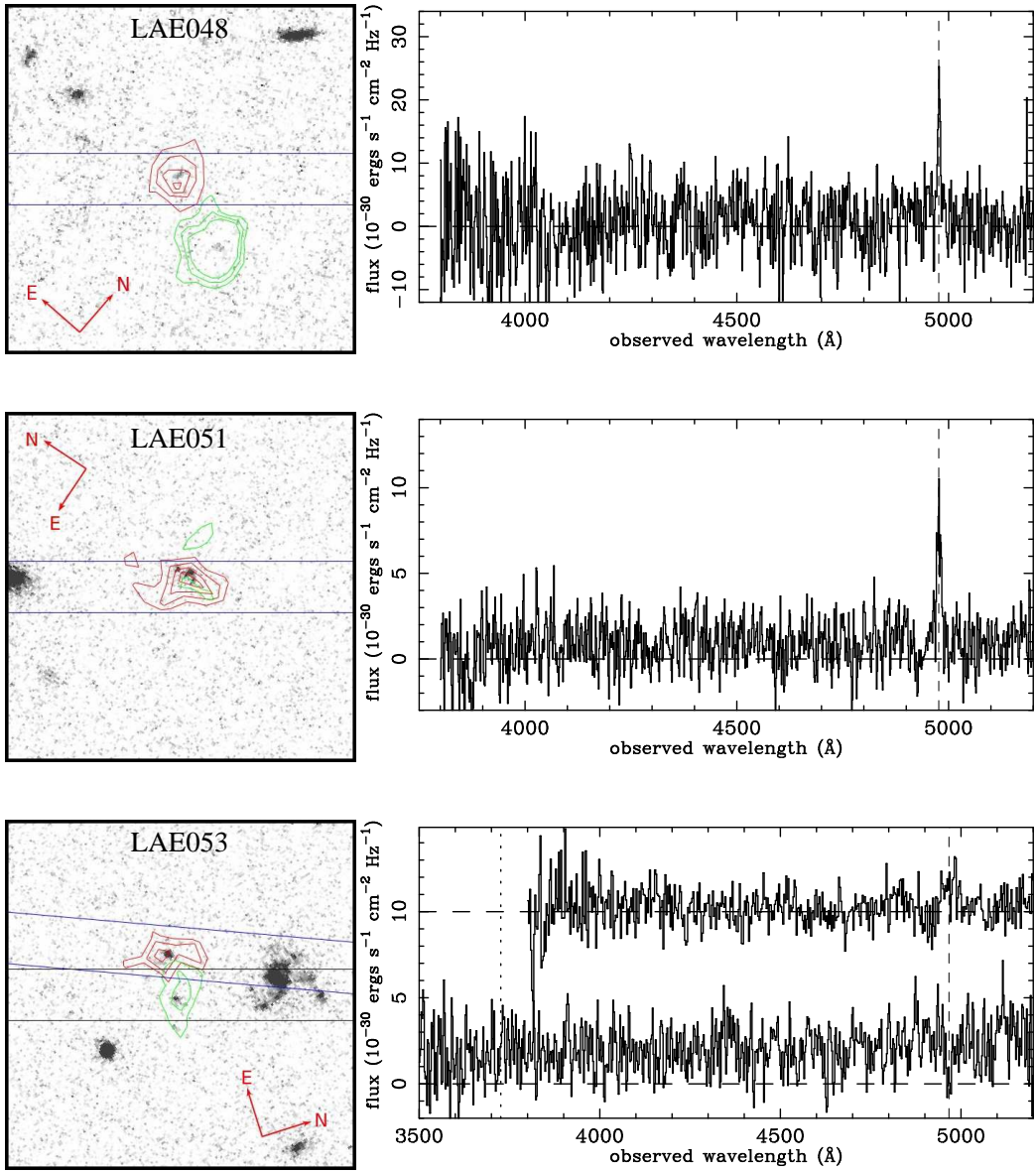}
\caption{\small 
Same as Figure \ref{f:3laes1}, but for LAE048, LAE051, and LAE053.
For LAE053, we show both the shallow- (offset, upper) and deep-mask
(lower) spectra.  \lya\ appears in absorption in the deep-mask spectrum
of LAE053, as the slit was centered on the NB3640
detection which is offset by $\simeq0$\secpoint9 from the $LyA$
emission.  
\label{f:3laes4}}
\epsscale{1.0}
\end{figure*}

\begin{figure*}
\epsscale{0.9}
\centering
\plotone{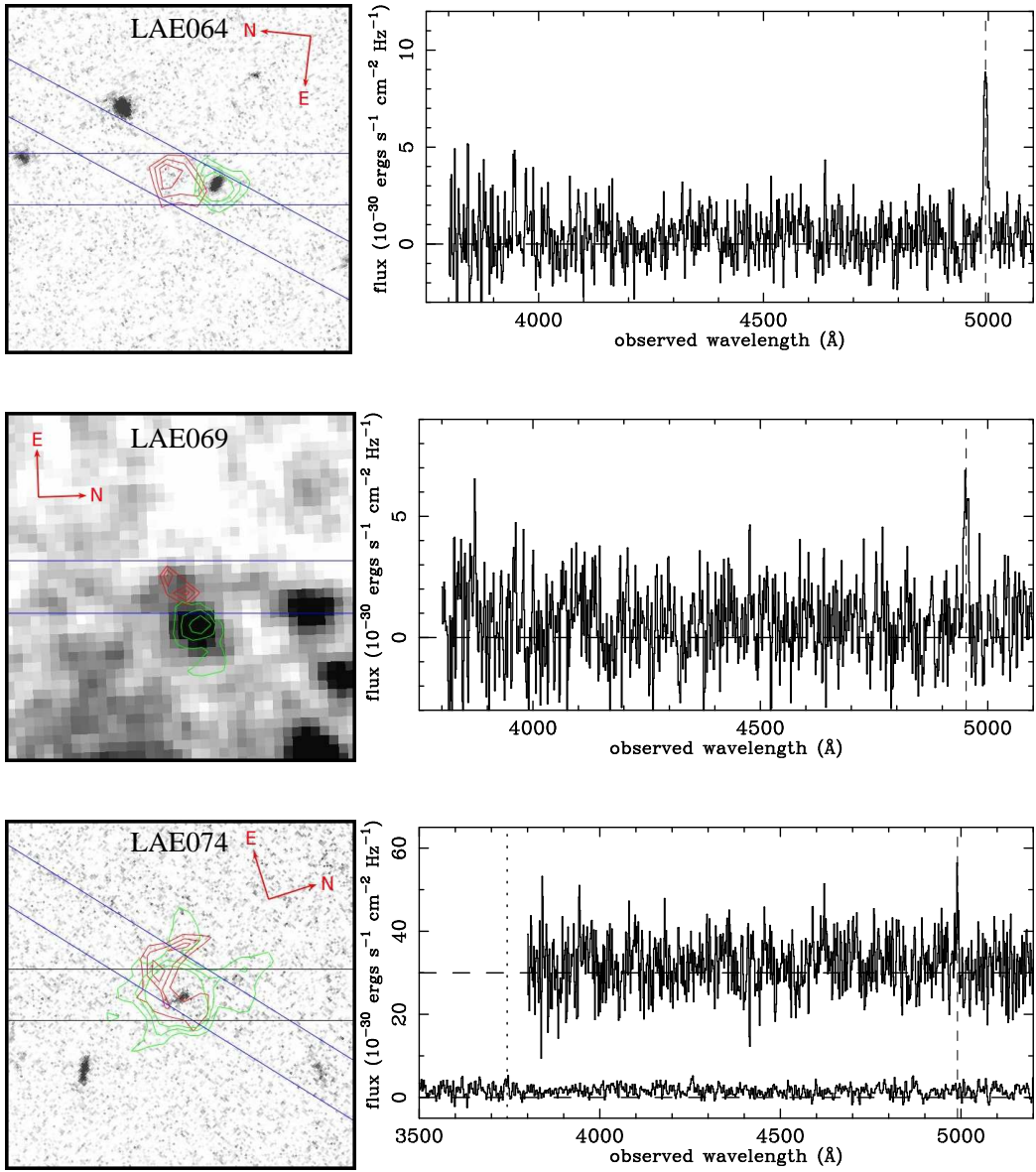}
\caption{\small 
Same as Figure \ref{f:3laes1}, but for LAE064, LAE069, and LAE074.
Our BV image is shown
for LAE069 as it does not have $HST$/ACS-F814W imaging.
For LAE074, we show both the shallow- (offset, upper) and deep-mask
(lower) spectra.  \lya\ emission is not detected in the deep-mask
spectrum.  As the slit was centered on the NB3640
detection the bulk of  the $LyA$ emission fell outside of the slit.  
\label{f:3laes5}}
\epsscale{1.0}
\end{figure*}

\begin{figure*}
\epsscale{0.9}
\centering
\plotone{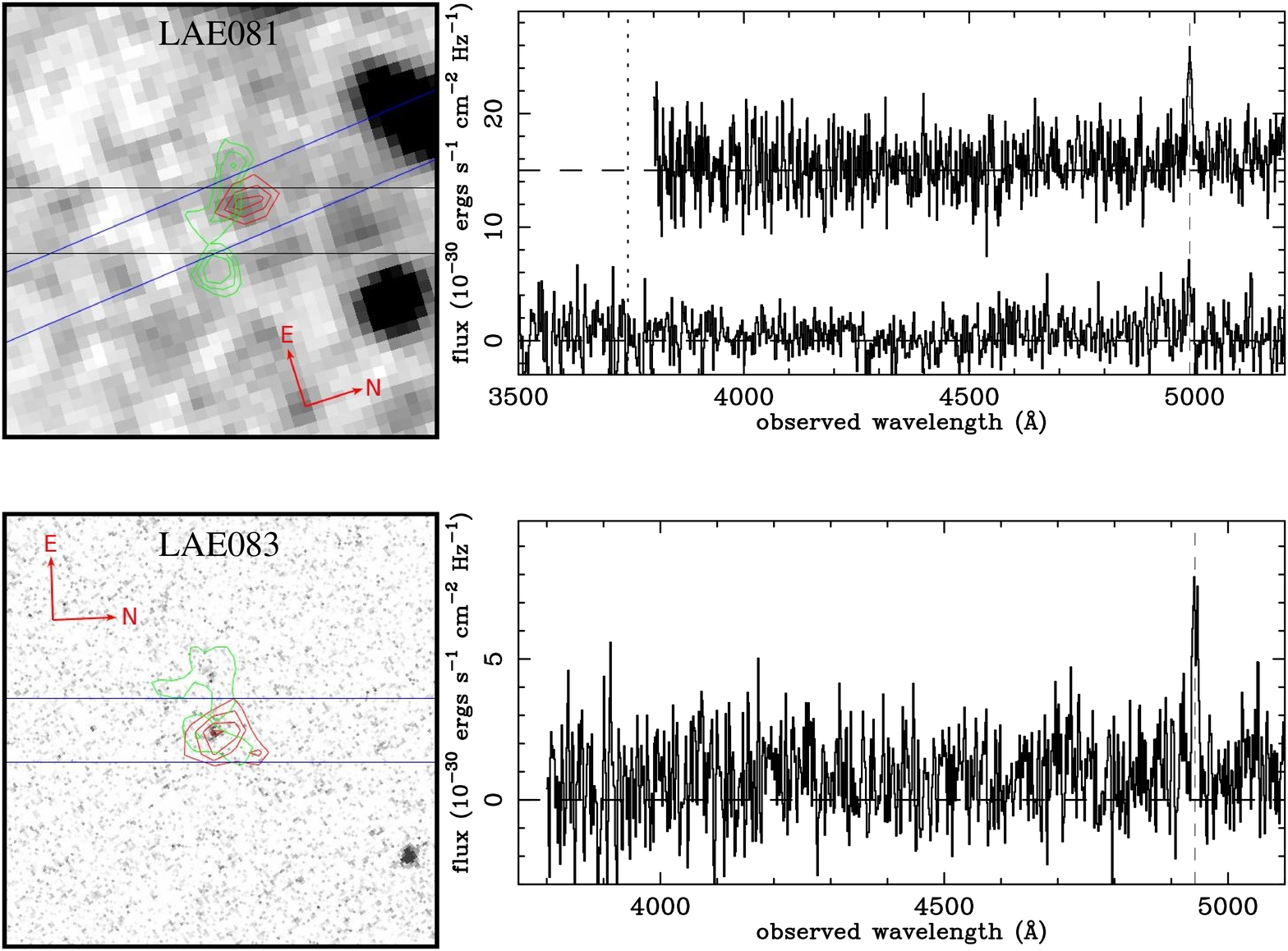}
\caption{\small 
Same as Figure \ref{f:3laes1}, but for LAE081 and LAE083.  For LAE081,
we show both the shallow- (offset, upper) and  deep-mask    
(lower) spectra and BV image as it does not have $HST$/ACS-F814W imaging. 
\label{f:3laes6}}
\epsscale{1.0}
\end{figure*}

 The region of the two-dimensional spectrum corresponding to redshifted \lya\
emission is shown in panel (d).  The \lya\ emission line
appears asymmetric, which can be seen more clearly in 
panel (c), where we have averaged over $\approx4$\AA\
(rest-frame) in the spectral direction, centered on the peak of the \lya\ 
emission line, and fitted a Gaussian profile (dotted curve) constrained only by
the data south of the spatial zero-point indicated by the figure axis.  
This fit highlights the asymmetric northwestern
extension, which is spatially coincident
with the northwestern clump seen in the $HST$ imaging and the NB3640 flux in
our Keck imaging.  The spatial coincidence of the \lya\ and NB3640
emitting region implies that the northwestern clump is also very likely
to be emitting \lya\ at $z\simeq3.091$, which in turn indicates that
the NB3640 flux is indeed escaping LyC emission.

Panel (e) shows the spatial profile for the region of the
two-dimensional spectrum just
below the redshifted Lyman-limit (green histogram),
averaged in the spectral direction over $\lambda \simeq
880$\AA\ $- 910$\AA\ in the rest-frame, as well as that of the \lya\
region (black histogram, re-scaled by factor of 26 for ease of
comparison).  Again, the NB3640 flux is spatially aligned with the
extended \lya, suggesting that it is actually LyC flux escaping from a
galaxy at $z=3.09$.
We extracted one-dimensional spectra in two locations in the two-dimensional
spectrum.  The extraction apertures are shown in panel (d).  
Figure~\ref{f:md46spec} displays the resulting one-dimensional
spectra.  The black spectrum, which has been smoothed with a 5-pixel
boxcar, represents the northwestern extraction corresponding to the 
extended \lya\ emission and the detected NB3640 flux.  The red 
spectrum, which has been
scaled down by a factor of 5.6 such that the average flux level
redward of \lya\ emission is equal to that of the black spectrum,
represents the southern extraction, corresponding to the 
central clump, which is undetected in our NB3640 imaging.  The spectrum of the central 
clump appears to exhibit a strong break at the Lyman limit, the
position of which is indicated by
the vertical dashed line.  In contrast, the northwestern clump appears to
lack a sharp break at the Lyman limit.  We also note that several \lya\
forest absorption lines appear to be present in both spectra (e.g., at
$\lambda\simeq 4720$\AA, 4860\AA, 4925\AA, etc.)
suggestive of the continua originating at similar emission redshifts.
However, the poor signal to noise ratios prevent quantitative analysis
of any perceived correlation.

We conclude that the $z=3.091$ LBG MD46 contains three distinct
(rest-frame) UV-bright regions.  The south-eastern region is a \lya\ absorber as
seen from our viewing perspective, while the central and northwestern regions exhibit \lya\
in emission.  Of these latter regions, the fainter region (in both continuum and
line emission) is more compact and can be seen in ionizing
continuum along our sightline.  Thus we retain MD46 in our
LyC-emitting LBG sample.

\medskip
\centerline{\it C14}
\smallskip

We obtained spectra of the LBG candidate C14 on two of the shallow
masks.  Each spectrum exhibits an emission line at $\lambda
\simeq 5130$\AA, which we attribute to \lya\ at $z=3.220$.  C14 is
detected in our NB3640 image (Figure~\ref{f:lbgs}).  In the $HST$/ACS
image, shown in the top panel of Figure~\ref{f:3lbgs1}, C14 breaks
into a brighter and a 
fainter clump, with the NB3640 emission appearing to span the two
clumps.   As no other emission features were robustly detected in either
spectrum, we retain C14 as a possible LyC-emitting galaxy.  We note,
however, that the spectrum corresponding to the slit position 
that is better aligned with both clumps (not shown) is of relatively
poor quality, and therefore our limits on 
the absence of other emission features from the fainter clump are
relatively weak. 

\medskip
\centerline{\it M2}
\smallskip

We obtained spectra of the LBG candidate M2 on two of the shallow
masks.  Both spectra exhibit an emission line at $\lambda
\simeq 5334$\AA, which we attribute to \lya\ at $z=3.388$.  M2 is
detected in our NB3640 image (Figure~\ref{f:lbgs}).  In the $HST$/ACS
image, shown in the middle panel of Figure~\ref{f:3lbgs1}, M2 breaks into
a brighter and a 
fainter clump, with the NB3640 emission appearing to span the two
clumps.   No other emission features were robustly detected in either
spectrum.  Thus, we retain M2 as a possible LyC-emitting galaxy.

\medskip
\centerline{\it M29}
\smallskip

We obtained a spectrum of the LBG candidate M29 on 
one of the shallow masks.  The spectrum does not exhibit emission
lines.  However, we identify interstellar absorption lines from OI,
SiIV, SiII, and CIV corresponding to a redshift of $z=3.228$.  M29 is
detected in our NB3640 image (Figure~\ref{f:lbgs}).  Two clumps of
emission are seen in the $HST$/ACS image, shown in the bottom panel of
Figure~\ref{f:3lbgs1}.
The NB3640 emission only covers the more compact clump.   It is
possible that both clumps are at $z=3.228$, with the NB3640 emission
emanating from a compact star-forming clump.  Alternatively, the more
diffuse source may be at high redshift while the compact source associated
with the NB3640 emission is in the foreground.  Lacking any direct
evidence to favor either scenario, we retain M29 as a possible
LyC-emitting galaxy and account for the possibility of contamination
with our Monte Carlo simulation (Section~\ref{s:monte}).

\subsubsection{LBGs With Evidence For Contamination}
\label{s:lbg_contam}

\medskip
\centerline{\it MD32}
\smallskip

The $HST$/ACS image and our deep-mask spectrum of MD32 ($z=3.102$) are shown in
the top panel of Figure~\ref{f:3lbgs2}.  Our spectrum exhibits flux below the
redshifted Lyman limit, and contains no evidence for foreground
contaminants.  However, in the NIRSPEC observations the slit
was aligned along the direction of elongation of MD32.  The resulting two-dimensional
spectrum exhibits an additional line within $\sim$0\secpoint7 of MD32
which, if [OIII] $\lambda5007$, would correspond to a redshift of
$z=2.88$.  Due to this evidence of contamination, we remove MD32 from
our list of possible LyC-emitting galaxies.   

\medskip
\centerline{\it M5}
\smallskip

We obtained spectra of the LBG candidate M5 with both the deep mask and
one of the shallow masks.  Each spectrum exhibits an emission line at $\lambda
\simeq 5260$\AA, which we attribute to \lya\ at $z=3.327$.  M5 is
detected in our NB3640 image (Figure~\ref{f:lbgs}).  Two clumps of
emission are observed in the $HST$/ACS image, shown in the middle panel of
Figure~\ref{f:3lbgs2}.
The NB3640 emission appears to span the two clumps.   However, our
deep-mask spectrum of M5, also shown in Figure~\ref{f:3lbgs2} along with
the shallow-mask spectrum, reveals the presence of an
additional, bluer, emission line at $\lambda\simeq4040$\AA, indicating the presence of a
lower-redshift interloper.  It is noteworthy, however, that the
detected NB3640 flux covers both clumps of continuum emission.  Thus,
if the interloper is associated with only one of the clumps, it may be
that M5 is indeed being detected in 
LyC emission.  Nonetheless, we conservatively remove M5 from 
our list of possible LyC detections.

\medskip
\centerline{\it C49}
\smallskip

Although we did not obtain a new LRIS spectrum of C49 ($z=3.163$), it was
observed with NIRSPEC.  As shown in Figure~\ref{f:lbgs}, the $HST$/ACS
image reveals that C49 comprises two distinct clumps.  The NIRSPEC slit was positioned to
capture the spectra of both of these clumps.  While the spectrum of
the southern clump confirms the redshift $z=3.16$, that of the
northern clump, which is spatially closer to the NB3640 detection, has
an additional emission line which,  if [OIII] $\lambda5007$, would
correspond to a redshift of $z=2.97$.  We
therefore remove C49 from our list of possible LyC-emitting galaxies.   

\subsubsection{Unconfirmed LBG redshift}
\label{s:aug96M16}

\medskip
\centerline{\it aug96M16}
\smallskip

The $HST$/ACS image of aug96M16 is shown in the bottom panel of
Figure~\ref{f:3lbgs2}.  As with MD46, the $HST$ imaging reveals the
emission from aug96M16 to comprise several distinct
clumps.  The detected NB3640 flux is coincident with the central
region of emission that spans the 1\secpoint2 width of our LRIS slit.
Only the western-most clump ($\approx 1$\secpoint3 west of the
slit edge) was  
resolved as a distinct clump, separate from LBG candidate aug96M16, in the
ground-based data originally used to select LBG 
candidates based on $U-G$ and $G-\cal{R}$ colors in the SSA22a field
\citep{steidel2003}.  This western-most clump has a previously
determined spectroscopic redshift $z=3.285$, 
which was also (erroneously) attributed to aug96M16 in \citet{nestor2011}.  The slit
position in our LRIS deep mask covered neither the western-most nor
eastern-most clump ($\approx 0$\secpoint75 east of the
slit edge).  The slit position of
our NIRSPEC spectrum (Siana et al., in preparation; see
Section~\ref{sect:data-spec}) of aug96M16 was aligned to cover all
of the emission clumps,
however.  Using the NIRSPEC data, we determine a redshift
$z=3.09$ for the eastern-most clump and confirm the redshift of the
western-most clump as $z=3.29$.  The large difference in redshift
between the the eastern-most and western-most clumps  indicates that,
although both are at $z>3.055$, they are physically unrelated.  Our
deep-mask spectrum of the central clumpy region, which is coincident
with the NB3640 detection, is also shown in   
Figure~\ref{f:3lbgs2}.  We are unable to determine a redshift for
this region from either our deep mask or NIRSPEC spectra.  Therefore,
although it was photometrically selected as an LBG candidate and is within
$\sim 1$\arcsec\ on the sky to a spectroscopically-confirmed galaxy at
$z=3.09$, we conservatively remove aug96M16 from our list of possible
LyC-emitting galaxies as well as from the parent $z\ge3.055$ LBG sample.   

\subsubsection{Summary of LBGs}
As discussed above, our sample began with 10 LBGs with
possible LyC detections.  Our new data includes spectra
of seven of these 10 objects.  In these spectroscopic data we find evidence
for the presence of a foreground object in close proximity on the sky
to three of these LBGs -- MD32, C49, and
M5 -- and are unable to confirm the redshift of the source associated
with the NB3640 flux in a fourth -- aug96M16.  Of the seven LBGs
having NB3640 detections with new
spectroscopic data, we retain as possible LyC emission NB3640
detections in four: MD46, C14, M2, and M29, as well as two detections for
which we do not have new data: D17 and C16.  Thus, our LBG sample
now contains of a total of six putative LyC-leaking galaxies from a
parent sample of 41 LBGs (i.e., excluding aug96M16; see
Section~\ref{s:aug96M16}).  Notably, we are able to study the spatial 
distribution of the \lya\ and NB3640 emission in detail for MD46, and
find compelling evidence that the NB3640 emission is escaping LyC flux.  

It is possible that some of the 32 LBGs with no NB3640 detections
have levels of escaping LyC flux that are below our detection limit of
$m_{\mathrm{NB3640}} \sim 27.3$.  To investigate this possibility, 
we stacked cut outs of the NB3640 image centered on the locations of
these 32 LBGs.  We detect no flux in the stacked image down to a
3$\sigma$ limit on the  average magnitude of $m_{\mathrm{NB3640}} =27.96$.  

\subsection{LAEs}
\label{s:laedet}
As above, in this section we discuss the individual LAEs with NB3640
detections.  We begin with an overview of the 17 systems
which we retain as possible LyC-leaking galaxies.  We next discuss the
three LAEs for which we find evidence for the presence of a foreground
interloper.  We then discuss the six LAEs with NB3640 detections for which
we were unable to confirm redshifts and are thus not included in our
sample, and conclude the section with a summary of our LAE NB3640
detections.

\subsubsection{LAE Lyman-continuum Candidates}

As indicated in Table~\ref{t:3}, we spectroscopically confirm redshifts of $3.070
\le z \le 3.108$ with no evidence for contamination for 17 LAEs with
NB3640 detections.  The images and spectra for these LAEs are shown
in Figures~\ref{f:3laes1}~$-$~\ref{f:3laes6}.  We briefly discuss each
individual field in the appendix.  Of particular note is LAE053, for
which the $LyA$ and NB3640 emission are offset by $\simeq 0$\secpoint9.  
The F814W image reveals that each of the $LyA$ and NB3640 detections
is coincident with one of two distinct emission clumps.  Our shallow- and
deep-mask slits cover only the clump associated 
with the $LyA$ and NB3640 fluxes, respectively.  In the shallow-mask
spectrum, we detect \lya\ in
emission at $z=3.090$.  In the deep-mask spectrum, however, we 
detect an absorption feature at the same
wavelength as the emission line in the shallow-mask spectrum.  We
interpret this feature as \lya\ absorption at $z=3.090$, which is compelling
evidence that the clump associated with the NB3640 emission is also at 
$z=3.090$ and, in turn, that the NB3640 flux is indeed LyC emission.
In contrast, however, we caution that for four of
the candidates shown in Figures~\ref{f:3laes1}$-$\ref{f:3laes6}
(LAE021, LAE046, LAE048 and LAE069), the position of the NB3640 flux
was not covered by 
any of the slit spectra.  Thus we are unable to search directly for 
evidence of foreground contamination being responsible for the
detected NB3640 flux.  Additionally, the F814W
images of LAE046 (Figure \ref{f:3laes3}) and LAE048 (Figure
\ref{f:3laes4}) reveal that the NB3640 and $LyA$ fluxes, respectively, are 
coincident with sources of non-ionizing continuum that are significantly
($\ga 1$\secpoint5) spatially offset from each other, suggesting the sources are
unrelated.  In the absence of direct spectroscopic evidence confirming
contamination for these individual systems, 
we ran a Monte Carlo simulation (described in Section~\ref{s:monte})
to statistically address the possibility
of contamination in the entire sample of 17 detections.  In this
simulation, NB3640 detections with large 
offsets are likely to be rejected as foreground interlopers.  Thus,
while individually the conclusions that LAEs such as LAE046 and LAE048
are LyC-leaking galaxies should be taken with caution, our statistical
results are robust to the possibility of foreground contamination.

\subsubsection{LAEs With Evidence For Contamination}
\label{s:laecont}

We find evidence for the presence of a foreground galaxy in the
spectra of three of our LAEs, which we present below.  
Additionally, one of the LAE candidates from \citet{nestor2011}, LAE034, is at
a low enough redshift that it experiences contamination of its NB3640
flux from non-ionizing UV radiation.  We have thus removed LAE034 from
our list of possible LyC-emitting galaxies as well as from the parent
$z\ge3.055$ LAE sample.  

\begin{figure*}
\epsscale{0.9}
\centering
\plotone{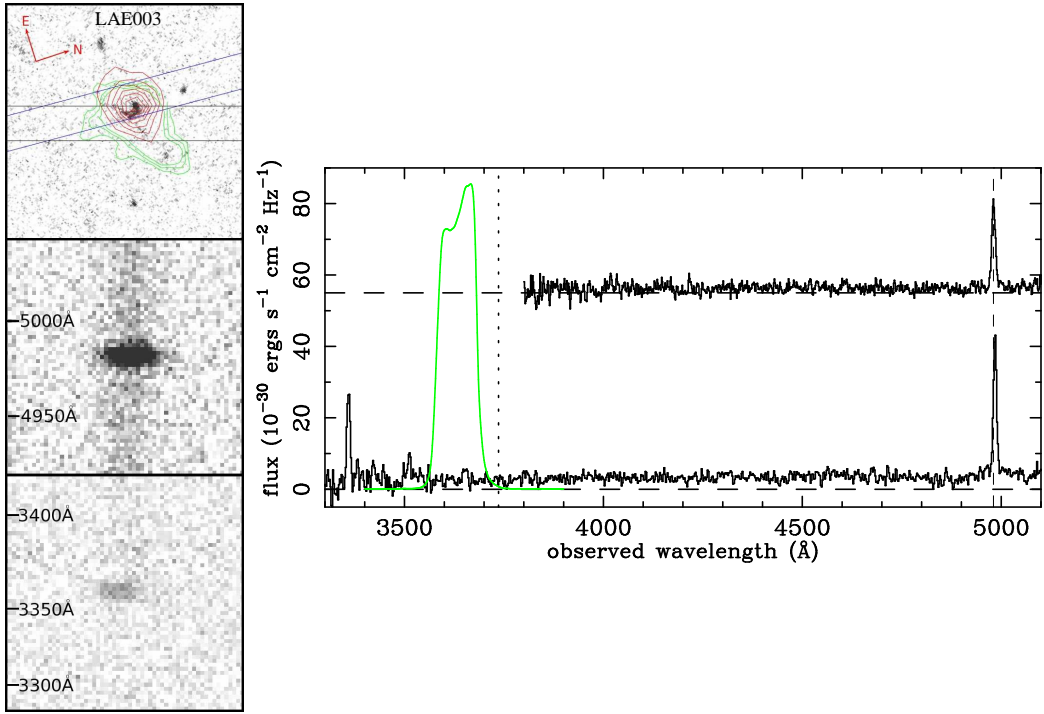}
\caption{\small 
The emission features of LAE003.   Left:
The top panel shows the {\it HST}/ACS-F814W image.  
The middle panel shows the two-dimensional spectrum in the redshifted \lya\ region,
while the bottom panel
shows a redder region of the spectrum, centered at the wavelength
($3359$\AA) of a bluer emission line.
Right: The extracted one-dimensional spectra of
LAE003.  Shown  are  both the shallow- (offset, upper) and deep-mask
(lower) spectra.  We identify the emission feature 
at $\lambda=4984$\AA\ as \lya\ at redshift $z=3.097$.  The corresponding Lyman
break is marked with a vertical dotted line.  The bluer emission line is
clearly detected at $\lambda=3359$\AA\ in the deep-mask spectrum.  The
green curve below the Lyman limit indicates the shape of the NB3640 filter.
\label{f:c1_0494a}}
\epsscale{1.0}
\end{figure*}

\begin{figure*}
\epsscale{0.9}
\centering
\plotone{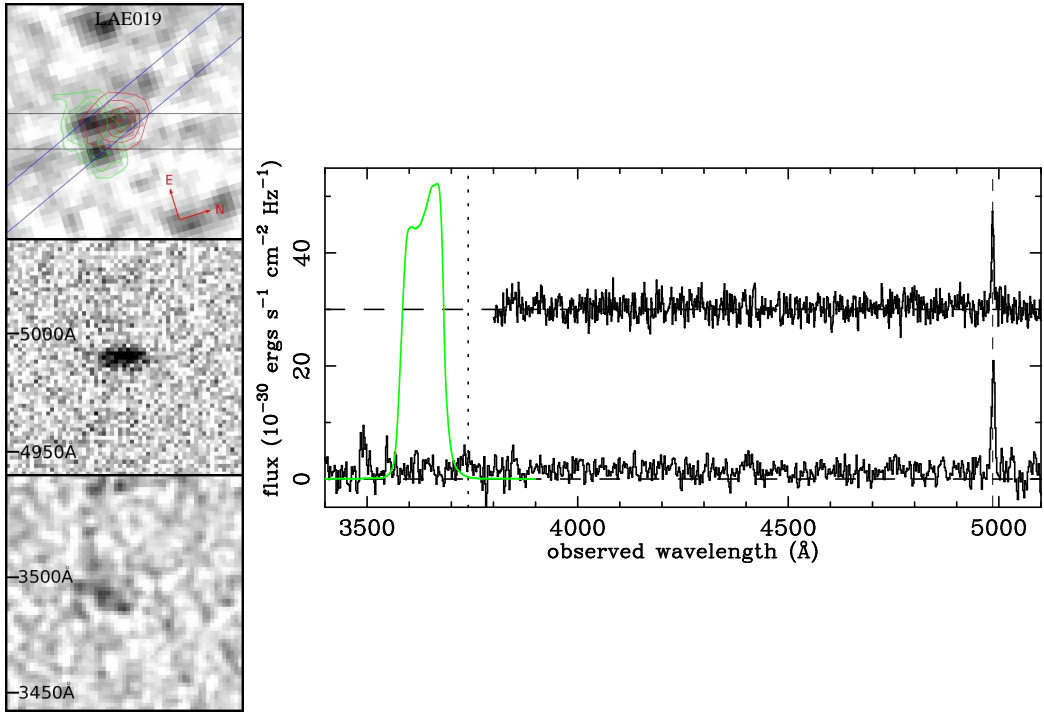}
\caption{\small 
The emission features of LAE019.  Left:
The top panel shows our BV image.  The middle and bottom panels
show the two-dimensional deep-mask spectrum in the regions of the redshifted \lya, and
the bluer emission line, respectively.  Right: The
extracted one-dimensional spectra of LAE019.  Shown  are  both the shallow- (offset, upper) and deep-mask
(lower) spectra.  We identify the emission feature 
at $\lambda=4985$\AA\ as \lya\ at redshift $z=3.101$.  The corresponding Lyman
break is marked with a vertical dotted line.  The bluer emission line can be
seen at $\lambda=3490$\AA\ in the deep-mask spectrum.
\label{f:nb942}}
\epsscale{1.0}
\end{figure*}
 
\begin{figure}
\epsscale{1.1}
\centering
\plotone{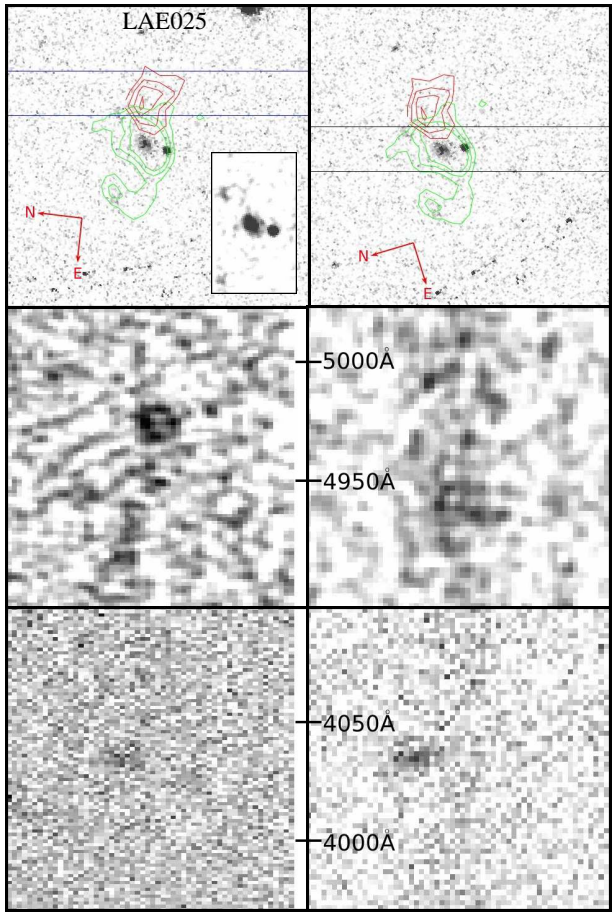}
\caption{\small 
The emission features of LAE025.  
The top panels show the {\it HST}/ACS-F814W image.  The boxes indicate
the positions of the slits in the shallow- (left) and deep- (right)
masks.  The inset in the top-left panel shows the central region without
contours after slight (2.35-pixel FWHM) smoothing.  The middle panels
show the two-dimensional spectra in the redshifted \lya\ region for the
corresponding slit spectra, with an emission line at $\simeq4974$\AA\
in the shallow-mask spectrum and, possibly, at
$\simeq4942$\AA\ in the deep-mask spectrum.  
The bottom panels show the two-dimensional spectra in the vicinity of the emission
line at $\simeq 4035$\AA.
\label{f:nb2082}}
\epsscale{1.0}
\end{figure}

\begin{figure}
\epsscale{1.15}
\centering
\plotone{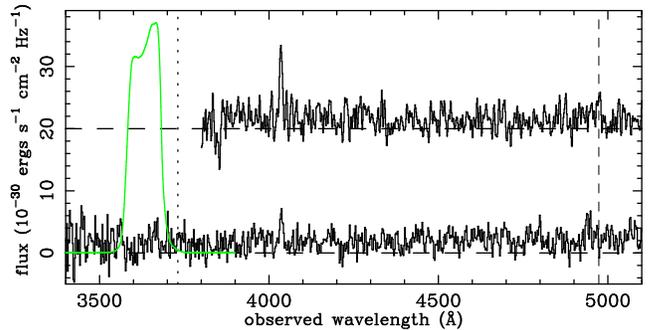}
\caption{\small 
The extracted shallow- (offset, upper) and deep-mask (lower) spectra
of LAE025.  We identify the
emission feature at $\lambda=4974$\AA\ in the shallow-mask spectrum as
\lya\ at redshift $z=3.091$.  The
corresponding Lyman break is marked with a vertical dotted line.  The bluer emission line 
at $\lambda=4035$\AA\ can be seen in both spectra.  The deep-mask
spectrum appears to exhibit \lya\ emission at $\lambda=4942$\AA,
corresponding to $z=3.065$.  Also shown in green, below the Lyman limit, is
the NB3640 filter transmission curve. 
\label{f:lae025spec}}
\epsscale{1.0}
\end{figure}

\medskip
\centerline{\it LAE003} 
\smallskip

LAE003 has the brightest NB3640 detection
($m_{\mathrm{NB3640}}=24.74$) of any of our LBGs 
or LAEs in \citet{nestor2011}.  The top-left panel of
Figure~\ref{f:c1_0494a} shows the F814W image indicating the relative
positions of the non-ionizing UV flux, the deep and shallow-mask slits, the \lya\
flux and the NB3640 flux.
The middle panel shows the region of
the two-dimensional spectrum corresponding to the expected location of the redshifted
\lya\ line.  We identify the line detected at $\lambda=4980$ as \lya,
indicating that LAE003 is at $z=3.097$.  

However, as can be seen in the lower
panel, which shows a bluer region of the two-dimensional spectrum, there is another
emission line spatially consistent with the LAE.  If this line is also
\lya\ emission, it corresponds to a redshift of $z=1.763$.  The
bluer line is very 
slightly offset to the south, consistent with the tail-like structure
seen in the $HST$/ACS image.  It is therefore possible that the
$z=3.097$ LAE003 is opaque below the Lyman limit and the NB3640
flux is non-ionizing continuum emerging from this lower-redshift galaxy.
It is noteworthy that the NB3640 flux is not centered on the portion of the
F814W flux that we associate with the lower-redshift interloper, and extends to
the north-west.  However, due to the clear presence of an interloper
we take the conservative approach and remove LAE003 from our list of 
possible LyC-leaking galaxies.  The one-dimensional extracted 
spectra are shown in the right panel of Figure~\ref{f:c1_0494a}.

LAE003 had previously been studied by \citet{inoue2011} who 
determined a spectroscopic redshift of $z=3.100$.  Their spectrum did
not extend blueward enough to detect the emission line at
$\lambda=3359$\AA, however.  As the NB3640 
\citep[NB359 in][]{inoue2011} flux is spatially coincident with the $R$-band and
$LyA$ flux for LAE003, these authors argue that the probability of foreground
contamination is small and therefore the NB359 flux is indeed LyC.  
They interpret the implied very high ratio of
ionizing to non-ionizing UV flux density of four objects, including
LAE003, in terms of very young
stellar populations with top-heavy initial mass functions.  This interpretation 
is invalidated for LAE003 by the discovery of the low-redshift
interloper in our deep-mask spectrum.  The case of LAE003
highlights the importance of obtaining spectra extending as
far to the blue as possible for properly interpreting possible
LyC-leaking galaxies. 

\medskip
\centerline{\it LAE019 } 
\smallskip

We do not have $HST$ imaging of LAE019.  The top panel of
Figure~\ref{f:nb942} shows our $BV$ image with the deep and shallow-mask slits
indicated.  The \lya\ emission centroid is slightly offset to the
north in both the imaging and two-dimensional deep-mask spectrum (middle panel), while
the NB3640 flux centroid is slightly offset to the south in our
imaging.  The deep-mask spectrum of LAE019, shown in
Figure~\ref{f:nb942}, also exhibits a bluer emission 
feature at $\lambda\simeq 3490$\AA.  If this feature is \lya\ 
emission, it would indicate the presence of an interloper at
$z=1.872$.  The bottom panel shows the two-dimensional spectrum in the
$\lambda\simeq 3490$\AA\ region.  This lower-redshift emission line
is spatially consistent with the NB3640 flux, indicating that the NB3640
flux is likely due to the interloper.  Therefore, we remove
LAE019 from our list of possible LyC-leaking galaxies. 

\medskip
\centerline{\it LAE025} 
\smallskip

We obtained spectra of LAE025 in both shallow and 
deep masks.  The top panels of Figure~\ref{f:nb2082} show the
$HST$/ACS image, with the corresponding slit positions (shallow left;
deep right).  The inset in the top-left panel shows the central region
without contours, after smoothing by a Gaussian kernel with
FWHM$=2.35$~pixels (0\secpoint1).  The bulk of the 
NB3640 flux is associated with the central clump.  We detect flux
in neither the F814W nor $BV$ images at the location of the
$LyA$ flux centroid.  
The middle-left panel of Figure~\ref{f:nb2082} shows the region of
the two-dimensional shallow-mask spectrum corresponding to the
expected location of 
the redshifted \lya\ line.  We detect a strong emission line spatially 
coincident with the $LyA$ flux, which we identify as \lya\ at
$z=3.091$.  However, the non-ionizing UV flux detected in the $HST$/ACS
image that is coincident with the NB3640 flux is just off of the
slit.  The middle-right panel shows the same spectral region of the
deep-mask spectrum, for which the slit was centered on the NB3640
detection.  No obvious emission at $\lambda\simeq4974$\AA\ is 
seen, as might be expected if the region coincident with the NB3640
flux was also at $z=3.091$.  Based on possible detections of \lya\
emission at  $\lambda\simeq4942$\AA\ and perhaps corresponding SiII
interstellar absorption, we tentatively assign a redshift of
$z=3.065$.  This is a high enough redshift such that the NB3640 filter
is still opaque to radiation longward of the Lyman limit.  It should
be noted, however, that the LAE at $z=3.091$ for which we searched for corresponding
NB3640 flux would be unrelated to the $z=3.065$ object (the relative
velocity between the two \lya\ lines being $\simeq
1800$~km\,s$^{-1}$) associated with the NB3640 flux;  their proximity
on the sky would be coincidental.

Further complicating this system is the presence of an additional,
much bluer line at $\lambda\simeq 4035$\AA, which is detected in both
spectra (Figure~\ref{f:lae025spec}).  It is spatially
consistent with the faint, low surface-brightness flux detected in the
F814W image to the north-northwest of the central clump, which falls in both
slits.  If identified as \lya, it implies a redshift of $z=2.319$.  

With the data at hand it is difficult to either confirm or refute the
claim that the detected NB3640 flux is ionizing continuum.
Nonetheless, considering (a) there is evidence for a nearby foreground
object, and (b) our identification of the redshift of the blob
associated with the NB3640 emission is tentative, we 
remove LAE025 from our list of possible LyC detections.

\subsubsection{Unconfirmed LAE Candidates with NB3640 Detections}
\label{s:uncon}

We were unable to confirm the redshifts of six of our LAE candidates
with NB3640 detections.  Of these six, we did not observe one,
LAE084.  The other five, LAE077, LAE087, LAE096, LAE101 and LAE102 all
have relatively small photometrically-estimated REWs and upper-limits
to their predicted detection significance (Section~\ref{s:sample} and
Table~\ref{t:3}) between $SL \simeq 1$ -- 6.  Furthermore, their
fluxes in the $LyA$ image are relatively diffuse
\citep[see][Figure~4]{nestor2011}.  We remove these six systems from
our statistical sample, but are not able to rule them out as
$z\simeq3.09$ galaxies with escaping LyC flux.

\subsubsection{Summary: LAEs}
Our original sample of LAE candidates from \citet{nestor2011}
contained 27 sources with NB3640 detections.  Our new dataset includes
spectra of 26 of these 27 sources.  We were able to identify an
emission line that we attribute to \lya\ in 21 of the 26, although one
source (LAE034) is at a redshift that is too low for inclusion in our
sample.  The other five 
sources with NB3640 detections may also be LAEs at $z\simeq3.1$ as
our data are not of sufficient sensitivity to detect the expected \lya\ emission
line.  However, we conservatively remove these five systems from our
statistical sample.  We also remove three LAEs with NB3640 detections that show
evidence for foreground interlopers in the spectroscopic data.  Thus,
our current statistical sample of 91 spectroscopically confirmed $z\ge
3.055$ LAEs contains 17 objects with NB3640 detections and no evidence
for contamination of their NB3640 flux by foreground interlopers.
We stacked the NB3640 images of the 71 LAEs having no NB3640 detections
and detected no flux down to an 3$\sigma$ limit on the
average magnitude of $m_{\mathrm{NB3640}} = 28.39$.  

\begin{figure}
\epsscale{1.15}
\plotone{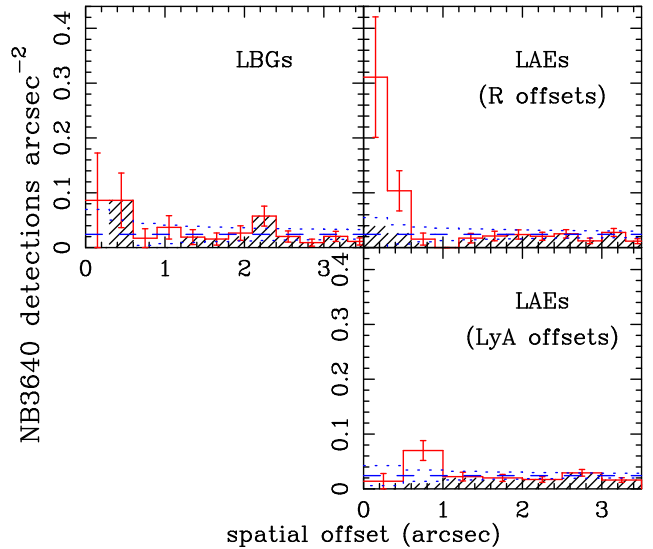}
\caption{\small 
The radial surface density of NB3640 detections around galaxies in our
LBG and LAE samples.  The solid red histograms include all
  detected sources; the subset of these sources associated
  with obvious neighbors (which we have
  excluded as possible LyC detections, see \S\ref{s:contam}) are
  represented by the hatched region.  The dashed lines indicate the
  global surface density of sources in our NB3640 magnitude range and
  thus represent the expected levels of contamination, while the dotted
  lines represent the expected 1$\sigma$ scatter in the contamination.   
  The excess surface density at low-offsets indicates that some of
  our low-offset LBG NB3640 detections and many
  of the low-offset LAE NB3640
  detections are physically associated with the $z\simeq 3.09$ sources
  and not random foreground interlopers.  The top panels use the displacement
  of the LyC centroid from that of the $R$ band detection (or $BV$
  or NB4980 for LAEs undetected in $R$).
  The bottom panel uses displacements from the $LyA$ detections of LAEs.
  As the NB3640 detections tend to be more spatially coincident with
  the non-ionizing UV continuum than with the $LyA$ emission, the
  significance of the excess of surface density is greater when using
  the $R$ band offsets.
\label{f:radial}}
\epsscale{1.}
\end{figure}

\section{ACCOUNTING FOR LOWER-REDSHIFT INTERLOPERS AND IGM ABSORPTION}
\label{s:monte}

As in \citet{nestor2011}, we use the retained NB3640 detections in our samples 
to estimate the contributions to the cosmic ionizing background from
LBGs and LAEs at $z\sim3$.  However, we must first account for two effects.
Although we have removed from our samples targets with  
evidence of foreground galaxies contaminating their NB3640 flux
measurements, the non-detection
of emission or absorption features belonging to a lower-redshift system is not
sufficient to rule out the presence of an interloper.  Such an
interloper may, for example, lack features strong enough to be
detected in our data, lie at a redshift such that no strong
spectral features fall within our wavelength coverage, or not be
covered by the position of the slit.
Not accounting for such interlopers would
result in an overestimation of \elyc.  At the same time, the ionizing
flux from LyC-leaking galaxies will experience an unknown amount of 
absorption by neutral gas in the intervening IGM, which will decrease
the observed value of \elyc\ relative to the intrinsic value.  We account for
these two effects in a statistical manner using a pair of Monte
Carlo simulations, which make use of the global surface density of 
NB3640 detections in the relevant magnitude range, and the
observed \lya\ forest statistics at $z\simeq3$.  The
procedures are similar to those used by \citet{shapley2006} and
\citet{nestor2011}, but employ updated methods
and statistics.  We briefly summarize each method here.

\subsection{Contamination Simulation}
\label{s:contam}
The centroids of the NB3640 detections in our samples are generally
offset from the corresponding 
centroids of the non-ionizing UV emission.  Many of these
NB3640 detections, despite their relatively close proximity on the sky
to LBGs and LAEs, have clear associations with unrelated 
neighboring sources and thus have already been rejected as possible
LyC emission by \citet{nestor2011}.
We have now also removed from our samples those galaxies with
unconfirmed redshifts, and have 
rejected NB3640 detections displaying evidence for contamination in their
LRIS or NIRSPEC spectra.  In Figure~\ref{f:radial} we show the resulting surface
density of the total (open histogram) and rejected (hatched region) number of
NB3640 detections, as a function of offset, for both the LBG and LAE samples.  
For the LAEs, we present surface densities computed using the offsets of
the NB3640 detections from both the UV continuum (i.e., $R$-band, or
$BV$ for LAE081 which is undetected in $R$) and $LyA$-band centroids.
The dashed line in Figure~\ref{f:radial} indicates the global
surface density of sources in the range of NB3640 magnitudes spanned
by our detections, $\rho_S = 0.024$~arcsec$^{-2}$ over $25 \le
m_{\mathrm{NB3640}} \le 27.25$, which corresponds to the expected level of contamination.
Both the LBG and LAE samples have, at relatively small ($\la 1$\arcsec) offsets, an
excess of NB3640 detection surface density above the 
expected foreground level.  In particular, the excess for the LAEs
using $R$-band offsets is strikingly significant.
In order to quantify the number NB3640 detections that may be due to
foreground interlopers, in addition to those already rejected,
we performed a Monte Carlo simulation using the observed surface
densities of sources and predicted levels of contamination.

To account for the increased probability that NB3640 detections at
larger offsets are contaminants, we considered each individual
detection in turn, computing the contamination probability based on 
offset in the following manner.  For the $k^{\mathrm{th}}$ detection, we
constructed an annulus, having area $A^k$,
encompassing the detection and centered on the non-ionizing UV flux
centroid.  We applied this same annulus to each of
the $N$ galaxies in the sample ($N=41$ or 91 for the LBG or LAE
samples, respectively) and determined the number of previously
rejected detections, $n_{rej}^k$, and putative LyC detections,
$n_{LyC}^k$, contained within the $N$ annuli.  We then constructed
the probability distribution for the number of expected random foreground
interlopers, $n_{fore}^k$, in the regions spanned by the annuli:
\begin{equation}
P(n_{fore}^k) = \left(\begin{array}{c}N\\n_{fore}\end{array}\right)\,p^{n_{fore}}\,(1-p)^{(N-n_{fore})},  
\end{equation}
where $p=A^k\times\rho_S$ is the probability that a given LBG or LAE has
a random foreground contaminant in $A^k$
\citep[see, e.g.,][]{vanzella2010a,nestor2011}.   In each realization of
the Monte Carlo simulation, we randomly chose a value for $n_{fore}^k$
from $P(n_{fore}^k)$.  Of these $n_{fore}^k$ interlopers,
$n_{rej}^k$ had already been accounted for.  Thus the number of {\it additional}
interlopers predicted to lie within the $N$ annuli is $n_{fore}^k - n_{rej}^k$.  The 
NB3640 detection in question was flagged as an interloper if $n_{fore}^k -
n_{rej}^k \ge n_{LyC}^k$, and was retained if $n_{fore}^k - n_{rej}^k \le 0$.
Otherwise, we randomly determined if
the NB3640 detection was to be flagged as an interloper
based on a probability $= (n_{fore}^k - n_{rej}^k) / n_{LyC}^k$.  
We then proceeded to the next (i.e., $k^{\mathrm{th}}+1$) detection
and repeated the entire process.  Once each
possible LyC detection in the sample had been considered, we
recorded the total predicted number of 
additional interlopers.  The simulation was repeated for a total of
1000 iterations to determine the expected number and uncertainty in the
average number of uncontaminated NB3640 detections.

\begin{figure}
\epsscale{1.0}
\plotone{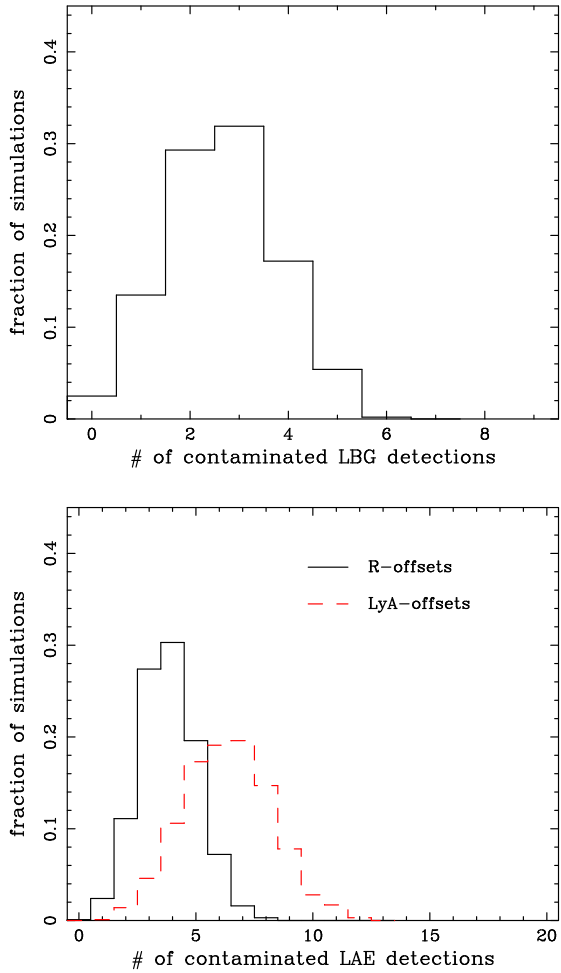}
\caption{\small 
The distribution of the number of contaminated 
NB3640 detections as predicted
by our Monte Carlo simulations, excluding the three LBGs and three
LAEs showing evidence for foreground contamination in their spectra. 
The upper panel shows the results for the LBGs, which have a total of
six possible LyC detections, while the lower panel shows the results
for the LAEs, which have a total of 17 possible LyC detections.  In
the simulations, NB3640 detections with larger spatial offsets from
the corresponding non-ionizing UV flux centroids are more likely to be
flagged as interlopers.  For the LAE sample, we considered offsets
from both the $R$ band (black solid histogram) and $LyA$ (red dashed
histogram) flux centroids. 
\label{f:csim}}
\epsscale{1.0}
\end{figure}

The distribution of the number of NB3640 detections flagged
as interlopers in each realization of the simulation is shown in
Figure~\ref{f:csim} for the LBG and LAE samples.  The widths of the distributions
depend slightly on the size of annular apertures used in the
simulations due to NB3640 detections stochastically entering and exiting
the apertures when their sizes were varied.  The variability in the
distribution widths was small and did not depend systematically on the
aperture sizes.  We used a circle of radius
equal to the seeing FWHM in the NB3640 image (0\secpoint8) for 
detections with offsets $\le$0\secpoint4.  For detections at larger offsets, the
radii of the annuli were set such that all values of $A^k$ were
equal.  Our simulation
suggests that $2.6 \pm 1.2$ of the 6 possible LBGs with LyC detections 
are contaminated by foreground sources.  Together with the three
LBGs showing evidence for contamination in their spectra discussed in
Section~\ref{s:lbg_contam}, the resulting contamination
rate is $62\pm13\%$ (i.e., 5.6$\pm$1.2 of 9).
The contamination-corrected detection rate for the sample as a whole
is $8\pm3$\% ($3.4\pm1.2$ of 41). 

Using $R$-band offsets, our simulation suggests that $3.8 \pm 1.3$
of the 17 possible LAEs with LyC detections
are contaminated.  Together with the three LAEs with contamination
discussed in Section~\ref{s:laecont},
the resulting contamination rate for the LAE sample is $34\pm7$\%
($6.8\pm1.3$ of 20) and the 
contamination-corrected detection rate is $15\pm1$\% ($13.2\pm1.3$ of 91).
If we instead use $LyA$ offsets, the predicted number additional
contaminants becomes $6.4 \pm 1.9$, the contamination rate becomes
$47\pm10$\% ($9.4\pm1.9$ of 20) and the detection rate becomes
$12\pm2$\% ($10.6\pm1.9$ of 91). 

In addition to estimating the number of foreground interlopers, our
simulation computes the (contamination-corrected) average LyC and $R$
magnitudes and uncertainties.  The uncertainties include sample
variance computed by first randomly reassigning individual 
magnitudes based on the measured magnitude and error,
assuming Gaussian magnitude uncertainties determined from
our photometric simulations, and then bootstrap
re-sampling each data set.  When computing the sample-average LyC
magnitudes, we 
alternately assumed that NB4630 non-detections and detections flagged
as interlopers had flux levels equal to zero or the maximum average
magnitude consistent with the (1$\sigma$) limits set by our stacking analysis.
The resulting contaminated-corrected sample-average NB3640$-R$ colors are listed in
Table~\ref{t:4}.

\subsection{IGM Simulation}
To investigate the attenuation of escaping LyC flux by absorption from
the IGM, we used the observed column
density and frequency distributions of the \lya\ forest over the
relevant redshift range, $1.7 \le z \le z_{source}$, to model 500 random
sightlines through the IGM for each LBG and LAE redshift.
We then computed
the attenuation of the continuum flux in the NB3640 filter due
to the randomly generated neutral clouds.  In this
manner we
determined the fraction of flux transmitted, $t^j$, for each of the 500
sightlines.  The
process was identical to that described in \citet{nestor2011} except
that we made use of updated estimates of the IGM opacity (Rudie et
al.\ 2012, in prep.) and a more 
precise treatment of the higher-order Lyman absorption lines from each
cloud.  As the galaxies in our sample lie either in or behind the
SSA22a protocluster, it may be expected that the IGM in the proximate
foreground of our sources is either more opaque due to the mass-overdensity
of the region, or less opaque due to the overdensity of ionizing
emissivity.  In either scenario, the difference in average IGM opacity
should manifest in differences in rest-frame UV colors that
span \lya\ emission and/or the Lyman limit.  We compared the
$U-G$ and $G-R$ colors for our sources at $3.06 \le z
\le 3.12$ with those of field LBGs in the same redshift range and found no
statistical differences in their distributions, suggesting that any
such effect is small in relation to our other uncertainties.

For each of the $N$ redshifts in a sample, the expectation value for the fraction of
transmitted flux, $\left< t(z) \right>$, is simply the average of the
500 simulated $t^j(z)$ values.  For a single galaxy at $z\simeq3.09$, the
distribution of $t$ is very broad, ranging 
from $\approx0-60$\% \citep[see, e.g., Figure 8 of][]{nestor2011}.  The uncertainty in $t(z)$ is
thus a major uncertainty in our estimate of the contribution to \elyc\
from each individual source.  The LBG and LAE {\it sample-average}  LyC flux
values, however, can be corrected with much less uncertainty.
We define the sample-average transmission as 
\begin{equation}
\bar{t}_{sample} \equiv  \frac{\sum_{i=1}^N t(z_i) F_i}{\sum_{i=1}^N F_i},
\label{eqn:trans}
\end{equation}
where $F_i$ are the $N$ individual intrinsic (i.e., prior to any IGM
absorption) NB3640 fluxes, and $t(z_i)$ are the
actual IGM transmission values.  As before, $N = 41$ or 91 for
the LBG or LAE samples, respectively.  Although both the individual $F_i$ and
$t(z_i)$ values are unknown, as the two sets are independent 
the expectation value for $\bar{t}_{sample}$ is simply
the average of the $N$ values of $\left< t(z) \right>$, and is thus independent of
the values of $F_i$ and $t(z_i)$.   The uncertainty in
$\bar{t}_{sample}$, $\sigma_{\bar{t}}$, does depend on the $t(z)$ and $F$
distributions, however.  While we have an accurate model for the probability
distributions for the $t(z_i)$ values, the $F_i$ values are poorly
constrained.  In \citet{nestor2011}, we estimated $\sigma_{\bar{t}}$
by randomly drawing the $N$ values of $t(z_i)$ from the corresponding $t^j(z_i)$
values to compute $\bar{t}_{sample}$.  We computed
$\bar{t}_{sample}$ in this manner 1000 times, and equated 
$\sigma_{\bar{t}}$ to the standard deviation of the resulting
$\bar{t}_{sample}$ values.  That procedure is equivalent to assuming 
the $N$ values of $F_i$ are all equal, and will underestimate
$\sigma_{\bar{t}}$ for more realistic distributions of $F$.  

To improve upon our estimation of $\sigma_{\bar{t}}$, we assumed an
exponentially decreasing function for the intrinsic flux probability
distributions: $p(F) \propto e^{-F/\beta}$.  This choice of
parametrization has the advantage that the results are only mildly
sensitive to the $e$-folding parameter, $\beta$.  To determine the
best choice for $\beta$, we convolved $p(F)$ with the $t(z)$
distributions for a range of $\beta$ values.
We then used the resulting attenuated-flux probability distributions to compute the
likelihood of our contamination-corrected LBG and LAE NB3640 data sets, retaining
the value of $\beta$ that maximized these likelihoods.  To
determine $\sigma_{\bar{t}}$, we randomly
selected the $N$ $F_i$ values from the maximum likelihood 
exponential distributions.  For each flux value, we
also randomly selected $t(z_i)$ from one of the sets of $t^j(z_i)$
values.  These $F_i$ and $t(z_i)$ values were used with Equation~\ref{eqn:trans} to
determine $\bar{t}_{sample}$.  This process was repeated 1000 times,
and $\sigma_{\bar{t}}$ was set equal to the
standard deviation in the 1000 simulated $\bar{t}_{sample}$ values.

We found that an exponential form for $p(F)$, when convolved with our
modeled IGM transmission values,  resulted in a  
qualitatively acceptable match to our observed NB3640 fluxes.
However, as the actual $F$ distributions are only poorly constrained by the
data, it is likely that other functional forms are also able to match
the observations and 
may result in different estimates of $\sigma_{\bar{t}}$.  In
particular, if the true $p(F)$ distributions contain a spike at zero flux
(as suggested by our non-detection of NB360 flux in the stacked images
of the non-detections), we would still be underestimating
$\sigma_{\bar{t}}$.  Nonetheless, as our exponential model represents an
improvement over the past method which effectively assumed $p(F)=$
constant, and the error budget in \elyc\ is 
not dominated by the uncertainty in $\bar{t}_{sample}$, we adopt the
$\sigma_{\bar{t}}$ values determined with an exponential $p(F)$.

For our LBG sample, we found $\bar{t}_{sample} = 0.298 \pm 0.040$, and
for our LAE sample we found $\bar{t}_{sample} = 0.320 \pm 0.027$.  We
used these values to correct the sample-average NB3640 magnitudes and
NB3640$-R$ colors.  These colors can be expressed in terms of the
sample-average  UV-to-LyC flux-density ratios, $\eta \equiv
\left<F_{UV}/F_{LyC}\right>$.  After applying the contamination and
IGM corrections, we find $\eta_{LBG}=18.0^{+34.8}_{-7.4}$ for our LBG
sample, and $\eta_{LAE} =3.7^{+2.5}_{-1.1}$ for our LAE sample.   These
values are consistent with, though larger than, those estimated by
\citet{nestor2011}: $\eta_{LBG}=11.3^{+10.3}_{-5.4}$ for LBGs
and $\eta_{LAE}=2.2^{+0.9}_{-0.6}$ for LAEs.  The relative
uncertainties in our updated estimates of $\eta$ are larger than those
in our previous estimates, due to an
improved treatment of the errors.   The updated samples also 
contribute to the increased relative uncertainties, as the newly confirmed LBGs
have, on average, larger $R$-band photometric uncertainties than those
in the previous sample, and the 
spectroscopically-confirmed LAE sample used here is smaller than the
full photometrically-selected LAE sample.  Nonetheless, by refining
our samples and including empirical evidence for the presence or
absence of foreground contamination in individual systems, our revised
values represent improved estimates of $\eta$ for LBGs and LAEs at
$z\sim3$.  We list the colors and $\eta$ values in Table~\ref{t:4} for
our raw LBG and LAE samples, for those samples after application of
the contamination corrections, and after correcting for both
contamination and IGM absorption.  The contamination- and
IGM-corrected values are used below to estimate \elyc.

\section{RESULTS}
\label{s:results}
\subsection{Revised Estimate of $z \simeq 3.01$ LyC Emissivity}

One of the primary goals of this paper is to determine the global luminosity space
density of ionizing radiation, \elyc, contributed by star-forming galaxies at
$z\sim3$.   As the SSA22 protocluster 
represents an over-dense volume of the Universe, computing \elyc\
directly from our data would overestimate the space density of
ionizing flux.   However, the UV-continuum luminosity functions of
LBGs and LAEs have been measured over relatively 
representative volumes, at rest-frame effective wavelengths close to
those of our $R$-band data ($\lambda \sim 1600$\AA), by \citet{reddy2008} and \citet{ouchi2008},
respectively.  Thus, we can use the values determined for the sample-average UV-to-LyC
flux-density ratios together with the corresponding luminosity functions to determine \elyc:
\begin{equation}
\epsilon_{LyC} = \int \frac{1}{\eta}\, L\, \Phi\, dL,
\label{eqn:elyc}
\end{equation}
where $L$ refers to the {\it non}-ionizing UV continuum luminosity, and
$\eta$ may be a function of $L$.
When determined in this fashion, the value computed for \elyc\ will
depend on the choices of shape
and normalization for the luminosity function, and the range in luminosity over which
Equation~\ref{eqn:elyc} is integrated.  

\citet{reddy2008} give Schechter function parameters for the
$\lambda\sim1700$\AA\ UV-continuum luminosity function of LBGs at $z\sim3$ of
$\phi^*_{{LBG}}=1.66 \times 10^{-3}$~Mpc$^{-3}$,
$M^{*}_{{LBG}} = -20.84$, and $\alpha _{{LBG}} = -1.57$.
For LAEs at $z\sim3.1$, \citet{ouchi2008} determine
$\lambda\sim1600$\AA\ UV-continuum luminosity function parameters of
$\phi^*_{{LAE}}=0.56 \times 10^{-3}$~Mpc$^{-3}$, 
$M^{*}_{{LAE}} = -19.8$, and $\alpha _{{LAE}} = -1.6$.  However, their 
luminosity function is determined for LAEs having REW $\gtrsim 64$\AA, while our
sample has REW $\gtrsim 20$\AA.  From the REW distribution determined
by \citet{gronwall2007}, we estimate that LAEs represented by our
sample are a factor of $\approx 1.8$ more common at $z\sim3$ than
those described by \citet{ouchi2008}.  We therefore adopt $\phi^*_{{LAE}} =
1.01 \times 10^{-3}$~Mpc$^{-3}$ for our LAE sample.\footnote{Although
  the UV luminosity function of LAEs having 20\AA\ $\le$ REW $\la$
  64\AA\ (which compose $\sim$ half of LAEs with REW $\ge 20$\AA) may,
  in principle, have different values of $M^*$ and $\alpha$ compared
  to LAEs with REW $\ge$ 64\AA, we proceed under the assumption that the
  \citet{ouchi2008} values for $M^*$ and $\alpha$ apply for our entire
  range of REW.} 

\begin{figure}
\epsscale{1.15}
\plotone{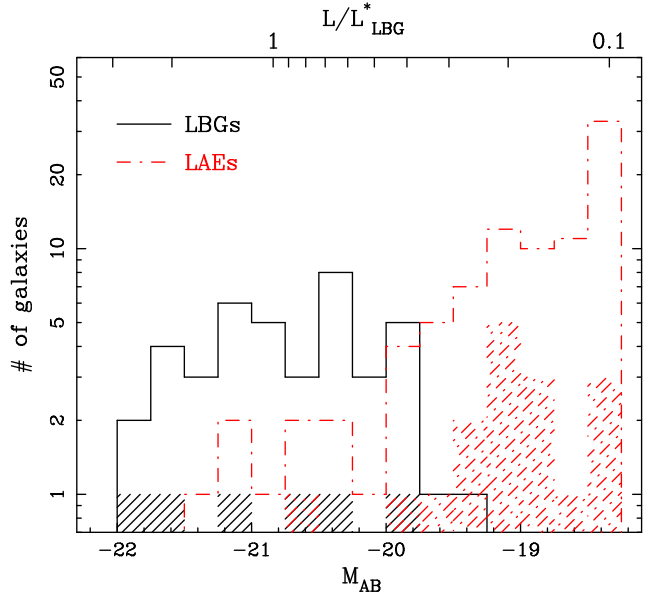}
\caption{\small The rest-frame $\lambda \sim 1600$
\AA\ absolute magnitude distribution for our LBG (solid black) and LAE
(dash-dot red) samples.  The hatched regions indicate galaxies with
NB3640 detections.  The faintest bin in the LAE histogram also
contains the 10 LAEs fainter than our detection limit of $M_{AB} =
-18.3$.  
The top axis indicates the corresponding luminosity
relative to the LBG characteristic luminosity $L^*_{LBG}$.  Our LBG
sample is dominated by galaxies brighter than $M_{AB}=-20$,
corresponding to $L \simeq 0.46 L^*_{LBG}$, while our LAE sample is
dominated by sources fainter than $M_{AB}=-20$.
\label{f:lf}}
\epsscale{1.}
\end{figure}

The absolute magnitude distributions for our LBG and LAE samples are
shown in Figure~\ref{f:lf}.  Spectroscopic confirmation of
color-selected LBGs in the SSA22a field was limited to ${\cal R} \le
25.5$.  Consequently, the bulk of our LBG sample is brighter than 
$M_{AB}=-20$.  In contrast, our LAE sample, which includes
galaxies down to our photometric limit of $R=27.3$ (or $M_{AB} \simeq
-18.3$ at $z\simeq 3.09$), is dominated by galaxies
fainter than $M_{AB}=-20$.  Thus, we first compute the contributions
to \elyc\ from LBGs and LAEs separately, and only from sources within the
UV-continuum luminosity 
ranges over which we determined the values of $\eta$: $M_{AB} \le
-20.0$ or our LBG sample, and $-20 < M_{AB} \le -18.3$ for our LAE
sample.  The absolute magnitude limits $M_{AB} = -20.0$ and $-18.3$ correspond
to  $L_{min} = 0.46 L^*_{{LBG}}$ and $0.1 L^*_{{LBG}}$, respectively. 
With these magnitude ranges, the parameters for the respective luminosity
functions discussed above, and 
$\eta_{LBG} = 18.0^{+34.8}_{-7.4}$ for LBGs and $\eta_{LAE} =
3.7^{+2.5}_{-1.1}$ for LAEs as determined in 
Section~\ref{s:monte},  we find \elyca$ =
5.2^{+3.6}_{-3.4}\times10^{24}$~ergs~s$^{-1}$~Hz$^{-1}$~Mpc$^{-3}$ for
$M_{AB} \le -20.0$ and  \elycb$ =
6.7^{+3.6}_{-3.4}\times10^{24}$~ergs~s$^{-1}$~Hz$^{-1}$~Mpc$^{-3}$ for $-20 <
M_{AB} \le -18.3$.  

We now turn our attention to the total contribution to \elyc\ from
{\it all} star-forming galaxies brighter than $M_{AB} = -18.3$ at $z\sim3$.
First, however, we must address the principle source of the large
(factor of $\sim5$)
difference between the values determined for $\eta$ in
our LBG and LAE samples.
As the LBG and LAE samples are dominated by galaxies with relatively
bright and faint UV-continuum luminosities, respectively, it may be
that $\eta$ has a strong dependence on galaxy luminosity.  In this
scenario, which we refer to as our luminosity-dependent $\eta$ model,
we identify $\eta_{LBG}$ with {\it all} star forming galaxies having $M_{AB}
\le -20.0$ and and $\eta_{LAE}$ with {\it all} star forming galaxies
having $-20 < M_{AB} \le -18.3$.  Here we make the
approximation that the \citet{reddy2008} luminosity function, which is
determined from color-selected galaxies, is a complete census of all
star-forming galaxies with $M_{AB} \le -18.3$; i.e., that there are no
sources brighter than this limit that contribute to \elyc\ at $z\sim3$
with UV colors such that they would not be selected as LBGs.  With
these assumptions we can integrate Equation~\ref{eqn:elyc} piecewise:
\begin{eqnarray}
\nonumber \epsilon_{LyC}^{lum.-dep.} & = & \frac{1}{\eta_{LAE}}\,\int_{0.1L^*}^{0.46L^*} L\,\Phi_{LBG}\,dL  \\
 & + & \frac{1}{\eta_{LBG}}\, \int_{0.46L^*}^{\infty} L\, \Phi_{LBG}\, dL,
\label{eqn:elyc2}
\end{eqnarray}

finding 
$\epsilon_{LyC}^{lum.-dep.}=32.2^{+12.0}_{-11.4}$~ergs~s$^{-1}$~Hz$^{-1}$~Mpc$^{-3}$, 
with contributions from sources fainter
and brighter than $M_{AB}=-20$ of
$27.0^{+11.4}_{-10.9} \times 10^{24}$~ergs~s$^{-1}$~Hz$^{-1}$~Mpc$^{-3}$ and 
$5.2^{+3.6}_{-3.4} \times 10^{24}$~ergs~s$^{-1}$~Hz$^{-1}$~Mpc$^{-3}$, respectively.
This value of $\epsilon_{LyC}^{lum.-dep.}$ is $\approx 60$\% of that estimated
in \citet{nestor2011}, primarily due to our refined values
for \elbg\ and \elae.

Alternatively, the large difference in $\eta$ between the LBG and LAE
samples may be driven by some property of star-forming galaxies
associated with LAEs.   In this scenario, which we refer to as our LAE-dependent $\eta$ model, 
we can estimate the total value of \elyc\ by assuming \elae\ holds for all
LAEs over the full range $M_{AB} \le -18.3$, and \elbg\
is representative of all LBGs with $M_{AB} \le -18.3$ that are not
also LAEs.  The fraction of
LBGs that are not LAEs is determined from the LBG and LAE space densities,
which can be computed by integrating
the respective LBG and LAE luminosity functions, $\rho = \int_{L_{min}}^{\infty}
\Phi\, dL$.\footnote{In this simple model, we do not account for the
  observed luminosity dependence of the fraction of LBGs that are also
  LAEs  \citep[see, e.g.,][]{stark2010}.}  Setting $L_{min} = 0.1L^*_{{LBG}}$, we estimate that
23\% of LBGs are also LAEs with REW $\ga 20$\AA, which is consistent
with past results at $z\sim3$
\citep{steidel2000,shapley2003,kornei2010}.  Here we have made the
approximation that LAEs comprise a sub-sample of LBGs; i.e., that all
LAEs at $z\sim3$ with $M_{AB} \le -18.3$ would meet the
color-selection criteria used by \citet{reddy2008}.
Equation~\ref{eqn:elyc}
then becomes:
\begin{eqnarray}
\nonumber \epsilon_{LyC}^{LAE-dep.} & = & \frac{0.77}{\eta_{LBG}}\, \int_{0.1L^*}^{\infty} L\, \Phi_{LBG}\, dL \\
 & + & \frac{1}{\eta_{LAE}}\, \int_{0.1L^*}^{\infty} L\, \Phi_{LAE}\, dL,
\label{eqn:elyc3}
\end{eqnarray}
 resulting in 
$\epsilon_{LyC}^{LAE-dep.}=16.8^{+6.9}_{-6.5} \times 10^{24}$~ergs~s$^{-1}$~Hz$^{-1}$~Mpc$^{-3}$, with
contributions from non-LAE and LAE galaxies of
$8.2^{+5.8}_{-5.5} \times 10^{24}$~ergs~s$^{-1}$~Hz$^{-1}$~Mpc$^{-3}$ and  
$8.6^{+3.7}_{-3.5} \times 10^{24}$~ergs~s$^{-1}$~Hz$^{-1}$~Mpc$^{-3}$, respectively.
This value of $\epsilon_{LyC}^{LAE-dep.}$ is similar to that estimated
for just LBGs with $L\ge 0.1 L^*$ in \citet{nestor2011}, and $\approx 30$\%
of the total value of \elyc\ estimated by \citet{nestor2011}.
This difference is due to the combination of our improved measurements
of \elbg\ and \elae\ and the reduced contribution from sources with low
values of $\eta$ in our LAE-dependent model.  

It is difficult, given the present data, to distinguish which of the
luminosity-dependent or LAE-dependent $\eta$\ models is more
appropriate.  In \citet{nestor2011}, we investigated differences in
the average \lya\ emission strengths of systems with NB3640 detections
compared
to those without detections by creating stacks of LRIS spectra of our
LBGs, and by comparing the distributions of
photometrically-estimated \lya\ REWs of our LAEs.  For both LBGs and LAEs, we found that
systems with NB3640 detections tend to have weaker \lya\ emission
relative to those not detected in NB3640.  We
repeated these tests using our refined LBG and LAE samples and
excluding systems showing evidence for contamination in their
spectra, and found trends consistent with our previous results. 
Thus, it is unlikely that \lya\ emission strength has a direct
influence on the average value of $\eta$.  However, 
the LyC properties of $z\sim3$ galaxies may be dependent on some other
property or combination of properties, such as star formation surface density, stellar
population age, metallicity, etc., that is more typical of the LAEs in
our sample compared to our LBGs, independent of UV-continuum luminosity.
For example, in a sample of local LBG analogs, \citet{heckman2011}
present evidence for small covering factors for optically thick
neutral gas in the most compact galaxies in their sample.
Indeed, in the available $HST$/ACS F814W
imaging, the LAEs in our sample do appear, qualitatively, more 
compact on average than the LBGs.  
Nonetheless, as our LBG and LAE samples are largely distinct in 
magnitude range, future work including including fainter LBGs and
larger LAE samples is needed to clearly differentiate between the two
proposed scenarios.

We summarize the various contributions to \elyc\ in
Table~\ref{t:5}.  The uncertainties in the values of \elyc\ are
dominated by the uncertainties in our estimates $\eta$, which are
large compared to the errors in the first moments of the luminosity
functions.  The uncertainties
in \elyc\ do not, however, include systematic errors arising from our 
choice of luminosity functions.  For example, \citet{reddy2009} find a
steeper faint-end slope at $z\sim3$  previous works \citep[see
discussion in][]{reddy2009}.  Furthermore, in estimating $\eta$\ we
have only accounted for the contribution from galaxies brighter than
$0.1L^*$, i.e., the luminosity range over which we have empirical
estimates of $\eta$.  We note that, for the
\citet{reddy2008} luminosity function, galaxies fainter than $0.1L^*$
contribute $\sim40$\% of the luminosity density at $1600$\AA.

Given the spectral shape of the ionizing continuum flux and the mean
free path to ionizing radiation in the IGM, $\lambda_{\mathrm{mfp}}$, the value of \elyc\
implies a corresponding (proper) hydrogen photoionization rate in 
the IGM, $\Gamma_{\mathrm{H\,{\sc I}}}$.  If we assume a power law
form for the LyC flux density such that $f_{\nu} \propto \nu^{-\alpha}$, then
\begin{equation}
\Gamma_{\mathrm{H\,{\sc I}}} =
\frac{(1+z)^3\,\sigma_{\mathrm{H\,{\sc I}}}\, \lambda_{\mathrm{mfp}}}{h\,(3 + \alpha)}\, \epsilon_{LyC}^{all,\,L\ge0.1L^*}.
\end{equation}
where $h$ is Planck's constant, and $\sigma_{\mathrm{H\,{\sc
      I}}} = 6.3 \times 10^{-18}$~cm$^{-2}$ is the atomic hydrogen
photoionization cross section at the Lyman limit.  Following
\citet{nestor2011}, we adopt $\alpha = 3$ and $\lambda_{\mathrm{mfp}}
= 75.6$~Mpc.  For the two estimates of 
\elyc\ determined above, we find $\Gamma_{\mathrm{H\,{\sc
      I}}}^{lum.-dep.} = 
2.7^{+1.0}_{-0.9} \times 10^{-12}$~s$^{-1}$ in our luminosity-dependent
$\eta$ model, and $\Gamma_{\mathrm{H\,{\sc I}}}^{LAE-dep.} =
1.4\pm0.6 \times 10^{-12}$~s$^{-1}$ in our LAE-dependent
$\eta$ model.\footnote{More recent estimates of the mean free
  path of ionizing photons at $z\sim3$ suggest
  $\lambda_{\mathrm{mfp}} = 63$~Mpc in the IGM, and
  $\lambda_{\mathrm{mfp}} = 49$~Mpc 
 if the opacity contributed by the circum-galactic medium of
LBGs is included (G.~Rudie, private communication).  Adopting these
values would lead to estimates of $\Gamma_{\mathrm{H\,{\sc I}}}$ that
are 83\% and 65\%, respectively, of the values presented here.}  We
list these values of $\Gamma_{\mathrm{H\,{\sc 
      I}}}$ in Table~\ref{t:6}, together with  
estimates of $\Gamma_{\mathrm{H\,{\sc I}}}$ in the $z\sim3.1$ \lya\ forest by
\citet{meiksin2004}, \citet{bolton2007}, and \citet{faucher2008}.
The values we obtain for $\Gamma_{\mathrm{H\,{\sc I}}}$ are
larger than the values reported in the literature by a factor of
$\approx 2-4$, although they are consistent given the uncertainties
with the literature values at $\sim 2 \sigma$ and $\sim 1 \sigma$, for
the luminosity- and LAE-dependent $\eta$\ assumptions,
respectively. 

\subsection{Escape Fractions}
It is often of interest to estimate the fraction of LyC radiation 
produced by star formation that escapes into the IGM,
$f_{esc}^{LyC}$.  This value can be useful for constraining models of  
star formation histories, star formation feedback in the interstellar
medium (ISM), etc.  Additionally, \fesc\ can 
provide a useful parametrization in reionization models.
The value of \fesc\ can be estimated empirically by 
\begin{equation}
f_{esc}^{\mathrm{LyC}} =
  \frac{\eta^{\mathrm{stars}}}{\eta}\, f_{esc}^{\mathrm{UV}},
\end{equation}
where $\eta^{\mathrm{stars}}$ is the
intrinsic ratio of UV-to-LyC luminosity
densities in star-forming regions, and $f_{esc}^{\mathrm{UV}}$ is the 
escape fraction of non-ionizing UV radiation.  \citet{reddy2008} have estimated
$f_{esc}^{\mathrm{UV}} \approx 20$\% in LBGs at $z\sim3$.  
LAEs, however, exhibit bluer average UV continua
and smaller UV attenuation, with $f_{esc}^{\mathrm{UV}} \approx 30$\%
\citep[e.g.,][]{ouchi2008,kornei2010,blanc2011}.  The term 
$\eta^{stars}/\eta$, which is often referred to as the relative escape
fraction, is a measure of the attenuation of the UV flux relative
to that of the LyC flux.   
In this work we have estimated $\eta$ for samples of LBGs and LAEs.
The value $\eta^{stars}$, however, must be obtained
from spectral synthesis models, and depends on
the ages of the stellar populations.  Even for a given star-formation
history, various models produce differing
values for $\eta^{stars}$.   To illustrate the dependencies of the
inferred value of \fesc\ on different models and parameters, we
determined $\eta^{stars}$ using the stellar population synthesis
models of \citet[][BC03]{bc2003}, as well as
a recent set of model integrated spectral energy distributions
produced by The Binary Population and Spectral Synthesis code 
\citep[BPASS,][]{eldridge2009}, which includes the effects of massive binary
stars and nebular emission.  For
each of the models we adopted a constant SFR and two different
metallicites ($Z=0.004$ and 0.020) and measured $\eta^{stars}$, using
our NB3640 and $R$ filter passbands, at three ages ($T =
10^6, 10^7$ and $10^8$~yr).   The model values we obtain for
$\eta^{stars}$ range from $\simeq 1.3$ for the $T=10^6$~yr, $Z=0.004$ population
in the BPASS models, to $\simeq 6.4$ for the  $T=10^8$~yr, $Z=0.02$
population in the BC03 models.  The complete set of modeled
$\eta^{stars}$ values are given in Table~\ref{t:7}.

For a given combination of $\eta$ and \fuv, the modeled values of
$\eta^{stars}$ imply a corresponding \fesc.  We show these \fesc\
values in Table~\ref{t:7} for each value of $\eta^{stars}$, assuming
$\eta_{LBG}=18.0$ for LBGs and $\eta_{LAE}=3.7$ for LAEs,
and adopting, in turn, \fuv $= 0.2$, 
\fuv $= 0.3$, and the limiting dust-free case where \fuv $=1$.
The \fesc\ values vary by factors of $\sim 5 -7$ for a given $\eta$
and \fuv.  They vary from 1\% for LBGs
in the youngest \fuv $= 0.2$ BPASS model, to $\sim50$\% for LAEs in the
oldest \fuv $= 0.3$ BC03 
models, and reach greater than unity for several of the dust-free LAE models.
For a given \fuv and $\eta$, \fesc\ varies most with 
stellar population age, although the choice of BC03 or BPASS model also affects the
inferred value of \fesc\ by as much as a factor of $\sim2$.
Over the range investigated, metallicity has only a small effect on \fesc.
There are other model inputs that we have not varied,
such as the stellar initial mass function or star-formation history,
that are also likely to affect $\eta^{stars}$ and therefore the 
determination of \fesc.  With these caveats in mind, 
LBGs at $z\sim3$ have a median age of $\sim 300$~Myr \citep{kornei2010}
with metallicities approaching $\sim$~solar
\citep{pettini2001,shapley2004,maiolino2008} and \fuv$\simeq0.2$,
suggesting $f_{esc}^{LyC\mathrm{, LBG}} \simeq 5 
- 7$\%.  High-redshift LAEs, in contrast, have typical ages $\sim 20$~Myr
\citep{gawiser2007}, low metallicity \citep{finkelstein2011,nakajima2012}, and
\fuv$\simeq0.3$, suggesting
$f_{esc}^{LyC\mathrm{, LAE}} \simeq 10 - 30$\%.
In general, we caution against the naive use of LyC escape fractions in
reionization models without consideration of the values of
$\eta^{stars}$ and $f_{esc}^{\mathrm{UV}}$ used in their
determination.  

In Table~\ref{t:4} we also list the values of $\eta$ for only galaxies with LyC
detections.   Our sample of LBGs with LyC
detections has $\eta = 1.7^{+1.7}_{-0.8} \approx \eta^{stars}$
for most of our modeled $\eta^{stars}$ values, with the exception of
the older BC03 models, implying a $\sim$~unity relative escape
fraction in LBGs with LyC detections.  Note that, as LyC radiation is
more susceptible than non-ionizing UV radiation to attenuation by gas
and dust, $\eta \approx \eta^{stars}$ implies little attenuation and
therefore a large value of $f_{esc}^{\mathrm{UV}}$.  
For the LAEs with NB3640 detections,
our IGM-corrected value $\eta = 0.4^{+0.2}_{-0.1}$ is inconsistent with all
of our modeled values of $\eta^{stars}$ at more than $3 \sigma$ (note
that  it is unphysical to have $\eta > \eta^{stars}$).  We can bring
our LAE detections-only value for $\eta$ in line with the model
predictions if we assume that we can only detect LyC in our LAE sample
along fortuitously clear IGM sightlines, and therefore can neglect the
IGM correction, resulting in $\eta = 1.4^{+0.7}_{-0.4}$.  This scenario, however, is very
unlikely.\footnote{The average of the largest 12\% (equal to the LAE
  LyC detection rate) of our modeled IGM transmissions is 0.566.
  Even adopting this value for the IGM correction we find $\eta=0.79$,
which is still significantly below our lowest predicted value of $\eta^{stars}$.}
Deciphering the correct interpretation of these surprising low values
of $\eta$ is a primary goal of our ongoing work, which includes
$HST$/UVIS LyC and UV imaging of many of our NB3640 detections.
Better multi-wavelength constraints on the stellar populations of
NB3640-detected LAEs will also be a key component of determining their
underlying nature.

In Section~\ref{s:contam} we found a LyC detection rate of $\sim8$\% in
our LBG sample, and $\sim12-15$\% in our LAE sample.  While it is not
clear if the LyC properties of a ``typical'' $z\sim3$ LBG or LAE are
similar to that of the average system, these rates do
indicate the solid angle over which LyC radiation escapes, at a level
above our detection limit, averaged over all galaxies in each sample.
In \citet{nestor2011}, we proposed a ``blow out'' model in which feedback
from regions of dense star formation clear portions of the ISM of gas
and dust.  When viewed along favorable sightlines, such regions appear
to have large escape fractions, as we find in our LBGs and LAEs with
LyC detections.  Galaxies that either have failed to sufficiently
remove their ISM over significant solid angle, or are viewed along
unfavorable sightlines, will appear to have negligible escape fractions.  The strong
LyC-flux upper-limits in systems without NB3640 detections, derived
from our stacking analysis, are consistent with this picture.  Additionally,
in the subsample of our LAEs having $HST$ imaging,\footnote{The number of LBGs with NB3640
  detections and $HST$ imaging is too small to make meaningful
  comparisons to the sample of LBGs without NB3640 detections.} we
find no significant difference in the distributions of sizes or
surface brightnesses for sources with and without NB3640 detections.
The similarity of these properties between NB3640 detected and
non-detected LAEs is
consistent with a scenario in which viewing angle is a 
significant factor in the ability to detect escaping LyC.

\section{DISCUSSION and SUMMARY}
\label{s:summary}
Directly studying galaxies in the ionizing continuum
is a difficult endeavor.  The dearth of QSOs above $z\sim 3$ suggests
that, at high redshift, the ionization balance in the IGM is maintained by
LyC flux escaping from star-forming galaxies.  However, the 
increasingly opaque IGM makes the detection of any escaping LyC flux
unlikely above $z\sim3.5$ \citep[see, e.g.,][]{vanzella2012}.  
Below $z\sim2.4$, the redshifted Lyman limit falls below the atmospheric cutoff
requiring observations from space to detect ionizing flux.  Current observations at
$z\approx1.3$ sampling rest-frame $\lambda
\sim 700$\AA\ have resulted only in upper limits to \elyc\
\citep{siana2007,siana2010}.  The non-detection of LyC emission at
$z\sim1$ together with 
the apparent need for a galaxy contribution to \elyc\ at high redshift
implies that \elyc\ evolves strongly over $z\sim 1 - 3$ \citep[see, e.g., Figure~9
of][]{nestor2011,inoue2006}.  We note, however, that if the average UV to LyC flux-density ratio $\eta$
is luminosity-dependent, the non-detection of LyC in star
forming galaxies at $z\sim1.3$ could be due in part to a selection
bias, as the galaxies observed by \citet{siana2010} are preferentially
bright.  

In any case, short of a large space-based program, the window
$2.5 \la z \la 3.5$ offers the best opportunity to identify and study
LyC-leaking galaxies.  As this and other recent works have shown,
however, difficulties remain even at $z\sim3$.  While the IGM
is more transparent to ionizing radiation at $z\sim3$ compared to
higher redshift, there remains an unknown level of IGM attenuation of
escaping LyC flux.  The only practical way to correct for the
attenuation is by applying a statistical correction, determined through
modeling of the IGM, to the sample-average LyC properties.  In this work, we have improved upon
past efforts to estimate this correction by using an updated estimate of the IGM opacity at
$z\sim3$, including a more precise treatment of the higher-order Lyman
absorption lines, and developing a new method to obtain a more realistic estimate of the
transmission uncertainties.  For the sample-average IGM correction
method to be effective, we require
a relatively large statistical sample.  This necessity presents a second
difficulty, as each galaxy requires a deep observation at a large observatory
to be detected in the UV and LyC at these redshifts.   Thus, 
we conducted our observations in a
single field that contains a large overdensity of galaxies in a narrow
range of redshift, using Keck and archival Subaru and $HST$ imaging
data, and new Keck/LRIS multi-object spectroscopy.  Our resulting LBG and LAE samples are the
largest of their types to have been imaged below the Lyman limit in an attempt to
identify LyC flux.  The samples presented here are improvements of
those in \citet{nestor2011}, through a $\sim 60$\% increase in the number of LBGs and
the spectroscopic confirmation of $\ga 90$\% of our LAE candidates.
Finally, perhaps
the most critical impediment to unambiguous identification of
leaking LyC at $z\sim3$ is the high sky surface density 
of foreground sources with observed-frame blue magnitudes
similar to expected $z\sim3$ LyC magnitudes.  Our new spectra 
mitigate this problem by allowing us to directly search for spectral
features from galaxies at lower redshift nearby on the sky (and thus
falling on the slit) to the $z\sim3$ galaxies.  As in
\citet{nestor2011}, we also statistically correct the samples for yet
unidentified foreground contaminants, here using a slightly refined
method.  Overall, in addition to clarifying the interpretation of the
NB3640 detections of several specific LBGs and LAEs,
our new data have led to improved estimates of the galaxy
contributions to the ionizing background flux and ionization rates.
Our main results include:

\begin{enumerate}
\item We obtained spectra of 41 LBG candidates and confirm redshifts
$z>2.45$ for 26.  Of these 26, 16 have $z>3.055$ such that our NB3640
filter is opaque above the redshifted Lyman break.  Together with the
25 previously identified $z>3.055$ LBGs in the field, our NB3640 image
samples below the Lyman limit for a total of 41 confirmed LBGs at $z >
3.055$.

\item We obtained spectra of 96 of the 110 LAE candidates from
\citet{nestor2011}.  We confirm redshifts $3.057 \le z \le 3.108$ for
87 of the candidates.  An additional four candidates, which are also
part of our LBG sample, have previously determined redshifts
$z>3.055$.  Thus, our our NB3640 image
samples below the Lyman limit for a total of 91 confirmed LAEs.

\item Nine galaxies in our LBG sample are detected in NB3640.  The
spectra of three of these detections contain evidence for the presence
of a lower-redshift interloper, and thus are removed from
consideration as LyC leakers.  Our Monte Carlo simulation suggests
that an additional $2.6\pm1.2$ of our NB3640 detections are
contaminated.  The resulting predicted contamination rate is
$62\pm13$\%, and the predicted LyC detection rate is $8\pm3$\%.

\item We detect 20 galaxies in our LAE sample in the NB3640 image.  The
spectra of three of these detections contain evidence for the presence
of a lower-redshift interloper, and thus are removed from
consideration as LyC leakers.  Our Monte Carlo simulation suggests
that an additional $6.4\pm1.9$ of our NB3640 detections are
contaminated.  The resulting predicted contamination rate is
$47\pm10$\%, and the predicted LyC detection rate is $12\pm2$\%.

\item Using the sample-average NB3640$-R$ colors, we determined the
sample-average UV-to-LyC flux-density ratios $\eta =
18.0^{+34.8}_{-7.4}$ for our LBG sample and $\eta = 3.7^{+2.5}_{-1.1}$ for
our LAE sample.  We then used these ratios, together with the $z\sim3$ LBG
and LAE luminosity functions to estimate the contributions to \elyc\
from each sample, over the luminosity ranges for which $\eta$ was
constrained: 
$\epsilon_{LyC}^{LBG}=
5.2^{+3.6}_{-3.4} \times 10^{24}$~ergs~s$^{-1}$~Hz$^{-1}$~Mpc$^{-3}$ for $M_{AB} \le
-20$ and  \elycb$ =
6.7^{+3.6}_{-3.4} \times 10^{24}$~ergs~s$^{-1}$~Hz$^{-1}$~Mpc$^{-3}$ for $-20 <
M_{AB} \le -18.3$.  The total galaxy contributions are
$\epsilon_{LyC}^{lum.-dep.} = 32.2^{+12.0}_{-11.4} \times 10^{24}$~ergs~s$^{-1}$~Hz$^{-1}$~Mpc$^{-3}$ in our
luminosity-dependent $\eta$ model or
$\epsilon_{LyC}^{LAE-dep.}=16.8^{+6.9}_{-6.5} \times 10^{24}$~ergs~s$^{-1}$~Hz$^{-1}$~Mpc$^{-3}$ in our 
LAE-dependent $\eta$ model, for galaxies with $M_{AB} \le
-18.3$ which corresponds to $L \simeq 0.1 L^*_{LBG}$ at $z\sim3$.

\item The intergalactic hydrogen photoionization rates $\Gamma$ inferred by our
values of \elyc, $\Gamma_{\mathrm{H\,{\sc I}}}^{lum.-dep.} =
2.7^{+1.0}_{-0.9} \times 10^{-12}$~s$^{-1}$ in our luminosity-dependent
$\eta$ model, and $\Gamma_{\mathrm{H\,{\sc I}}}^{LAE-dep.} =
1.4\pm0.6 \times 10^{-12}$~s$^{-1}$ in our LAE-dependent
$\eta$ model,  are larger than but consistent with published values
for $\Gamma$ measured in the \lya\ forest.

\item We use two suites of stellar population synthesis models to
predict the intrinsic UV-to-LyC flux-density ratio in star-forming
regions, $\eta^{stars}$, finding values that range from $1.3 - 6.4$.
The inferred sample-average values of \fesc\ for our samples depend on the
assumed values of $\eta^{stars}$ and \fuv.  Despite the
uncertainty in these two values, our best estimates of \fesc\ are 
$\sim 5-7$\% for LBGs and $\sim 10-30$\% for LAEs.  In our ``blow
out'' model for LyC escape, we predict 
near-unity apparent escape fractions in galaxies with LyC detections,
and negligible escape fractions in other systems.
\end{enumerate}

Despite our significant progress, several outstanding issues remain.  In order to assess
the robustness of our contamination corrections, it is important to
determine unambiguously the LyC or interloper nature of as many of our
NB3640 detections as possible.  For many of our galaxies with NB3640
detections, our LRIS and/or NIRSPEC data enable us to associate a
redshift with one or more specific regions (or ``clumps'') of
non-ionizing UV flux.  With the data at hand, we are already able to
place high confidence on the association of NB3640 detections with $z
> 3.055$ clumps in at least two of our objects, MD46 and LAE053.
We expect to increase the number of such high-confidence LyC
identifications in the near future with our ongoing programs.  Notably,
the resolution afforded by our ongoing
$HST$/UVIS cycle 19 imaging program will allow the measurement of
$F_{UV}/F_{LyC}$ in the individual clumps, thus providing strong
constraints on the nature of the NB3640 detections.  $K$-band and/or
$Spitzer$/IRAC data, together with our rest-frame UV observations, will
enable investigations of the stellar populations of galaxies with and
without LyC detections.
It is also necessary to extend our investigation to other fields in order to
ensure that our results are not a phenomenon (or, e.g., a systematic
bias) unique to the SSA22a field.  To this end, we are in the process
of finalizing a similar study in an independent field at $z\sim2.85$
(Mostardi et al., in prep).  Finally, the recently commissioned near-IR
multi-object spectrograph MOSFIRE at the Keck observatory will allow the study of
rest-frame optical spectral features of galaxies with LyC detections.

\acknowledgements
We thank Gwen Rudie for providing us with updated measurements of the
$z\sim 2 - 3$ IGM opacity, and Gwen Rudie and Milan Bogosavljevi\'{c}
for their assistance in the collection of some of the data used in
this paper.  D.B.N., A.E.S.\ and K.A.K\ acknowledge
support from the David and Lucile Packard Foundation. C.C.S.\
acknowledges additional support from the John D.\ and Catherine T.\
MacArthur Foundation, the Peter and Patricia Gruber Foundation, and 
NSF grants AST-0606912 and AST-0908805.  
We wish to extend special thanks to those of Hawaiian ancestry on whose
sacred mountain we are privileged to be guests.  Without their
generous hospitality, most of the observations presented herein would
not have been possible.

\bibliographystyle{apj}

\clearpage

\appendix

\section{LAE Lyman-continuum Candidates}

\subsection{LAE010} 
We show the $HST$/ACS imaging and one-dimensional shallow-mask
spectrum of LAE010 in the top panel of Figure~\ref{f:3laes1}.  
An emission line is clearly detected at
$\lambda=4979$\AA, indicating a redshift of 
$z=3.096$.  The shallow-mask slit also covers the position of the
NB3640 detection.  No additional emission features are detected in
the spectrum.  Thus, we retain LAE010 as a possible LyC-leaking galaxy

\subsection{LAE016}
We show the $HST$/ACS imaging and one-dimensional shallow-mask
spectrum of LAE016 in the middle panel of Figure~\ref{f:3laes1}.  
An emission line is clearly detected at
$\lambda=4973$\AA, indicating a redshift of 
$z=3.091$.  The NB3640 detection is weak and offset by 0\secpoint8, but covered by
the slit.  As no additional emission features are detected in
the spectrum, we retain LAE016 as a possible LyC-leaking galaxy

\subsection{LAE018} 
The bottom panel of Figure~\ref{f:3laes1} shows the $HST$/ACS image of LAE018 as well as the
shallow and deep-mask spectra.  An emission line at 
$\lambda=4976$\AA\ is detected in both spectra, indicating
$z=3.093$.  The $LyA$ flux is centered on a compact source to the
north east, while the NB3640 flux extends over more diffuse
non-ionizing UV flux to the south and west.  Both slits cover the
NB3640 flux.  However, no other emission 
features were detected in either spectrum.  Thus, we retain LAE018 as
a possible LyC-leaking galaxy.

\subsection{LAE021} 
As we do not have $HST$ imaging of LAE021, we show our $BV$ image and
shallow-mask spectrum in the top panel of Figure~\ref{f:3laes2}.  An emission line is
clearly detected at $\lambda=4948$\AA, indicating a redshift of
$z=3.070$.  No other emission features are detected and we retain
LAE021 as a possible LyC-leaking galaxy.  We note,
however, that the NB3640 flux is offset by 1\secpoint4 from the $LyA$ flux, and is
not covered by the shallow-mask slit.  As with all of our retained NB3640
detections, we account for the possibility the NB3640 flux is due
to a foreground interloper in our Monte Carlo simulation.

\subsection{LAE028} 
We obtained spectra of LAE028 on both shallow and deep masks.
The middle panel of Figure~\ref{f:3laes2} shows the
$HST$/ACS image.  There is an offset of $\simeq0$\secpoint9 between the
$LyA$ and NB3640 flux 
centroids, each of which is centered on an individual clump.  An
emission line is detected at
$\lambda\simeq4973$\AA\ in both spectra indicating a
redshift of $z=3.088$. 
No additional emission or absorption features are detected in either
spectra.  We therefore retain
LAE028 as a possible LyC-leaking galaxy.  

\subsection{LAE038} 
We obtained spectra of LAE038 in the deep mask as well as two of the
shallow masks.  The $HST$/ACS image, the deep-mask spectrum and one of
the shallow-mask spectra are shown in the
bottom panel of Figure~\ref{f:3laes2}.  An emission line at
$\lambda\simeq4983$\AA, indicating $z=3.099$, is detected in each of
the spectra.  In all three of the
spectra, the \lya\ emission line is offset from the continuum by
$\simeq0\secpoint5$.  The directions and magnitude of the offsets are
consistent with the continuum emanating from the location of NB3640
flux and the \lya\ emission from the $LyA$ flux.  No additional lines 
are detected in any of the three spectra.  We therefore retain
LAE038 as a possible LyC-leaking galaxy.  

\subsection{LAE039} 
We show the $BV$ image and shallow-mask spectrum of
LAE039 in the top panel of Figure~\ref{f:3laes3}.  We identify an
emission line at $\lambda=4978$\AA, indicating $z=3.095$.
No additional lines are detected.  We therefore retain
LAE039 as a possible LyC-leaking galaxy.  

\subsection{LAE041} 
We obtained spectra of LAE041 in the deep mask as well as two of the
shallow masks.  The $HST$/ACS image, the deep-mask spectrum and one of
the shallow-mask spectra are shown in the
middle panel of Figure~\ref{f:3laes3}.  We identify an emission line at
$\lambda\simeq4983$\AA, indicating $z=3.066$.  No additional lines 
are detected in any of the three spectra.  We therefore retain
LAE041 as a possible LyC-leaking galaxy.  

\subsection{LAE046} 
We show our $BV$ image and
shallow-mask spectrum of LAE046 in the bottom panel of
Figure~\ref{f:3laes3}.  An emission line is detected at
$\lambda=4984$\AA, indicating a redshift of $z=3.100$.  No other
emission features are detected and we retain LAE046 as a possible
LyC-leaking galaxy.  We note, however, that the NB3640 flux is offset
by 1\secpoint8 from the $LyA$ flux, and is not covered by the
shallow-mask slit.  As with all of our retained NB3640
detections, we account for the possibility the the NB3640 flux is due
to a foreground interloper in our Monte Carlo simulation.

\subsection{LAE048} 
We show the $HST$/ACS image and
shallow-mask spectrum of LAE048 in the top panel of Figure~\ref{f:3laes4}.  An emission line is
detected at $\lambda=4977$\AA, indicating a redshift of
$z=3.094$.  No other emission features are detected and we retain
LAE048 as a possible LyC-leaking galaxy.  We note,
however, that the NB3640 flux is offset by 1\secpoint7 from the $LyA$ flux, and is
not covered by the shallow-mask slit.  As with all of our retained NB3640
detections, we account for the possibility the the NB3640 flux is due
to a foreground interloper in our Monte Carlo simulation.

\subsection{LAE051} 
We show the $HST$/ACS imaging and shallow-mask
spectrum of LAE051 in the middle panel of Figure~\ref{f:3laes4}.  
An emission line is clearly detected at
$\lambda=4977$\AA, indicating a redshift of 
$z=3.094$.  The shallow-mask slit also covers the position of the
NB3640 detection.  No additional emission features are detected in
the spectrum.  Thus, we retain LAE051 as a possible LyC-leaking galaxy

\subsection{LAE053} 
We obtained spectra of LAE053 in one of the shallow masks and the
deep mask.  The bottom panel of Figure~\ref{f:3laes4} shows the
$HST$/ACS image and extracted spectra.  There is an offset of
$\simeq0$\secpoint9 between the  $LyA$ and NB3640 flux 
centroids, each of which is centered on an individual clump.  Each
clump and corresponding $LyA$ or NB3640 flux is covered by only one of
the masks.  A relatively broad emission line is detected at
$\lambda\simeq4973$\AA\ in the shallow-mask spectrum, indicating a
redshift of $z=3.088$.  An absorption feature is detected in the deep
mask at $\lambda\simeq4967$\AA\ indicating \lya\ in absorption at
$z=3.085$.  This absorption feature suggests that the NB3640 source is
also at $z\simeq 3.09$.  No other emission or absorption features are
detected in either spectra.  We therefore retain
LAE053 as a possible LyC-leaking galaxy.  

\subsection{LAE064} 
We show the $HST$/ACS imaging and the combined shallow-mask
spectrum of LAE064 in the top panel of Figure~\ref{f:3laes5}.  
An emission line is clearly detected at
$\lambda=4994$\AA, indicating a redshift of 
$z=3.108$.  Both slits also cover the position of the
NB3640 detection, which is offset by 1\secpoint0 from the $LyA$ flux.
The extraction apertures for the spectra shown were centered on the
\lya\ line, which was offset from faint
continuum emission consistent in magnitude ($\sim$1\arcsec) and direction from the
position of the NB3640 flux.
No additional emission features are detected in 
either (one- or two-dimensional) spectra.  Thus, we retain LAE064 as a
possible LyC-leaking galaxy. 

\subsection{LAE069} 
We show our $BV$ image and
shallow-mask spectrum of LAE069 in the middle panel of Figure~\ref{f:3laes5}.  An emission line is
detected at $\lambda=4953$\AA, indicating a redshift of
$z=3.074$.  No other emission features are detected and we retain
LAE069 as a possible LyC-leaking galaxy.  We note,
however, that the NB3640 flux is offset from the $LyA$ flux by 0\secpoint9, and is
not covered by the shallow-mask slit. 

\subsection{LAE074} 
The bottom panel of Figure~\ref{f:3laes5} shows the $HST$/ACS image of
LAE074 as well as the shallow and deep-mask spectra.  An emission line at 
$\lambda=4990$\AA\ is detected in the shallow-mask spectrum, indicating
$z=3.093$.   While the \lya\ emission line is not detected in the
deep-mask spectrum, the slit covers only part of the diffuse $LyA$
flux flux.  No other emission 
features were detected in either spectrum.  Thus, we retain LAE074 as
a possible LyC-leaking galaxy.

\subsection{LAE081} 
The top of Figure~\ref{f:3laes6} shows the $HST$/ACS image of
LAE081 as well as the shallow and deep-mask spectra.  An emission line at 
$\lambda=4989$\AA\ is detected in both spectra, indicating
$z=3.104$.   Both slits cover the bulk of the $LyA$ and NB3640 flux.  No other emission 
features were detected in either spectrum.  Thus, we retain LAE081 as
a possible LyC-leaking galaxy.

\subsection{LAE083} 
The bottom of Figure~\ref{f:3laes6} shows the $HST$/ACS image of
LAE083 as well as the shallow-mask spectrum.  An emission line at 
$\lambda=4942$\AA\ is detected, indicating
$z=3.065$.   The slit also covers the bulk of the NB3640 flux.  No other emission 
features were detected.  Thus, we retain LAE083 as
a possible LyC-leaking galaxy.

\begin{deluxetable*}{ccccccccc}
\tablewidth{0pt} 
\tabletypesize{\scriptsize}
\tablecaption{The $z>3.055$ LBG sample.}
\tablehead{
\colhead{ID} & \colhead{RA} & \colhead{Dec} & \colhead{$z_{em}$\tablenotemark{a}} & \colhead{$z_{abs}$\tablenotemark{b}} & \colhead{$R$} & \colhead{NB3640} & \colhead{$\Delta_R$\tablenotemark{c}} & \colhead{$\frac{F_{UV}}{F_{LyC}}_{obs}$\tablenotemark{d}}  \\
 & \colhead{(J2000)} & \colhead{(J2000)} & & & & & & 
}
\startdata
C3 & 22:17:32.53 & 00:10:57.4 & 3.098 & 3.091 & 24.59 & \nodata & \nodata & $>12.1$ \\
C4 & 22:17:38.91 & 00:11:02.0 & 3.076 & \nodata & 24.28 & \nodata & \nodata & $>16.2$ \\
C7 & 22:17:24.60 & 00:11:31.3 & \nodata & 3.062 & 24.06 & \nodata & \nodata & $>19.8$ \\
C9 & 22:17:28.29 & 00:12:12.3 & 3.071 & 3.072 & 25.84 & \nodata & \nodata & $>3.8$ \\
C11 & 22:17:25.68 & 00:12:35.4 & 3.101 & 3.096 & 24.21 & \nodata & \nodata & $>17.3$ \\
C12 & 22:17:35.29 & 00:12:47.9 & 3.107 & 3.095 & 24.22 & \nodata & \nodata & $>17.0$ \\
C14 & 22:17:34.04 & 00:12:51.3 & 3.220 & \nodata & 25.67 & 26.43 & 0\secpoint1 & 2.0 $\pm$ 0.5 \\
C15 & 22:17:26.13 & 00:12:55.4 & 3.094 & \nodata & 25.02 & \nodata & \nodata & $>8.2$ \\
C16 & 22:17:31.95 & 00:13:16.3 & \nodata & 3.065 & 23.62 & 26.43 & 1\secpoint9 & $13.3 \pm 4.0$ \\
C24 & 22:17:18.94 & 00:14:45.4 & 3.103 & 3.096 & 24.19 & \nodata & \nodata & $>17.6$ \\
C26 & 22:17:39.54 & 00:15:15.6 & 3.178 & \nodata & 25.01 & \nodata & \nodata & $>8.2$ \\
C28 & 22:17:21.13 & 00:15:27.7 & 3.076 & \nodata & 24.87 & \nodata & \nodata & $>9.3$ \\
C30 & 22:17:19.29 & 00:15:45.0 & 3.104 & 3.097 & 23.70 & \nodata & \nodata & $>27.5$ \\
C32 & 22:17:25.63 & 00:16:12.9 & 3.294 & 3.292 & 23.64 & \nodata & \nodata & $>29.2$ \\
C35 & 22:17:20.23 & 00:16:52.5 & \nodata & 3.098 & 24.06 & \nodata & \nodata & $>19.7$ \\
C39 & 22:17:20.99 & 00:17:09.5 & 3.076 & \nodata & 25.04 & \nodata & \nodata & $>8.0$ \\
C47 & 22:17:20.24 & 00:17:32.5 & 3.075 & 3.065 & 23.78 & \nodata & \nodata & $>25.7$ \\
C48 & 22:17:18.58 & 00:18:16.7 & 3.090 & 3.079 & 24.57 & \nodata & \nodata & $>12.3$ \\
C49 & 22:17:19.81 & 00:18:18.8 & 3.163 & 3.149 & 23.81 & 26.84\tablenotemark{e} & 0\secpoint4 & \nodata \\
C50 & 22:17:37.68 & 00:18:21.2 & \nodata & 3.086 & 25.01 & \nodata & \nodata & $>8.2$ \\
D3 & 22:17:32.40 & 00:11:33.6 & 3.074 & 3.066 & 23.92 & \nodata & \nodata & $>22.4$ \\
D17 & 22:17:18.86 & 00:18:17.0 & 3.090 & 3.070 & 24.29 & 27.00 & 0\secpoint9 & $12.2 \pm 5.1$ \\
M2 & 22:17:33.51 & 00:11:10.8 & 3.388 & 3.386 & 25.19 & 25.99 & 0\secpoint8 & 2.1 $\pm$ 0.4 \\
M5 & 22:17:26.07 & 00:11:33.1 & 3.327 & \nodata & 25.71 & 26.31\tablenotemark{e} & 0\secpoint4 & \nodata \\
M6 & 22:17:28.13 & 00:11:40.5 & 3.180 & \nodata & 25.65 & \nodata & \nodata & $>4.6$ \\
M8 & 22:17:25.10 & 00:11:56.8 & 3.064 & 3.062 & 24.56 & \nodata & \nodata & $>12.5$ \\
M10 & 22:17:26.80 & 00:12:21.3 & 3.103 & 3.095 & 24.50 & \nodata & \nodata & $>13.1$ \\
M11 & 22:17:31.77 & 00:12:51.3 & 3.107 & 3.103 & 25.22 & \nodata & \nodata & $>6.8$ \\
M13 & 22:17:31.46 & 00:12:55.2 & 3.107 & \nodata & 25.38 & \nodata & \nodata & $>5.9$ \\
M14 & 22:17:39.05 & 00:13:30.1 & 3.091 & \nodata & 25.20 & \nodata & \nodata & $>6.9$ \\
M19 & 22:17:36.90 & 00:15:00.9 & \nodata & 3.082 & 24.96 & \nodata & \nodata & $>8.6$ \\
M20 & 22:17:34.40 & 00:15:02.9 & 3.109 & 3.096 & 25.08 & \nodata & \nodata & $>7.7$ \\
M25 & 22:17:31.49 & 00:16:31.2 & 3.098 & 3.091 & 24.79 & \nodata & \nodata & $>10.1$ \\
M28 & 22:17:31.66 & 00:16:58.0 & 3.094 & 3.088 & 24.75 & \nodata & \nodata & $>10.4$ \\
M29 & 22:17:37.40 & 00:17:08.8 & \nodata & 3.228 & 24.93 & 26.76 & 1\secpoint0 & 5.4 $\pm$ 1.3 \\  
M31 & 22:17:36.87 & 00:17:12.4 & 3.099 & \nodata & 25.70 & \nodata & \nodata & $>4.4$ \\
M34 & 22:17:33.80 & 00:17:57.2 & \nodata & 3.084 & 25.41 & \nodata & \nodata & $>5.7$ \\
MD14 & 22:17:37.91 & 00:13:43.9 & \nodata & 3.097 & 24.49 & \nodata & \nodata & $>13.4$ \\
MD23 & 22:17:28.01 & 00:14:29.6 & 3.085 & 3.075 & 24.34 & \nodata & \nodata & $>15.2$ \\
MD32 & 22:17:23.70 & 00:16:01.6 & 3.102 & \nodata & 25.14 & 25.51\tablenotemark{e} & 0\secpoint4 & \nodata \\
MD46 & 22:17:27.28 & 00:18:09.7 & 3.091 & 3.080 & 23.49 & 25.22 & 1\secpoint0 & $4.9 \pm 0.7$ \\
aug96M16\tablenotemark{f} & 22:17:30.86 & 00:13:10.8 & \nodata & \nodata & 24.47 & 25.23 & 0\secpoint7 & $2.0 \pm 0.4$ 
\enddata
\vspace{0.1cm}
\tablenotetext{a}{\mbox{Ly$\alpha$} emission redshift.}
\tablenotetext{b}{Interstellar absorption redshift.}
\tablenotetext{c}{Spatial offset between the centroids of $R$-band and NB3640 emission.}
\tablenotetext{d}{Observed ratio and uncertainty in non-ionizing UV and LyC flux densities, inferred from the NB3640$-R$ color.  This value has not been corrected for either contamination by foreground sources or IGM absorption.}
\tablenotetext{e}{Spectrum contains evidence for the presence of a foreground object.  Thus, the NB3640 flux is likely contaminated by non-ionizing UV flux from the interloper.}
\tablenotetext{f}{We were unable to determine the redshift of the LBG aug96M16, which in \citet{nestor2011} was erroneously associated with a nearby galaxy with $z_{em} = 3.298$.  We have removed it from our sample but include it in this table for completeness.}
\label{t:one}
\end{deluxetable*}

\begin{deluxetable}{cccccc}
\tablewidth{0pt} 
\tabletypesize{\scriptsize}
\tablecaption{LBGs with new redshifts $z<3.055$.}
\tablehead{
\colhead{ID} & \colhead{RA} & \colhead{Dec} & \colhead{$z_{em}$\tablenotemark{a}} & \colhead{$z_{abs}$\tablenotemark{b}} & \colhead{$R$} \\
 & \colhead{(J2000)} & \colhead{(J2000)} & & &
}
\startdata
C8 & 22:17:27.64 & 00:11:59.1 & 2.459 & \nodata & 26.42  \\
C19 & 22:17:38.36 & 00:14:16.4 & 2.993 & \nodata  & 24.84 \\
D5 & 22:17:34.65 & 00:12:33.6 & 2.699 & 2.690 & 25.24 \\
D10 & 22:17:38.92 & 00:14:32.1 & \nodata & 2.605 & 24.61 \\
M3 & 22:17:35.81 & 00:11:16.5 & 2.876 & \nodata & 25.27 \\
M16 & 22:17:24.66 & 00:14:07.8 & 3.021 & 3.018 & 24.88 \\
M35 & 22:17:32.67 & 00:18:05.0 & 2.938 & \nodata & 25.42 \\
MD11 & 22:17:34.94 & 00:13:22.7 & \nodata & 2.857 & 24.79 \\
MD42 & 22:17:35.83 & 00:17:19.8 & \nodata & 2.792 & 25.37 \\
MD45 & 22:17:25.36 & 00:18:04.2 & 3.025 & 3.022 & 24.39 
\enddata
\vspace{0.1cm}
\tablenotetext{a}{\mbox{Ly$\alpha$} emission redshift.}
\tablenotetext{b}{Interstellar absorption redshift.}
\label{t:2}
\end{deluxetable}

\begin{deluxetable*}{ccccccccc}
\tablewidth{0pt}
\tabletypesize{\scriptsize}
\tablecaption{Summary of LAE Data}
\tablehead{
\colhead{ID} & \colhead{RA} & \colhead{Dec}  & \colhead{$R$} & \colhead{NB3640} & \colhead{$\Delta_R$\tablenotemark{a}} & \colhead{$\Delta_{LyA}$\tablenotemark{a}} & \colhead{$\frac{F_{UV}}{F_{LyC}}_{obs}$\tablenotemark{b}} & \colhead{$z_{em}$\tablenotemark{c}} \\
 & \colhead{(J2000)} & \colhead{(J2000)} & & & & &  &
}
\startdata
001\tablenotemark{d,e} & 22:17:32.40 & 0:11:34.1 & 23.92 & $>27.3$ & \ldots & \ldots & $>22.4$ & 3.075 \\
002\tablenotemark{d,e} & 22:17:38.90 & 0:11:01.8 & 24.28 & $>27.3$ & \ldots & \ldots & $>16.2$ & 3.076 \\
003 & 22:17:24.79 & 0:17:17.4 & 24.42 & 24.74\tablenotemark{f} & 0\secpoint3 & 0\secpoint6 & \ldots & 3.097\\
004\tablenotemark{d,e} & 22:17:28.01 & 0:14:30.0 & 24.34 & $>27.3$ & \ldots & \ldots & $>15.2$ & 3.092 \\
005 & 22:17:35.86 & 0:15:59.4 & 25.65 & $>27.3$ & \ldots & \ldots & $>4.6$ & 3.096 \\
006 & 22:17:24.80 & 0:11:16.8 & $>27.0$ & $>27.3$ & \ldots & \ldots & \ldots & 3.071 \\
007 & 22:17:27.78 & 0:17:36.9 & $>27.0$ & $>27.3$ & \ldots & \ldots & \ldots & 3.089 \\
008\tablenotemark{d,e} & 22:17:21.11 & 0:15:28.0 & 24.87 & $>27.3$ & \ldots & \ldots & $>9.3$ & 3.076 \\
009\tablenotemark{d} & 22:17:28.29 & 0:12:12.3 & 25.84 & $>27.3$ & \ldots & \ldots & $>3.8$ & 3.071 \\
010 & 22:17:20.38 & 0:18:04.2 & 25.77 & 26.74 & 0\secpoint3 & 0\secpoint3 & $2.4\pm1.1$ & 3.096 \\
011 & 22:17:33.85 & 0:12:14.9 & 26.07 & $>27.3$ & \ldots & \ldots & $>3.1$ & 3.102 \\
012\tablenotemark{d,e} & 22:17:31.69 & 0:16:57.6 & 24.75 & $>27.3$ & \ldots & \ldots & $>10.5$ & 3.094 \\
013 & 22:17:27.18 & 0:16:21.7 & 25.98 & $>27.3$ & \ldots & \ldots & $>3.4$ & 3.095 \\
014 & 22:17:19.25 & 0:14:50.9 & 25.82 & $>27.3$ & \ldots & \ldots & $>3.9$ & 3.067 \\
015 & 22:17:21.84 & 0:12:12.7 & $>27.0$ & $>27.3$ & \ldots & \ldots & \ldots & 3.067 \\
016 & 22:17:35.61 & 0:18:00.2 & 26.24 & 26.91 & 0\secpoint9 & 0\secpoint8 & $1.8\pm0.9$ & 3.091 \\
017 & 22:17:25.40 & 0:17:16.8 & 26.22 & $>27.3$ & \ldots & \ldots & $>2.7$ & 3.105 \\
018 & 22:17:39.01 & 0:17:26.4 & 26.25 & 25.69 & 0\secpoint1 & 1\secpoint0 & $0.6\pm0.2$ & 3.093 \\
019 & 22:17:26.15 & 0:13:20.1 & 25.70 & 26.24\tablenotemark{f} & 0\secpoint4 & 0\secpoint9 & \ldots & 3.101 \\
020 & 22:17:37.33 & 0:16:31.4 & 25.45 & $>27.3$ & \ldots & \ldots & \ldots & \ldots\tablenotemark{g} \\
021 & 22:17:18.77 & 0:15:18.1 & $>27.0$ & 27.17 & 1\secpoint3 & 1\secpoint4 & $<1.8$ & 3.070 \\
022 & 22:17:19.68 & 0:11:49.4 & 26.11 & $>27.3$ & \ldots & \ldots & $>3.0$ & 3.066 \\
023 & 22:17:31.73 & 0:16:06.9 & 24.91 & $>27.3$ & \ldots & \ldots & $>9.1$ & 3.101 \\
024 & 22:17:34.17 & 0:16:09.7 & 26.73 & $>27.3$ & \ldots & \ldots & $>1.7$ & 3.096 \\
025 & 22:17:36.74 & 0:16:28.8 & 25.54 & 25.85\tablenotemark{f} & 0\secpoint3 & 1\secpoint2 & \ldots & 3.091\\
026 & 22:17:18.96 & 0:12:00.8 & 26.59 & $>27.3$ & \ldots & \ldots & $>1.9$ & 3.092 \\
027 & 22:17:24.94 & 0:17:17.3 & 26.25 & $>27.3$ & \ldots & \ldots & $>2.6$ & 3.101 \\
028 & 22:17:31.80 & 0:17:17.9 & 25.50 & 26.71 & 0\secpoint3 & 0\secpoint9 & $3.1\pm1.3$ & 3.088 \\
029\tablenotemark{d} & 22:17:31.49 & 0:12:55.0 & 25.38 & $>27.3$ & \ldots & \ldots & $>5.9$ & 3.107 \\
030 & 22:17:21.75 & 0:11:38.8 & $>27.0$ & $>27.3$ & \ldots & \ldots & \ldots & 3.096 \\
031 & 22:17:33.63 & 0:17:15.1 & 26.37 & $>27.3$ & \ldots & \ldots & $>2.3$ & 3.092 \\
032 & 22:17:26.61 & 0:13:18.1 & 26.61 & $>27.3$ & \ldots & \ldots & $>1.9$ & 3.101 \\
033 & 22:17:37.50 & 0:14:08.3 & 26.62 & $>27.3$ & \ldots & \ldots & $>1.9$ & 3.076 \\
034 & 22:17:23.41 & 0:16:35.4 & 25.42 & 25.76\tablenotemark{h} & 0\secpoint5 & 0\secpoint0 & $1.4\pm0.4$ & 3.044\\
035 & 22:17:27.03 & 0:13:13.2 & $>27.0$ & $>27.3$ & \ldots & \ldots & \ldots & 3.093 \\
036 & 22:17:22.25 & 0:11:55.1 & $>27.0$ & $>27.3$ & \ldots & \ldots & \ldots & 3.084 \\
037 & 22:17:20.96 & 0:18:07.3 & 25.61 & $>27.3$ & \ldots & \ldots & $>4.7$ & 3.089 \\
038 & 22:17:34.77 & 0:15:41.3 & 26.17 & 25.82 & 0\secpoint1 & 0\secpoint7 & $0.7\pm0.3$ & 3.099 \\
039 & 22:17:24.08 & 0:11:31.7 & 26.48 & 26.77 & 0\secpoint4 & 0\secpoint8 & $1.3\pm0.7$ & 3.095 \\
040 & 22:17:31.93 & 0:13:08.5 & $>27.0$ & $>27.3$ & \ldots & \ldots & \ldots & 3.105 \\
041 & 22:17:24.54 & 0:15:06.7 & 25.97 & 25.94 & 0\secpoint4 & 0\secpoint5 & $1.0\pm0.4$ & 3.066 \\
042 & 22:17:21.50 & 0:17:04.7 & 25.50 & $>27.3$ & \ldots & \ldots & $>5.2$ & 3.072 \\
043 & 22:17:21.65 & 0:12:23.4 & 26.24 & $>27.3$ & \ldots & \ldots & $>2.7$ & 3.071 \\
044 & 22:17:36.41 & 0:12:51.0 & $>27.0$ & $>27.3$ & \ldots & \ldots & \ldots & 3.059 \\
045 & 22:17:35.97 & 0:16:30.2 & $>27.0$ & $>27.3$ & \ldots & \ldots & \ldots & 3.094 \\
046 & 22:17:21.47 & 0:14:54.6 & $>$27.0 & 26.43 & 1\secpoint8 & 1\secpoint8 & $<0.8$ & 3.100 \\
047 & 22:17:36.05 & 0:15:06.9 & $>27.0$ & $>27.3$ & \ldots & \ldots & \ldots & 3.096 \\
048 & 22:17:27.37 & 0:16:51.5 & 26.73 & 26.00 & 2\secpoint2 & 1\secpoint7 & $0.5\pm0.2$ & 3.094 \\
049 & 22:17:39.29 & 0:16:10.5 & $>27.0$ & $>27.3$ & \ldots & \ldots & \ldots & 3.094 \\
050 & 22:17:24.56 & 0:15:56.8 & $>27.0$ & $>27.3$ & \ldots & \ldots & \ldots & 3.082 \\
051 & 22:17:33.72 & 0:15:04.9 & 26.26 & 27.21 & 0\secpoint3 & 0\secpoint7 & $2.4\pm1.3$ & 3.094 \\
052 & 22:17:36.84 & 0:13:17.2 & 26.63 & $>27.3$ & \ldots & \ldots & $>1.8$ & 3.102 \\
053 & 22:17:34.70 & 0:16:33.4 & 26.53 & 26.98 & 0\secpoint6 & 0\secpoint9 & $1.5\pm0.8$ & 3.090 \\
054 & 22:17:39.05 & 0:11:33.9 & 26.25 & $>27.3$ & \ldots & \ldots & $>2.6$ & 3.096 \\
055 & 22:17:35.80 & 0:11:50.0 & 26.67 & $>27.3$ & \ldots & \ldots & $>1.8$ & 3.072 
\enddata
\vspace{0.1cm}
\label{t:3}
\end{deluxetable*}

\setcounter{table}{2}
\begin{deluxetable*}{ccccccccc}
\tablewidth{0pt}
\tabletypesize{\scriptsize}
\tablecaption{Summary of LAE Data (\emph{Continued})}
\tablehead{
\colhead{ID} & \colhead{RA} & \colhead{Dec}  & \colhead{$R$} & \colhead{NB3640} & \colhead{$\Delta_R$\tablenotemark{a}} & \colhead{$\Delta_{LyA}$\tablenotemark{a}} & \colhead{$\frac{F_{UV}}{F_{LyC}}_{obs}$\tablenotemark{b}} & \colhead{$z_{em}$\tablenotemark{c}} \\
 & \colhead{(J2000)} & \colhead{(J2000)} & & & & &  &
}
\startdata
056 & 22:17:22.42 & 0:17:20.7 & 26.86 & $>27.3$ & \ldots & \ldots & $>1.5$ & 3.073 \\
057 & 22:17:25.40 & 0:10:58.3 & 26.84 & $>27.3$ & \ldots & \ldots & $>1.5$ & 3.070 \\
058 & 22:17:19.61 & 0:15:38.4 & $>27.0$ & $>27.3$ & \ldots & \ldots & \ldots & \ldots\tablenotemark{i}\\
059 & 22:17:24.98 & 0:12:30.0 & 25.31 & $>27.3$ & \ldots & \ldots & $>6.2$ & 3.096 \\
060 & 22:17:28.19 & 0:11:17.1 & 26.61 & $>27.3$ & \ldots & \ldots & $>1.9$ & 3.065 \\
061 & 22:17:34.10 & 0:15:40.2 & $>27.0$ & $>27.3$ & \ldots & \ldots & \ldots & 3.101 \\
062 & 22:17:22.87 & 0:14:41.7 & 26.53 & $>27.3$ & \ldots & \ldots & $>2.0$ & 3.057 \\
063 & 22:17:23.32 & 0:15:52.9 & 26.55 & $>27.3$ & \ldots & \ldots & $>2.0$ & 3.099 \\
064 & 22:17:35.42 & 0:12:14.6 & 26.05 & 26.74 & 0\secpoint1 & 1\secpoint0 & $1.9\pm0.9$ & 3.108 \\
065 & 22:17:28.15 & 0:14:36.4 & 26.91 & $>27.3$ & \ldots & \ldots & $>1.4$ & 3.102 \\
066 & 22:17:20.86 & 0:15:11.8 & 26.64 & $>27.3$ & \ldots & \ldots & $>1.8$ & 3.066 \\
067 & 22:17:36.26 & 0:13:11.7 & 26.40 & $>27.3$ & \ldots & \ldots & $>2.3$ & 3.106 \\
068 & 22:17:18.37 & 0:17:26.1 & $>27.0$ & $>27.3$ & \ldots & \ldots & \ldots & 3.072 \\
069 & 22:17:18.96 & 0:11:12.0 & 24.62 & 27.22 & 0\secpoint2 & 0\secpoint9 & $10.9\pm5.1$ & 3.074 \\
070 & 22:17:39.28 & 0:14:00.2 & 25.98 & $>27.3$ & \ldots & \ldots & $>3.4$ & 3.090 \\
071 & 22:17:21.61 & 0:12:20.5 & 26.77 & $>27.3$ & \ldots & \ldots & $>1.6$ & 3.072 \\
072 & 22:17:31.24 & 0:17:32.1 & 27.00 & $>27.3$ & \ldots & \ldots & $>1.3$ & 3.084 \\
073 & 22:17:39.12 & 0:17:11.7 & 26.35 & $>27.3$ & \ldots & \ldots & $>2.4$ & 3.083 \\
074 & 22:17:36.47 & 0:12:54.8 & 26.16 & 25.52 & 0\secpoint1 & 0\secpoint7 & $0.6\pm0.2$ & 3.105 \\
075 & 22:17:22.97 & 0:11:25.8 & $>27.0$ & $>27.3$ & \ldots & \ldots & \ldots & \ldots\tablenotemark{i} \\
076 & 22:17:20.67 & 0:15:13.2 & $>27.0$ & $>27.3$ & \ldots & \ldots & \ldots & 3.066 \\
077 & 22:17:37.95 & 0:11:01.3 & 26.01 & 26.36 & 0\secpoint2 & 1\secpoint1 & \ldots & \ldots\tablenotemark{g} \\
078 & 22:17:37.68 & 0:16:48.3 & 25.95 & $>27.3$ & \ldots & \ldots & $>3.5$ & 3.090 \\
079 & 22:17:34.68 & 0:11:10.5 & $>27.0$ & $>27.3$ & \ldots & \ldots & \ldots & 3.103 \\
080 & 22:17:35.95 & 0:13:43.3 & $>27.0$ & $>27.3$ & \ldots & \ldots & \ldots & 3.104 \\
081 & 22:17:29.22 & 0:14:48.7 & $>$27.0 & 26.79 & 0\secpoint6\tablenotemark{j} & 0\secpoint7 & $<1.2$ & 3.104 \\
082 & 22:17:35.44 & 0:16:47.6 & $>27.0$ & $>27.3$ & \ldots & \ldots & \ldots & 3.087 \\
083 & 22:17:28.46 & 0:12:08.9 & 26.46 & 26.84 & 0\secpoint4 & 0\secpoint6 & $1.4\pm0.7$ & 3.065 \\
084 & 22:17:19.90 & 0:15:14.9 & $>$27.0 & 26.50 & 0\secpoint1 & 0\secpoint8 & \ldots & \ldots\tablenotemark{i} \\
085 & 22:17:30.86 & 0:14:38.2 & 26.94 & $>27.3$ & \ldots & \ldots & \ldots & \ldots\tablenotemark{i}\\
086 & 22:17:28.42 & 0:13:42.8 & $>27.0$ & $>27.3$ & \ldots & \ldots & \ldots & 3.100 \\
087 & 22:17:37.07 & 0:13:21.5 & $>$27.0 & 27.26 & 0\secpoint1\tablenotemark{j} & 1\secpoint6 & \ldots & \ldots\tablenotemark{g} \\
088 & 22:17:38.45 & 0:13:18.3 & $>27.0$ & $>27.3$ & \ldots & \ldots & \ldots & \ldots\tablenotemark{i}\\
089 & 22:17:38.54 & 0:15:22.5 & 26.59 & $>27.3$ & \ldots & \ldots & \ldots & \ldots\tablenotemark{i}\\
090 & 22:17:18.25 & 0:14:06.4 & $>27.0$ & $>27.3$ & \ldots & \ldots & \ldots & 3.095 \\
091 & 22:17:36.14 & 0:15:40.7 & $>27.0$ & $>27.3$ & \ldots & \ldots & \ldots & 3.095 \\
092 & 22:17:23.97 & 0:15:27.8 & $>27.0$ & $>27.3$ & \ldots & \ldots & \ldots & \ldots\tablenotemark{i}\\
093 & 22:17:27.48 & 0:13:57.5 & 26.35 & $>27.3$ & \ldots & \ldots & $>2.4$ & 3.106 \\
094 & 22:17:39.14 & 0:17:00.6 & $>27.0$ & $>27.3$ & \ldots & \ldots & \ldots & \ldots\tablenotemark{i}\\
095 & 22:17:37.19 & 0:13:28.0 & 26.17 & $>27.3$ & \ldots & \ldots & $>2.8$ & 3.094 \\
096 & 22:17:38.93 & 0:11:37.4 & 26.64 & 26.45 & 0\secpoint6 & 0\secpoint6 & \ldots & \ldots\tablenotemark{g} \\
097 & 22:17:27.11 & 0:14:08.7 & $>27.0$ & $>27.3$ & \ldots & \ldots & \ldots & 3.064 \\
098 & 22:17:24.01 & 0:13:19.5 & $>27.0$ & $>27.3$ & \ldots & \ldots & \ldots & \ldots\tablenotemark{g} \\
099 & 22:17:36.46 & 0:13:00.3 & $>27.0$ & $>27.3$ & \ldots & \ldots & \ldots & 3.102 \\
100 & 22:17:30.61 & 0:18:11.6 & $>27.0$ & $>27.3$ & \ldots & \ldots & \ldots & \ldots\tablenotemark{i}\\
101 & 22:17:25.33 & 0:17:22.5 & 26.89 & 26.58 & 0\secpoint2 & 1\secpoint4 & \ldots & \ldots\tablenotemark{g} \\
102 & 22:17:24.00 & 0:16:27.6 & 25.91 & 26.05 & 0\secpoint1 & 0\secpoint2 & \ldots & \ldots\tablenotemark{g} \\
103 & 22:17:19.40 & 0:15:26.1 & $>27.0$ & $>27.3$ & \ldots & \ldots & \ldots & 3.099 \\
104 & 22:17:37.66 & 0:12:55.5 & $>27.0$ & $>27.3$ & \ldots & \ldots & \ldots & 3.097 \\
105 & 22:17:35.46 & 0:12:23.9 & 26.81 & $>27.3$ & \ldots & \ldots & $>1.6$ & 3.104 \\
106 & 22:17:22.86 & 0:17:57.8 & 26.16 & $>27.3$ & \ldots & \ldots & \ldots & \ldots\tablenotemark{i}\\
107 & 22:17:20.96 & 0:14:46.7 & 26.57 & $>27.3$ & \ldots & \ldots & $>2.0$ & 3.075 \\
108 & 22:17:24.78 & 0:17:40.4 & 26.84 & $>27.3$ & \ldots & \ldots & $>1.5$ & 3.100 \\
109 & 22:17:23.98 & 0:17:57.8 & 26.77 & $>27.3$ & \ldots & \ldots & \ldots & \ldots\tablenotemark{g}\\
110 & 22:17:19.50 & 0:15:57.6 & $>27.0$ & $>27.3$ & \ldots & \ldots & \ldots & 3.074 
\enddata
\vspace{0.1cm}
\end{deluxetable*}

\setcounter{table}{2}
\begin{deluxetable*}{ccccccccc}
\tablewidth{0pt}
\tabletypesize{\scriptsize}
\tablecaption{Summary of LAE Data (\emph{Continued})}
\tablehead{
\colhead{ID} & \colhead{RA} & \colhead{Dec}  & \colhead{$R$} & \colhead{NB3640} & \colhead{$\Delta_R$\tablenotemark{a}} & \colhead{$\Delta_{LyA}$\tablenotemark{a}} & \colhead{$\frac{F_{UV}}{F_{LyC}}_{obs}$\tablenotemark{b}} & \colhead{$z_{em}$\tablenotemark{c}} \\
 & \colhead{(J2000)} & \colhead{(J2000)} & & & & &  &
}
\startdata
111 & 22:17:31.14 & 0:16:42.9 & $>27.0$ & $>27.3$ & \ldots & \ldots & \ldots & 3.096 \\
112 & 22:17:32.72 & 0:15:54.2 & $>27.0$ & $>27.3$ & \ldots & \ldots & \ldots & 3.096 \\
113 & 22:17:24.80 & 0:13:26.9 & $>27.0$ & $>27.3$ & \ldots & \ldots & \ldots & 3.069 \\
114 & 22:17:34.50 & 0:14:20.0 & $>27.0$ & $>27.3$ & \ldots & \ldots & \ldots & 3.103 \\
115 & 22:17:33.46 & 0:17:01.2 & $>27.0$ & $>27.3$ & \ldots & \ldots & \ldots & 3.093 \\
116 & 22:17:28.00 & 0:12:14.2 & $>27.0$ & $>27.3$ & \ldots & \ldots & \ldots & 3.096 \\
117 & 22:17:39.08 & 0:12:01.9 & $>27.0$ & $>27.3$ & \ldots & \ldots & \ldots & 3.098 \\
118 & 22:17:35.28 & 0:10:59.9 & 26.20 & 26.11 & 0\secpoint4 & 1\secpoint4 & \ldots & \ldots\tablenotemark{g} \\
119 & 22:17:25.63 & 0:12:47.8 & $>27.0$ & $>27.3$ & \ldots & \ldots & \ldots & 3.071 \\
120 & 22:17:26.76 & 0:10:59.8 & $>27.0$ & $>27.3$ & \ldots & \ldots & \ldots & 3.079 \\
121 & 22:17:26.44 & 0:15:27.5 & $>27.0$ & $>27.3$ & \ldots & \ldots & \ldots & 3.099 \\
122 & 22:17:38.19 & 0:14:03.7 & $>27.0$ & $>27.3$ & \ldots & \ldots & \ldots & 3.106 \\
123 & 22:17:35.06 & 0:17:26.0 & $>27.0$ & $>27.3$ & \ldots & \ldots & \ldots & \ldots\tablenotemark{i}\\
124 & 22:17:22.80 & 0:17:48.7 & $>27.0$ & $>27.3$ & \ldots & \ldots & \ldots & \ldots\tablenotemark{g} \\
125 & 22:17:38.02 & 0:14:03.6 & 26.79 & $>27.3$ & \ldots & \ldots & $>1.6$ & 3.101 \\
126 & 22:17:19.53 & 0:16:48.2 & $>27.0$ & $>27.3$ & \ldots & \ldots & \ldots & \ldots\tablenotemark{i}\\
127 & 22:17:36.91 & 0:11:27.1 & $>27.0$ & $>27.3$ & \ldots & \ldots & \ldots & 3.107 \\
128 & 22:17:23.43 & 0:16:07.4 & $>27.0$ & $>27.3$ & \ldots & \ldots & \ldots & 3.095 \\
129 & 22:17:22.28 & 0:10:57.9 & $>27.0$ & $>27.3$ & \ldots & \ldots & \ldots & 3.111 \\
130 & 22:17:32.84 & 0:16:48.8 & $>27.0$ & $>27.3$ & \ldots & \ldots & \ldots & 3.092 
\enddata
\vspace{0.1cm}
\tablenotetext{a}{Spatial offset between the centroids of $R$ or Ly$\alpha$ and NB3640 emission.}
\tablenotetext{b}{Observed ratio and uncertainty in non-ionizing UV and LyC flux densities, inferred 
from the NB3640$-R$ color.  This value has not been corrected for either contamination by 
foreground sources or IGM absorption.} 
\tablenotetext{c}{Spectroscopic redshift based on the observed wavelength of Ly$\alpha$ emission.}
\tablenotetext{d}{Objects that are also identified as LBGs. 001 is LBG D3; 002 is LBG C4;
004 is LBG MD23; 008 is LBG C28; 009 is LBG C9; 012 is LBG M28; 029 is LBG M13.}
\tablenotetext{e}{Objects with previously-known spectroscopic redshifts.}
\tablenotetext{f}{Spectrum contains evidence for the presence of a foreground galaxy.  Thus,
we do not consider the NB3640 detection to be LyC flux.}
\tablenotetext{g}{Objects with LRIS spectroscopic observations but without confirmed redshifts.}
\tablenotetext{h}{The redshift of this LAE is such that some non-ionizing UV flux will contribute
to the NB3640 detection, and therefore we exclude it from our LyC detection sample.}
\tablenotetext{i}{Objects without LRIS spectroscopic observations.}
\tablenotetext{j}{Offset determined from the centroid of the $BV$ detection.}
\end{deluxetable*}

\begin{deluxetable}{lcccc}
\tablewidth{0pt} 
\tabletypesize{\scriptsize}
\tablecaption{Sample average colors and UV to LyC flux density ratios.}
\tablehead{
\colhead{} & \multicolumn{2}{c}{LBGs} & \multicolumn{2}{c}{LAEs} \\
\colhead{Correction} & \colhead{$\left<\mathrm{NB3640}\right> - \left<R\right>$\tablenotemark{a}} &
\colhead{\fdr\tablenotemark{b}} & \colhead{$\left<\mathrm{NB3640}\right> - \left<R\right>$\tablenotemark{a}} & \colhead{\fdr\tablenotemark{b}} 
}
\startdata
&&&&\\
\multicolumn{5}{c}{Full Ensembles\tablenotemark{c}} \\
&&&&\\
none & $3.74^{+0.66}_{-0.42}$ & $31.5^{+26.4}_{-10.0}$	 & $2.13^{+0.36}_{-0.29}$ &  $7.1^{+2.8}_{-1.7}$ \\
contamination\tablenotemark{d} & $4.45^{+1.16}_{-0.56}$ &  $60.5^{+115.4}_{-24.2}$ & $2.66^{+0.56}_{-0.39}$ &  $11.6^{+7.8}_{-3.5}$ \\
IGM + contamination\tablenotemark{e} & $3.14^{+1.17}_{-0.57}$ & $18.0^{+34.8}_{-7.4}$ & $1.42^{+0.57}_{-0.40}$ & $3.7^{+2.5}_{-1.1}$ \\ \\
&&&&\\
\hline
&&&&\\
\multicolumn{5}{c}{Full Ensembles, maximum allowed flux for NB3640 non-detections\tablenotemark{f}} \\
&&&&\\
none & $3.40^{+0.43}_{-0.32}$ &  $23.0^{+11.0}_{-5.8}$ & $1.90^{+0.29}_{-0.25}$ &  $5.7^{+1.7}_{-1.2}$ \\
contamination\tablenotemark{d} & $3.84^{+0.48}_{-0.34}$ &  $34.4^{+19.3}_{-9.3}$ & $2.28^{+0.37}_{-0.29}$ &  $8.2^{+3.3}_{-1.9}$ \\
IGM + contamination\tablenotemark{e} & $2.53^{+0.51}_{-0.37}$ & $10.2^{+6.1}_{-3.0}$ & $1.04^{+0.38}_{-0.31}$ & $2.6^{+1.1}_{-0.6}$ \\
&&&&\\
\hline
&&&&\\
\multicolumn{5}{c}{NB3640 detections only} \\
&&&&\\
none & $1.77^{+0.67}_{-0.63}$ &  $5.1^{+4.4}_{-2.3}$	 & $0.29^{+0.41}_{-0.39}$ &  $1.3^{+0.6}_{-0.4}$ \\
contamination\tablenotemark{d} & $1.89^{+0.75}_{-0.66}$ &  $5.7^{+5.7}_{-2.6}$	 & $0.33^{+0.46}_{-0.42}$ &  $1.4^{+0.7}_{-0.4}$ \\
IGM + contamination\tablenotemark{e} & $0.57^{+0.77}_{-0.68}$ & $1.7^{+1.7}_{-0.8}$	 & $-0.91^{+0.47}_{-0.43}$ & $0.4^{+0.2}_{-0.1}$ 
\enddata
\vspace{0.1cm}
\tablenotetext{a}{Color determined from average NB3640 and $R$-band fluxes.  Uncertainties include individual flux and sample uncertainties.}
\tablenotetext{b}{Ratio and uncertainty in non-ionizing UV and LyC flux density inferred from $\left<\mathrm{NB3640}\right> - \left<R\right>$ color.}
\tablenotetext{c}{Color and flux-density ratio determined assuming NB3640 non-detections contribute zero LyC flux.}
\tablenotetext{d}{Color and flux-density ratio after statistically correcting sample for foreground contamination of NB3640 fluxes.}
\tablenotetext{e}{Color and flux-density ratio after correcting sample for foreground contamination and IGM absorption of NB3640 fluxes.}
\tablenotetext{f}{Color and flux-density ratio determined assuming NB3640 non-detections contribute the
maximum possible LyC flux consistent with our limits from stacking analysis.}
\label{t:4}
\end{deluxetable}

\begin{deluxetable}{rcccc}
\tablewidth{0pt} 
\tabletypesize{\scriptsize}
\tablecaption{Contributions to the ionizing backgrounds.}
\tablehead{
\colhead{} & \colhead{LF\tablenotemark{a}} & \colhead{$\eta$\tablenotemark{b}} & \colhead{Magnitude range\tablenotemark{c}} & \colhead{$\epsilon_{LyC}$\tablenotemark{d}} 
}
\startdata 
(i)   & LBG & $18.0^{+34.8}_{-7.4}$ & $M_{AB} \le -20.0$ &  $5.2^{+3.6}_{-3.4}$ \\
(ii)  & LAE & $3.7^{+2.5}_{-1.1}$    & $-20 < M_{AB} \le -18.3$ & $6.7^{+2.8}_{-2.7}$ \\
(iii) & LBG & $3.7^{+2.5}_{-1.1}$    & $-20 < M_{AB} \le -18.3$ & $27.0^{+11.4}_{-10.9}$ \\
(iv) & LBG & $18.0^{+34.8}_{-7.4}$ & $M_{AB} \le -18.3$ & $10.7^{+7.5}_{-7.1}$ \\
 (v) & LAE & $3.7^{+2.5}_{-1.1}$    & $M_{AB} \le -18.3$ & $8.6^{+3.7}_{-3.5}$ \\
& Total (lum.-dep.)\tablenotemark{e} & \nodata & $M_{AB} \le -18.3$ & $32.2^{+12.0}_{-11.4}$ \\
& Total (LAE-dep.)\tablenotemark{f} & \nodata & $M_{AB} \le -18.3$ & $16.8^{+6.9}_{-6.5}$ 
\enddata
\vspace{0.1cm}
\tablenotetext{a}{\mbox{Luminosity} function parameters used in Equations~\ref{eqn:elyc}$-\ref{eqn:elyc3}$: LBG from \citet{reddy2008}; or LAE from \citet{ouchi2008} scaled to include LAEs having REW$\ge20$\AA.}
\tablenotetext{b}{Sample average flux-density ratio used in Equations~\ref{eqn:elyc}$-\ref{eqn:elyc3}$.}
\tablenotetext{c}{Magnitude range over which the first moment of the luminosity function is determined.  $M_{AB} = -20.0$ and $-18.3$ correspond to $0.46L^*_{LBG}$ and $0.1L^*_{LBG}$, respectively.}
\tablenotetext{d}{Comoving specific emissivity of ionizing radiation in units of $10^{24}$~ergs~s$^{-1}$~Hz$^{-1}$~Mpc$^{-3}$.}
\tablenotetext{e}{Totals for our luminosity-dependent $\eta$ model, determined by summing rows (i) and (iii).}
\tablenotetext{f}{Total for our LAE-dependent $\eta$ model, determined by summing $0.77 \times$ row (iv) and row (v).}
\label{t:5}
\end{deluxetable}

\begin{deluxetable}{lc}
\tablewidth{0pt} 
\tabletypesize{\scriptsize}
\tablecaption{Estimates of the intergalactic hydrogen photoionization rate at $z\simeq3.1$.}
\tablehead{
\colhead{Reference} & \colhead{$\Gamma$ ($\times 10^{-12}$s$^{-1}$)}
}
\startdata 
\citet{meiksin2004} & $0.88^{+0.14}_{-0.12}$ \\
\citet{bolton2007} &  $0.86^{+0.34}_{-0.26}$ \\
\citet{faucher2008} & $0.63 \pm 0.08$ \\
This work, luminosity-dependent $\eta$ & $2.7^{+1.0}_{-0.9}$ \\
This work, LAE-dependent $\eta$ & $1.4^{+0.6}_{-0.6}$
\enddata
\vspace{0.1cm}
\label{t:6}
\end{deluxetable}

\begin{deluxetable}{cccccccccc}
\tablewidth{0pt} 
\tabletypesize{\scriptsize}
\tablecaption{Model dependent inferred values for $f_{esc}^{LyC}$.}
\tablehead{
\colhead{} & \colhead{} & \colhead{} &  \multicolumn{3}{c}{BC03 \tablenotemark{a}} & \colhead{} & \multicolumn{3}{c}{BPASS \tablenotemark{b}} \\
\cline{4-6} \cline{8-10}\\
\colhead{Age (yr) \tablenotemark{c}} & \colhead{$Z$ \tablenotemark{d}} &  \colhead{} & \colhead{$\eta^{stars}$ \tablenotemark{e}} & \colhead{$f_{esc}^{LyC\mathrm{, LBG}}$ \tablenotemark{f}} & \colhead{$f_{esc}^{LyC\mathrm{, LAE}}$ \tablenotemark{g}} &  \colhead{} & \colhead{$\eta^{stars}$ \tablenotemark{e}} & \colhead{$f_{esc}^{LyC\mathrm{, LBG}}$ \tablenotemark{f}} & \colhead{$f_{esc}^{LyC\mathrm{, LAE}}$ \tablenotemark{g}}  
}
\startdata 
&&&&&&&&&\\
\multicolumn{10}{c}{$f_{esc}^{UV} = 0.2$} \\
&&&&&&&&&\\
$10^6$  & 0.004    & & 1.98       &  0.02 & 0.11 & & 1.33      & 0.01 & 0.07 \\
\nodata  & 0.020    & & 1.90       & 0.02  & 0.10 & & 1.45      & 0.02 & 0.08 \\
$10^7$  & 0.004    & & 3.59       & 0.04 & 0.19 & & 2.10      & 0.02 & 0.11 \\
\nodata  & 0.020    & & 4.20       & 0.05 & 0.23 & & 2.70       & 0.03 & 0.15 \\
$10^8$  & 0.004    & & 6.17       & 0.07 & 0.33 & & 3.16      & 0.04 & 0.17 \\
\nodata  & 0.020    & & 6.38      & 0.07 & 0.34 & & 4.43       & 0.05 & 0.24 \\
\hline
&&&&&&&&&\\
\multicolumn{10}{c}{$f_{esc}^{UV} = 0.3$} \\
&&&&&&&&&\\
$10^6$  & 0.004    & & 1.98       &  0.03 & 0.16 & & 1.33      & 0.02 & 0.11 \\
\nodata  & 0.020    & & 1.90       & 0.03  & 0.15 & & 1.45      & 0.02 & 0.12 \\
$10^7$  & 0.004    & & 3.59       & 0.06 & 0.29 & & 2.10      & 0.04 & 0.17 \\
\nodata  & 0.020    & & 4.20       & 0.07 & 0.34 & & 2.70       & 0.05 & 0.22 \\
$10^8$  & 0.004    & & 6.17       & 0.10 & 0.50 & & 3.16      & 0.05 & 0.26 \\
\nodata  & 0.020    & & 6.38      & 0.11 & 0.52 & & 4.43       & 0.07 & 0.36 \\
\hline
&&&&&&&&&\\
\multicolumn{10}{c}{$f_{esc}^{UV} = 1.0$} \\
&&&&&&&&&\\
$10^6$  & 0.004    & & 1.98       &  0.11 & 0.54 & & 1.33      & 0.07 & 0.36 \\
\nodata  & 0.020    & & 1.90       & 0.11  & 0.97 & & 1.45      & 0.08 & 0.39 \\
$10^7$  & 0.004    & & 3.59       & 0.20 & $>1$ & & 2.10      & 0.12 & 0.57 \\
\nodata  & 0.020    & & 4.20       & 0.23 & $>1$ & & 2.70       & 0.15 & 0.73 \\
$10^8$  & 0.004    & & 6.17       & 0.34 & $>1$ & & 3.16      & 0.18 & 0.85 \\
\nodata  & 0.020    & & 6.38      & 0.35 & $>1$ & & 4.43       & 0.25 & $>1$ 
\enddata
\vspace{0.1cm}
\tablenotetext{a}{Stellar population synthesis models of \citet{bc2003}.}
\tablenotetext{b}{Stellar population synthesis models of \citet{eldridge2009}.}
\tablenotetext{c}{Time since turn on of constant star formation.}
\tablenotetext{d}{Metallicity of stellar population.}
\tablenotetext{e}{Ratio of intrinsic $F_{1600}$ to $F_{\mathrm{NB3640}}$ predicted by model.}
\tablenotetext{f}{Resulting LyC escape fraction for the given UV escape fraction, using $\eta_{LBG}$.}
\tablenotetext{g}{Resulting LyC escape fraction for the given UV escape fraction, using $\eta_{LAE}$.}
\label{t:7}
\end{deluxetable}

\end{document}